\documentclass[lettersize,journal]{IEEEtran}
\usepackage{amsmath,amsfonts} 

\usepackage{graphicx} 
\usepackage{subfigure}
\usepackage{xcolor}

\usepackage{array}
\usepackage[caption=false,font=normalsize,labelfont=sf,textfont=sf]{subfig}
\usepackage{textcomp}
\usepackage{stfloats}
\usepackage{url}
\usepackage{verbatim}
\usepackage{graphicx}
\usepackage{cite}
\usepackage{amsfonts,amsthm,amssymb}
\usepackage{mathrsfs,amsmath,arydshln,latexsym}
\usepackage{algpseudocode}
\usepackage{fancyhdr}  %
\usepackage{cases}
\usepackage{extarrows}
\usepackage{algorithm}
\usepackage{esint} 
\usepackage{tcolorbox}
\tcbuselibrary{breakable}
\tcbuselibrary{most}
\usepackage{cuted}
\usepackage{multicol}
\usepackage{makecell}
\usepackage{multirow}

\usepackage{nomencl}
\makenomenclature

\begin{document}

\title{ Electromagnetic Information Theory for Holographic MIMO Communications}

\author{Li Wei, Tierui Gong,~\IEEEmembership{Member,~IEEE,} Chongwen Huang,~\IEEEmembership{Member,~IEEE,} Zhaoyang Zhang,~\IEEEmembership{Senior Member,~IEEE,}  Wei E. I. Sha,~\IEEEmembership{Senior Member,~IEEE,}  Zhi Ning Chen, ~\IEEEmembership{Fellow,~IEEE,} Linglong Dai,~\IEEEmembership{Fellow,~IEEE,} M\'{e}rouane~Debbah,~\IEEEmembership{Fellow,~IEEE} and Chau~Yuen,~\IEEEmembership{Fellow,~IEEE}
		
		\thanks{L. Wei, T. Gong, and C. Yuen are with School of Electrical and Electronics Engineering, Nanyang Technological University, Singapore 639798 (e-mails: l\_wei@ntu.edu.sg, trgTerry1113@gmail.com, chau.yuen@ntu.edu.sg).}
		
		\thanks{C.~Huang, Z.~Zhang, and Wei E. I. Sha are with College of Information Science and Electronic Engineering, Zhejiang University, Hangzhou 310027, China (e-mails: \{chongwenhuang,  ning\_ming, weisha\}@zju.edu.cn).}
 
        \thanks{Zhi Ning Chen is with the Department of Electrical and Computer Engineering, National University of Singapore, Singapore 117583 (e-mail: eleczn@nus.edu.sg).}
        
        \thanks{L. Dai is with the Beijing National Research Center for Information Science and Technology and the Department of Electronic Engineering, Tsinghua University, Beijing 100084, China (e-mail: daill@tsinghua.edu.cn).}
		
		\thanks{M. Debbah is with KU 6G Research Center, Khalifa University of Science and Technology, P O Box 127788, Abu Dhabi, UAE (email: merouane.debbah@ku.ac.ae) and also with CentraleSupelec, University Paris-Saclay, 91192 Gif-sur-Yvette, France.}

	}


\maketitle

\begin{abstract} 
Holographic multiple-input multiple-output (HMIMO) utilizes a compact antenna array to form a nearly continuous aperture, thereby enhancing higher capacity and more flexible configurations compared with conventional MIMO systems,  making it attractive in current scientific research. Key questions naturally arise regarding the potential of HMIMO to surpass Shannon's theoretical limits and how far its capabilities can be extended. However, the traditional Shannon information theory falls short in addressing these inquiries because it only focuses on the information itself while neglecting the underlying carrier, electromagnetic (EM) waves, and environmental interactions.  To fill up the gap between the theoretical analysis and the practical application for HMIMO systems, we introduce electromagnetic information theory (EIT)  in this paper.  This paper begins by laying the foundation for HMIMO-oriented EIT, encompassing EM wave equations and communication regions. In the context of HMIMO systems, we present the resultant physical limitations, along with an overview of superdirective HMIMO and wideband MIMO systems. Field sampling and HMIMO-assisted oversampling are also discussed to guide the optimal HMIMO design within the EIT framework. To comprehensively depict the EM-compliant propagation process, we present the approximate and exact channel modeling approaches in near-/far-field zones. Furthermore, we discuss both traditional Shannon's information theory, employing the probabilistic method, and Kolmogorov information theory, utilizing the functional analysis, for HMIMO-oriented EIT systems. Our paper concludes by offering insights into future directions for HMIMO systems, covering the investigation of the mechanism of EM environment control and interactions, design and implementation of 3D superdirective HMIMO systems, efficient field sampling approaches, impacts of scatters and EM noise, accurate capacity region evaluation, and excitation/field encoding and modulation. Through these advancements, HMIMO systems are poised to revolutionize wireless communications with enhanced performance and adaptability.

\end{abstract}

\nomenclature{HMIMO}{Holographic multiple-input multiple-output}
\nomenclature{EM}{Electromagnetic}
\nomenclature{EIT}{Electromagnetic information theory }
\nomenclature{MIMO}{Multiple-input multiple-output }
\nomenclature{SIT}{Shannon’s information theory}
\nomenclature{KIT}{Kolmogorov’s information theory }
\nomenclature{TE}{Transverse
electric}
\nomenclature{TM}{Transverse magnetic }
\nomenclature{LoS}{Line-of-sight }
\nomenclature{NLoS}{Non-line-of-sight }
\nomenclature{SNR }{Signal-to-noise ratio }
\nomenclature{RF }{Radio-frequency }
\nomenclature{DOF}{Degrees-of-freedom }
\nomenclature{ULA}{Uniform linear array }
\nomenclature{UCA}{Uniform circular array }
\nomenclature{FTN}{Faster-than-Nyquist }
\nomenclature{MMSE }{Minimum mean square error}
\nomenclature{ ZF}{Zero-forcing }
\nomenclature{MRT}{Maximum ratio transmission } 
 \printnomenclature

\color{black}

\section{Introduction} 
The combination of electromagnetic (EM) theory and information content gradually becomes a hot topic for emerging technologies in wireless communications. 
Although EM wave interactions are well-understood in the EM community, the integration of EM basics in wireless communications is still in the early stages. In the majority of theoretical analyses in wireless communications, multiple assumptions are adopted to relieve the analytical and computational burdens, which may be far from the practical bound in applications. Such a situation has become even worse in recent years, as emerging technologies occur with the development of fabrication and low-profile materials, the assumptions valid in traditional communications may no longer hold, especially for compact antenna arrays and near-field communications. Therefore, exploring the fundamental limits of such scenarios forces us to go back and review the EM basics. By inspecting the physical limitations, e.g., quality factor and effective radiation gain, which are typically overlooked in performance analysis in wireless communications, a theoretical framework with a trade-off between numerical complexity and performance evaluation is anticipated to be provided.  
\subsection{Development of Holographic MIMO Systems}
\begin{figure}  
	\begin{center}
		{\includegraphics[width=.5\textwidth]{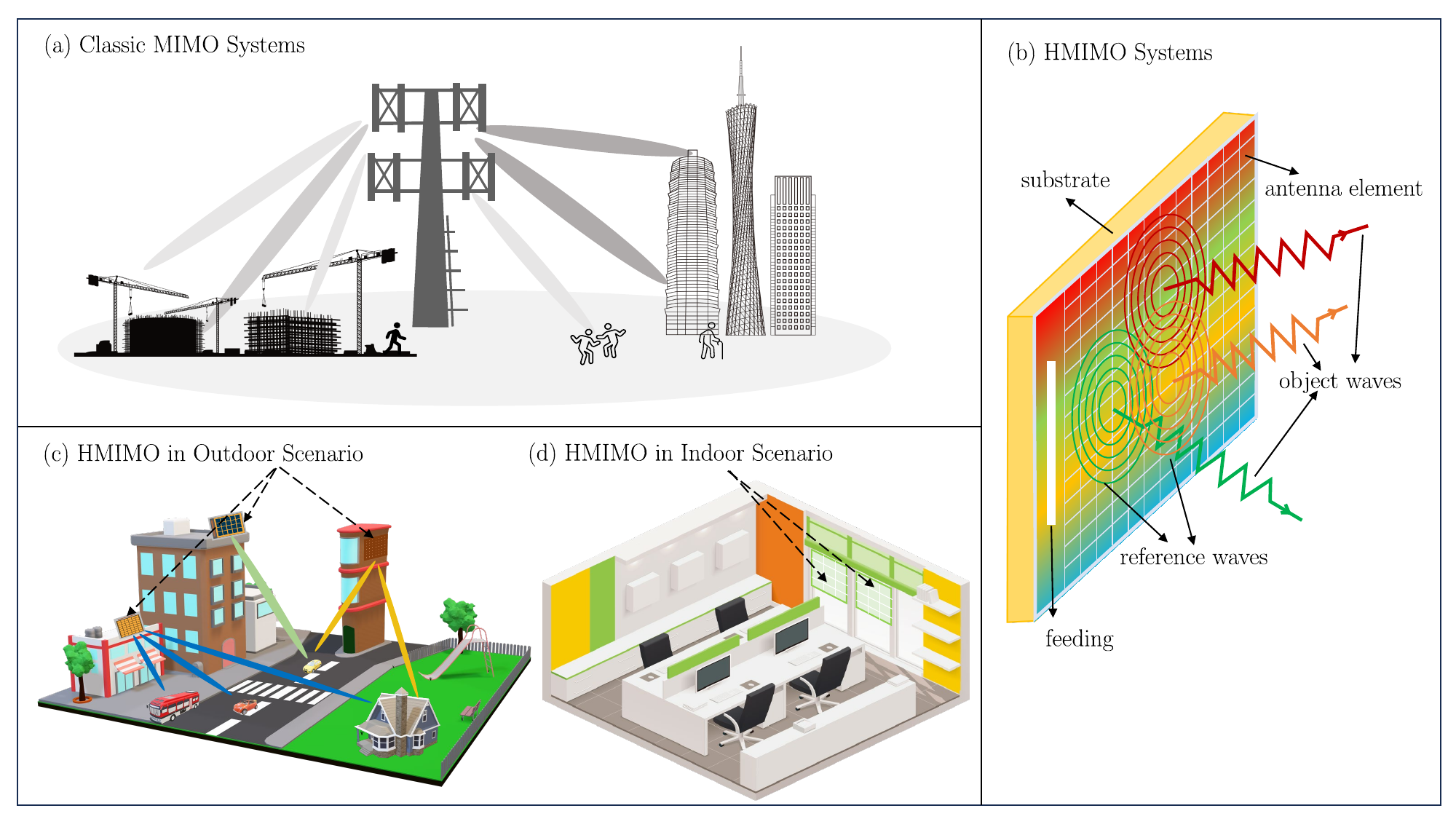}}  
		\caption{Depiction of MIMO and HMIMO (a) classic MIMO systems (b) HMIMO systems (c) HMIMO in outdoor scenarios, and (d) HMIMO in indoor scenarios.}  
		\label{fig:HMIMO} 
	\end{center}
\end{figure}

The traditional multiple-input multiple-output (MIMO) systems enable simultaneous and independent data transmission to serve a larger number of users, which is achieved by spatially separating multiple antennas with spacing larger than half of the wavelength to ensure channel independence, as shown in  Fig.~\ref{fig:HMIMO} (a). Since the number of mobile devices is growing dramatically and the communication scenarios have diversified these years, the traditional MIMO systems remain mainly two problems: one is the impracticality of accommodating multiple-antenna arrays in a constrained region, e.g., portable devices, and the other one is the high cost of achieving full coverage. To deal with the former problem, compact antenna arrays with few low-profile antennas (e.g., two or four antennas) are proposed to be housed in small terminals \cite{4447372,4012412,7953661}. To further support full coverage, a larger array size with more compact antenna elements is proposed, i.e., tightly coupled antenna arrays \cite{7120088,8605378}. However, the serious coupling effects and reduced radiation efficiency are unavoidable in compact antenna arrays. Therefore, some methods (e.g., employing decorrelation techniques, involving matching networks) are proposed to preserve MIMO channel capacity in compact antenna arrays. 

With the development of fabrication techniques and novel metamaterials, the concept of incorporating a larger number of low-cost and small-size antenna elements made of metamaterials in constrained areas is proposed, which is termed holographic multiple-input multiple-output (HMIMO) systems \cite{HuangHolographicMIMOSurfaces2020,10049818,Gong2023Holographic,10515204,10158690}. It should be noted that the HMIMO systems are in the conceptual stage for wireless communications, where the high-efficiency metamaterials-based HMIMO concept is still assumed in ideal conditions (e.g., perfect radiation energy efficiency).  The antenna aperture utilized in HMIMO systems comprises a multitude of electrically small antennas within a limited aperture area, thus forming a nearly continuous antenna surface. Furthermore, each antenna element is designed to be powerful in controlling EM waves with various expected responses. Such design modes facilitate HMIMO to manipulate EM waves at an unprecedented level, thereby possibly offering higher throughput than conventional MIMO systems for communication applications and potentially spurring a variety of non-communication applications, such as sensing, positioning, and imaging.

As shown in Fig.~\ref{fig:HMIMO} (b), HMIMO is mainly composed of radiators on the substrate fed by transmission lines, where the radiators are made of  low-power consumption metamaterials \cite{10049818}. By intelligently controlling the current distribution of all elements, the desired radiation patterns embedded in object waves are then generated. Compared with traditional MIMO systems, the fabrication cost and occupied space of antenna elements in HMIMO systems are much less. At the same time,  compared with the classic compact antenna arrays, HMIMO theoretically consists of dense antenna elements and can be implemented in arbitrary constrained sizes. Owing to these merits, HMIMO systems can be easily deployed both outdoors and indoors, as shown in Fig.~\ref{fig:HMIMO} (c)-(d), thus supporting more users in a seamless behavior. 
%To some degree, it is reasonable to regard HMIMO as the equivalent compact antenna arrays/tightly coupled arrays, but with some ideal assumptions in antenna configurations. 

However, there is a misalignment between the HMIMO in antenna engineering and wireless communications, yielding a mismatch between the applications and theories. For example, the antenna-dependent parameters (e.g., impedance matching bandwidth and radiation efficiency) are typically concerned in antenna engineering while the environment-dependent parameters (e.g., channel capacity and spectral efficiency) are mainly focused in wireless communications.  Since the performance of wireless communications is closely relevant to both transceivers and propagation environment, and the desirable parameters in wireless communications are different from those in antenna design, the physics-dependent information theory is necessary. The above concerns encourage us to review the basic theories in both the EM field and wireless communications, fusing these two fields in an interdisciplinary electromagnetic information theory (EIT).

\subsection{Electromagnetic Information Theory}
The superior capability of HMIMO systems in EM wave management is attractive for future communications, however, the conventional communication theories, established based on Shannon's information theory (SIT) \cite{6773024}, may fail to guide the system design, signal processing and characterize the fundamental limitation ascribed to the ignorance of physical effects of EM wave propagation. Furthermore, the emergence of near-field HMIMO communications, fueled by its substantial capacity improvements,  necessitates an effective performance analysis tool.  In short, an interdisciplinary framework incorporating wave- and information-theories, i.e., EIT, is urgently needed for HMIMO communications.

% Francia studied the informational content of the finite size antennas \cite{FranciaDirectivitySuperGainInformation1956}, 

The concept of EIT can be traced back to the 1940s  \cite{gabor1946theory, GaborCommunicationTheoryPhysics1953} when Gabor first indicated that communications are interpretations of physical effects. Shannon then proposed probabilistic models for communications in \cite{6773024} and Kolmogorov introduced functional sets for information measurement in   \cite{kolmogorov1959varepsilon}. Later, in the 1980s, Bucci and Franceschetti shaped the concept of spatial bandwidth and explored the degrees-of-freedom (DOF) of scattered fields   \cite{1144024,29386}, which were then extended to the arbitrary surfaces in the 1990s \cite{BucciRepresentationEM1998}. Subsequently, in the 2000s, Miller et al. investigated the DOF and power coupling strengths for optical systems \cite{miller2000communicating,piestun2000electromagnetic}; Poon et al. explored the DOF in multiantenna channels  \cite{1386525}; Migliore examined the DOF of the wave field in MIMO channels  \cite{MiglioreRoleNumberDegrees2006} and bridged electromagnetics and information theory via using the Kolmogorov approach  \cite{MiglioreElectromagneticsInformationTheory2008}; and Franceschetti et al. investigated information-theoretic limits of wireless communication problems in a deterministic structure \cite{FranceschettiCapacityWirelessNetworks2009}. In \cite{4685903}, the Shannon information capacity of space-time wireless channels is derived with physical constraints.   More works on the relationship between DOF of wireless networks and physics are discussed in \cite{FranceschettiWaveTheoryInformation2017}. Although a long history EIT possesses, the research interest is still growing in the 2020s as many authors were dedicated to this interdisciplinary discourse in multiple fields \cite{MiglioreShannonKolmogorovSpace2020,MiglioreWorldPhysicalLayer2021,YuanElectromagneticEffectiveDegree2022,WanMutualInformationElectromagnetic2023,JiExtraDoFNearField2023,10500751,vanwynsberghe2023universal,10012689,bjornson2024towards}.
 
Undoubtedly, EIT serves as an effective interdisciplinary framework to evaluate HMIMO communications with proper integration of information theory and EM wave theory \cite{MiglioreElectromagneticsInformationTheory2008,MiglioreShannonKolmogorovSpace2020,MiglioreWorldPhysicalLayer2021}. It models the communication systems by taking into account the physical effects of EM wave propagation, resulting in both probabilistic model and deterministic model, which is more physically consistent than the probabilistic only framework that is widely adopted in wireless communications. Particularly, EIT views the wireless channel as a continuous vector wave field excited by an impulse response, which is jointly described by time, frequency, space, polarization, and even orbital angular momentum with each dimension being capable of information-carrying. This interpretation of wireless channel enables EIT to capture the fully-dimensional information, and thus to assist communications for possibly providing higher multiplexing and diversity.   Overall, the emergence and development of EIT will potentially allow us to unveil the fundamental limits and perform system designs for HMIMO systems effectively and more realistically.

However, the involvement of EIT in HMIMO communications is still in its infancy. On the one hand, an integrated theoretical framework is still in development because the SIT and EM wave theory are normally investigated in two separate frameworks. On the other hand, the investigations on HMIMO communications encompass not only the propagation environment but also the interactions between antenna surfaces. As such, HMIMO-oriented EIT should consider all these aspects to offer a completed, physically consistent, and effective framework. 

To this aim, the physical limitations in HMIMO systems, especially those relevant to antenna designs, should be explored to shed more light on the interactions between EM waves and antenna surfaces.  The HMIMO-assisted oversampling techniques are also 
fascinating as an efficient method for improving spectral efficiency and capacity. Furthermore, the EM-compliant channel modeling in HMIMO systems, examining both the near-field and far-field characteristics, should be built to capture the essence of EM wave propagation, which becomes crucial for both theoretical analyses and system designs.    In the EIT framework, the two information measurements, i.e., Shannon capacity \cite{ShannonCommunicationPresenceNoise1949} and Kolmogorov $\epsilon$-capacity \cite{kolmogorov1959varepsilon}, stand out.  Shannon capacity is introduced as an effective measurement of information in the probabilistic model while Kolmogorov $\epsilon$-capacity evaluates the amount of information of signals with $\epsilon$ accuracy using functional analysis. Based on these two information measurements, the inherent capacity/DOF performance of HMIMO systems is discussed.

\begin{figure*}  
	\begin{center}
		{\includegraphics[width=.8\textwidth]{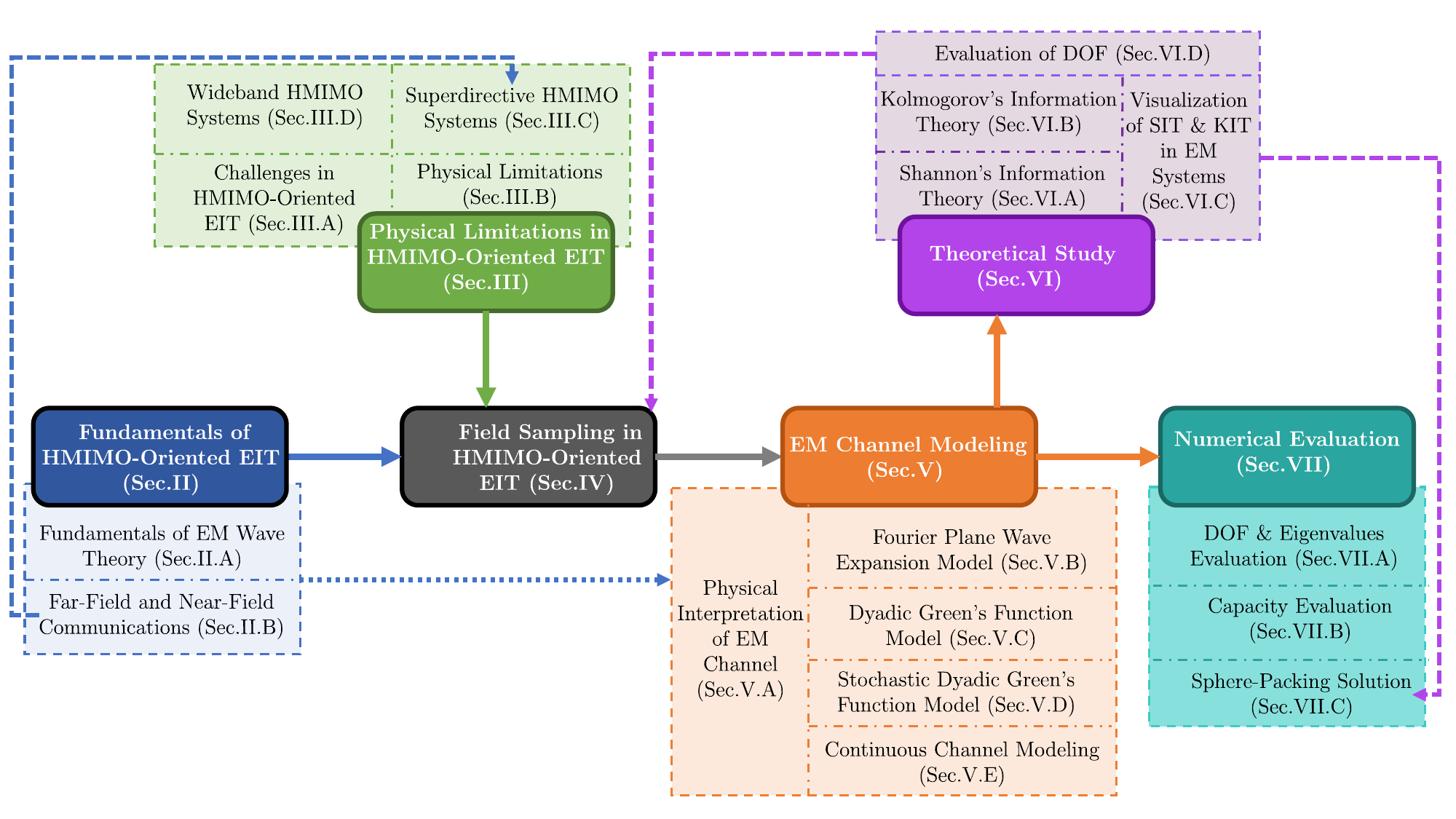}}  
		\caption{Main structure of the article.}  
		\label{fig:MainStructure_graph} 
	\end{center}
\end{figure*}

In this article, we present a comprehensive review of HMIMO-oriented EIT. The overall structure of the paper is illustrated in Fig.~\ref{fig:MainStructure_graph}.  The remaining sections of this article are organized as follows. In the ``Fundamentals of HMIMO-Oriented EIT" section, we first review the basics of EM equations and two field assumptions. In the ``Physical Limitations in HMIMO-Oriented EIT" section, we introduce challenges and physical limitations in super-directive HMIMO antenna design and wideband HMIMO systems, aiming at providing EM interpretation for MIMO systems, such as the directivity gain, quality factor, and embedded efficiency.  In the ``Field Sampling in HMIMO-Oriented EIT'' section, we introduce field sampling and HMIMO-assisted oversampling techniques.  In the ``Electromagnetic Channel Modeling'' section, we present different types of EM-compliant models. In the ``Theoretical Study" section, the probabilistic SIT and deterministic Kolmogorov's information theory (KIT) are presented to measure the EM information context, and the connection between these two theories is also discussed.  In the ``Numerical Evaluation", DOF and capacity analysis using SIT and KIT frameworks are evaluated in simulations. Subsequently, we put forward some potential future research directions in the ``Future Research Directions" section.  Finally, we conclude the article in the ``Conclusions'' section. 
 
\section{Fundamentals of HMIMO-Oriented EIT}  
%\section{Basics of EM Waves and Fields}  
 
\label{sec:EMbasics}
In this section, the fundamentals of EM wave theory are presented for capturing channel characteristics in HMIMO systems. In addition, the far field and near field wave interactions are discussed along with the emerging wireless communications.

\subsection{Fundamentals of EM Wave Theory} \label{subsec:EM equations}
The information-carrying media employed in wireless communications are EM waves, which can be completely depicted by Maxwell's equations, formulated by James Clerk Maxwell in 1865 \cite{maxwell1865viii}. Maxwell's equations accurately describe the interplay between electric sources and the generated EM fields following the differential forms \cite{chew1999waves}. Based on these equations, Green’s function and stochastic Green’s function are derived to characterize the EM fields distribution generated by sources.

\subsubsection{Dyadic Green's Function} 
Considering a point source $\mathbf{J}=\overline{{\mathbf{I}}}\delta(\cdot)$, where $\overline{{\mathbf{I}}}=\hat{x}\hat{x}+\hat{y}\hat{y}+\hat{z}\hat{z}$, one can express the solution of the vector wave equation of a point source as the dyadic Green's function, denoted by $\overline{{\mathbf{G}}}_0(\mathbf{r}, \mathbf{r}')$. The dyadic Green's function is an operator, which linearly relates the vector wave distribution to the source excitation. In essence, it characterizes the propagation of EM waves in free space, which corresponds to a linear, time-invariant, and homogeneous medium, as a linear system.

Specifically, the radiated electric field $\mathbf{E}(\mathbf{r})$ at the location $\mathbf{r}\in\mathbb{R}^{3}$ in free-space, due to the current $\mathbf{J}(\mathbf{r}')$ generated at the source location $\mathbf{r}'\in\mathbb{R}^{3}$,  is given by \cite{YuanElectromagneticEffectiveDegree2022,5991926}
\begin{equation} \label{equ:em_in_out}
	\mathbf{E}(\mathbf{r})\triangleq  -j \omega \mu  \int_{S} d s^{\prime}  {\overline{ {\mathbf{G}}}_0}\left(\mathbf{r}, \mathbf{r}^{\prime}\right) \cdot \mathbf{J}\left(\mathbf{r}^{\prime}\right),
\end{equation} 
where $S$ denotes the surface of the HMIMO aperture. The energy radiated by the antenna is closely related to the field strength in \eqref{equ:em_in_out}, while the field is determined by the controllable current distribution and the kernel function $ \overline{ {\mathbf{G}}}_0 (\mathbf{r},\mathbf{r}')$. Therefore, it is natural to deduce that the current distribution can be optimized to maximize the energy radiation for a given direction, which is detailed in Sec.~\ref{sec:antennaDesign}.

The point-to-point dyadic Green's function in free space is defined as \cite{ArnoldusRepresentationnearfieldmiddlefield2001}  
\begin{equation} \label{equ:dyadicGreen}
	\begin{aligned}
		&\overline{ {\mathbf{G}}}_0\left(\mathbf{r}, \mathbf{r}^{\prime}\right)\triangleq\left[\overline{ {\mathbf{I}}}+\frac{\nabla \nabla}{k_0^{2}}\right] g\left(\mathbf{r}, \mathbf{r}^{\prime}\right) ,
	\end{aligned}
\end{equation}
where $\overline{ {\mathbf{I}}}$ is the identity matrix, $\nabla \nabla g (\cdot)$ denotes the second-order derivative of function $g(\cdot)$ with respect to its argument, $k_0$ is also obtained as $k_0 \triangleq \frac{2\pi}{\lambda}$, %and the unit vector $\vec{\mathbf{r}}$ denotes the direction between the source point and radiated field. 
and $g\left(\mathbf{r}, \mathbf{r}^{\prime}\right)$ is the scalar Green's function, given by \cite{ArnoldusRepresentationnearfieldmiddlefield2001}  
\begin{equation} \label{equ:scalarGreen}
	g\left(\mathbf{r}, \mathbf{r}^{\prime}\right)\triangleq\frac{e^{-j k_0\left|\mathbf{r}-\mathbf{r}^{\prime}\right|}}{4 \pi\left|\mathbf{r}-\mathbf{r}^{\prime}\right|}.
\end{equation}
It should be noted that in a homogeneous medium, the propagation between two points is invariant under translation, consequently, the dyadic (scalar) Green’s function only depends on the source-field distance, i.e.,  $|\mathbf{r}-\mathbf{r}'|$ \cite{DeRosnyTheoryElectromagneticTimeReversal2010}. 
 
\subsubsection{Stochastic Dyadic Green's Functions} \label{subsubsec:StochasticGreen}
To characterize a more complex EM wave propagation environment involving phenomena like reflection, scattering, and diffraction, a statistical representation of the EM wave is essential in deriving the stochastic Green's function. As summarized in \cite{FranceschettiWaveTheoryInformation2017}, the central limit theorem is adopted to sum over all paths of real and imaginary parts in channel response, then the Rayleigh-distributed magnitude captures the wave’s attenuation, and the uniform phase represents the wave’s phase shift \cite{FranceschettiWaveTheoryInformation2017}. Subsequently, a similar method is adopted in the eigenfunction expansion method to model the stochastic Green's function \cite{LinStochasticGreenFunction2020,LinVectorialPropertyStochastic2022,LinPredictingStatisticalWave2023,LinStochasticGreenFunction2018}.  Specifically, the stochastic dyadic Green's function, which characterizes vectorial electromagnetic fields in the presence of random scatters, is represented in terms of eigenvalues $k_i^2$ and eigenvectors $\boldsymbol{\Psi}_i$  as \cite{LinPredictingStatisticalWave2023}
\begin{equation}
 \overline{{\mathbf{G}}}_{\text{S}}\left(\mathbf{r}, \mathbf{r}^{\prime}\right)=\sum_i \frac{\boldsymbol{\Psi}_i\left(\mathbf{r}, k_i\right) \otimes \boldsymbol{\Psi}_i\left(\mathbf{r}^{\prime}, k_i\right)}{k^2-k_i^2-j \frac{k^2}{\tilde{Q}}},
\end{equation}
where $\otimes$ indicates an outer product, and $\tilde{Q}$ is cavity quality factor.  The eigenfunctions are given by
\begin{equation} \label{equ:GreenFuncDecom}
    \Psi_i\left(\mathbf{r}, k_i\right)=\Psi_i^\theta\left(\mathbf{r}, k_i\right) \hat{\theta}+\Psi_i^\phi\left(\mathbf{r}, k_i\right) \hat{\phi},
\end{equation}
with
\begin{equation} \label{equ:eigenfunction}
    \begin{array}{r}
\Psi_i^\theta\left(\mathbf{r}, k_i\right) \simeq  \sum_{n=1}^{\infty}\left[a_n \cos \psi_n \cos \left(k_i \hat{\mathbf{e}}_n \cdot \mathbf{r}+\beta_n\right)\right] \\
\Psi_i^\phi\left(\mathbf{r}, k_i\right) \simeq \sum_{n=1}^{\infty}\left[a_n \sin \psi_n \cos \left(k_i \hat{\mathbf{e}}_n \cdot \mathbf{r}+\beta_n\right)\right],
\end{array}
\end{equation}
with $a_n$ being the $n$-th plane wave's amplitude, and  $\psi_n,\phi_n,\theta_n$ being polarization angle, azimuth angle, and elevation angle of the $n$-th plane wave, respectively. $\hat{\mathbf{e}}_n$ is the direction of the $n$-th plane wave.  

The randomness lies in constructing eigenfunctions and eigenvalues. The eigenvalue (i.e., eigenfrequency spectrum) is modeled by random matrix theory, and the eigenfunctions are approximated by probabilistic plane waves, as shown in \eqref{equ:eigenfunction}. Based on these two random variables, the eigenfunction expansion in \eqref{equ:GreenFuncDecom} is mathematically represented by a probabilistic model using the central limit theorem.   

Additionally, the sum in \eqref{equ:GreenFuncDecom} is composed of two parts, one consists of many terms in which the denominator has the property of $k_i^2\not \approx k^2-j \frac{k^2}{\Tilde{Q}}$, and the other consists of terms with $k_i^2 \approx k^2-j \frac{k^2}{\Tilde{Q}}$. Through such operations, the stochastic Green's function can be written as the sum of two parts: incoherent propagation resulted from diffuse multipath components, and the coherent propagation accounted for the direct paths between the transmitter and receiver \cite{LinNovelStatisticalModel2019, LinStatisticalAnalysisSpaceTime2021}.

\subsection{Far-Field and Near-Field Communications} \label{subsec:far_near_field}

\begin{figure}  
	\begin{center}
		{\includegraphics[width=0.5 \textwidth]{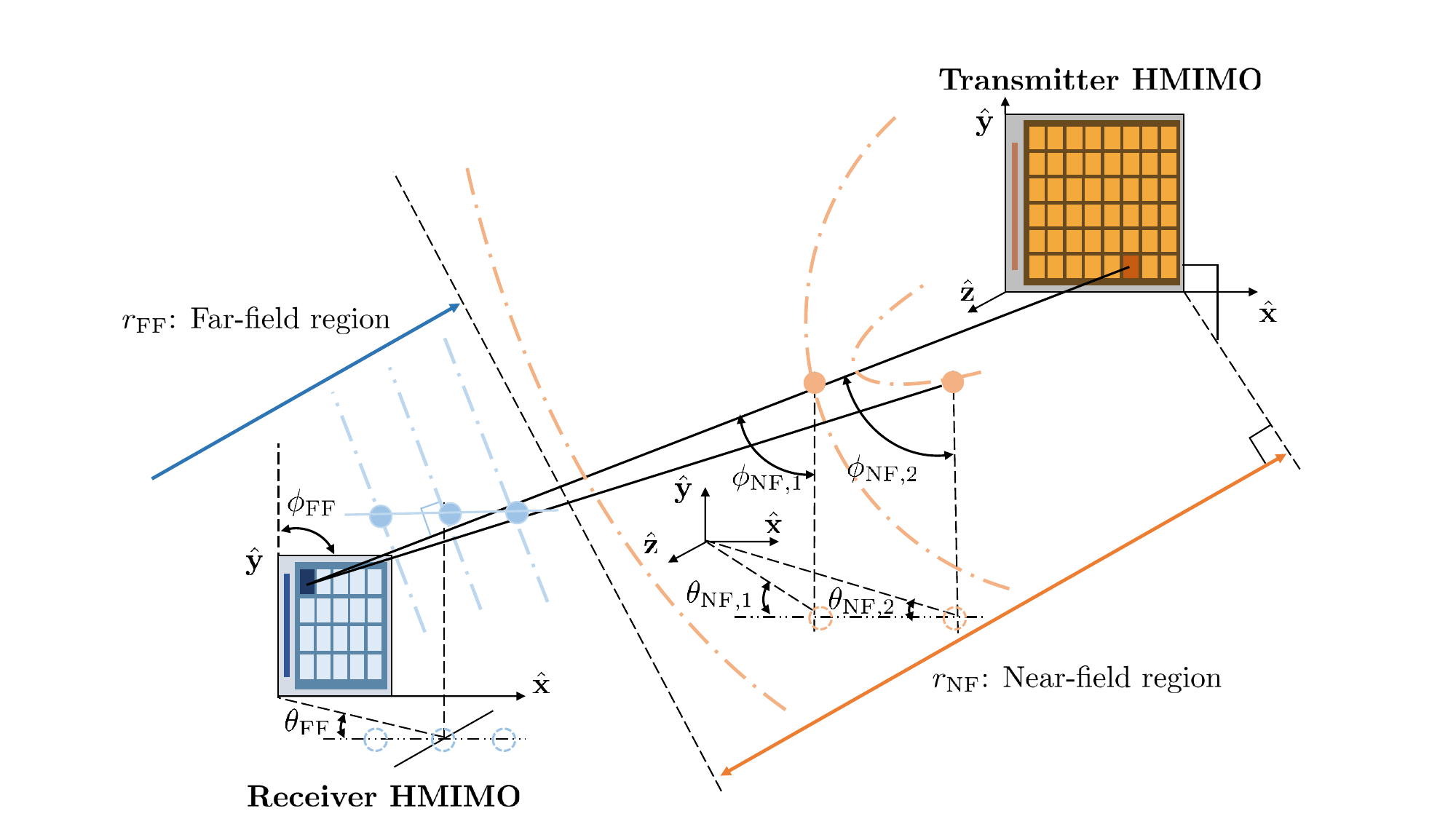}}  
		\caption{Planar wave and spherical wave model at the transmitter and receiver.}  
		\label{fig:PlaneSpherical_graph} 
	\end{center}
\end{figure}

In general, the communication region can be partitioned into two regions based on the Rayleigh distance, i.e., $r_{\mathrm{Rayleigh}} = \frac{2D^2}{\lambda}$ ($D$ is the antenna aperture). As illustrated in Fig.~\ref{fig:PlaneSpherical_graph}, the far field exists as the region beyond the Rayleigh distance, and conversely, the near field is found within the Rayleigh distance. The differences between far-field and near-field communications are presented below with a special focus on propagation wavefront, polarization diversity, and capacity.

\begin{itemize}
	\item{\textbf{Propagation Wavefront:}}
	In the far-field zone, the propagation wave exhibits the planar wavefront. For an HMIMO aperture with discrete antennas, the azimuth angle $\theta_{\mathrm{FF}}$ and the elevation angle $\phi_{\mathrm{FF}}$ are identical for all antennas. Conversely, within the near-field, the azimuth angle $\theta_{\mathrm{NF},i}$ and the elevation angle $\phi_{\mathrm{NF}, i}$ exhibit variations for each pair of transmit-receive antennas due to the spherical propagation wavefront \cite{Jeng-ShiannJiangSphericalwavemodelshortrange2005}.

\item{\textbf{Polarization Diversity:}} 
%The polarization domain can convey additional information bits, which is characterized by polarization diversity. 
The far-field and near-field waves exhibit distinct polarization diversity. Specifically, the fields may be in three orthogonal polarizations in near zones while there may be two orthogonally polarized components in the plane normal to the propagation direction in far-field zone. 

 %these three components are all prominent, resulting in the tri-polarization of EM waves. However, as the distance increases, the parallel component diminishes rapidly due to exponential decay, while the two orthogonal components decay at a relatively slower rate, ultimately leading to the dual-polarization of EM waves in the far-field region.  

\item{\textbf{Capacity:}}   
%The capacity gain provided by far-field and near-field communications is totally different.  
Near-field communications are able to accommodate substantially higher capacity demands compared to far-field communications,  as extensively demonstrated in \cite{Bohagensphericalvsplane2009,TarboushTeraMIMOChannelSimulator2021}. The capacity enhancement in the near-field region can be attributed to more information carried by extra DOF introduced due to the spherical wavefront and the extra polarization diversity. To elaborate further, near-field communications facilitate the support of more comprehensive information, encompassing both phase and distance information across different polarizations.  

\end{itemize}

Therefore, the near-field communications efficiently increase available spatial DOFs, strengthening information transfer in wireless communications, and making them appealing in some scenarios, especially for the massive MIMO technology. For example,  a $3200$-element extremely large-scale antenna array at $2.4$GHz in an array size of $2$m $\times 3$m generates nearly $200$m Rayleigh distance, which is larger than the radius of a typical 5G cell \cite{9903389}. With proper design, various communication scenarios will benefit from exploiting near-field effects, such as multi-user communications, accurate localization and focused sensing, and wireless power transfer with minimal energy pollution \cite{10068140}. Take near-field beam focusing as an example, both distance and phase information of spherical wavefront are exploited to focus the radiated energy in a specific spatial location, facilitating more efficient interference control and capacity-approaching near-field communications.

In brief, the near-field effects benefit information improvement due to the excited higher-order modes in evanescent waves. These higher-order modes' power exhibits rapid drop-off in the free space, therefore, the heavy-tail behavior of dominant spatial modes still occurs in the near field. Such a behavior can be explained in two ways, one from the perspective of real power combination carried by orthogonal modes on the observation surface \cite{5976389}, and the other is based on the phase transition behavior of singular values of the spatial band-limited field \cite{7101845}. Both methods yield the same conclusion in near-field effects on information gain.

Since the near-field information and the far-field information are correlated, it is natural to derive the relationship between these two fields \cite{LudwigNearfieldfarfieldtransformations1971,BucciNearfieldfarfieldtransformationspherical2003,OmiNearFieldFarFieldTransformation2016}. The work \cite{LudwigNearfieldfarfieldtransformations1971} employed the spherical wave expansion to express arbitrary fields via computing the near-field patterns by the far-field data, showing a remarkable numerical accuracy.  The work \cite{OmiNearFieldFarFieldTransformation2016} adopted the planar wave expansion iteratively during the near-field and far-field transformation process to reduce the number of unknowns and computational complexity. These works offer new insights into the reconstruction of a near-field region from far-field data, prompting the need for further explorations and investigations.

\color{black}

\section{Physical Limitations in HMIMO-Oriented EIT} \label{sec:antennaDesign}
\begin{figure}   [htp!]
	\begin{center}
		{\includegraphics[width=0.5 \textwidth]{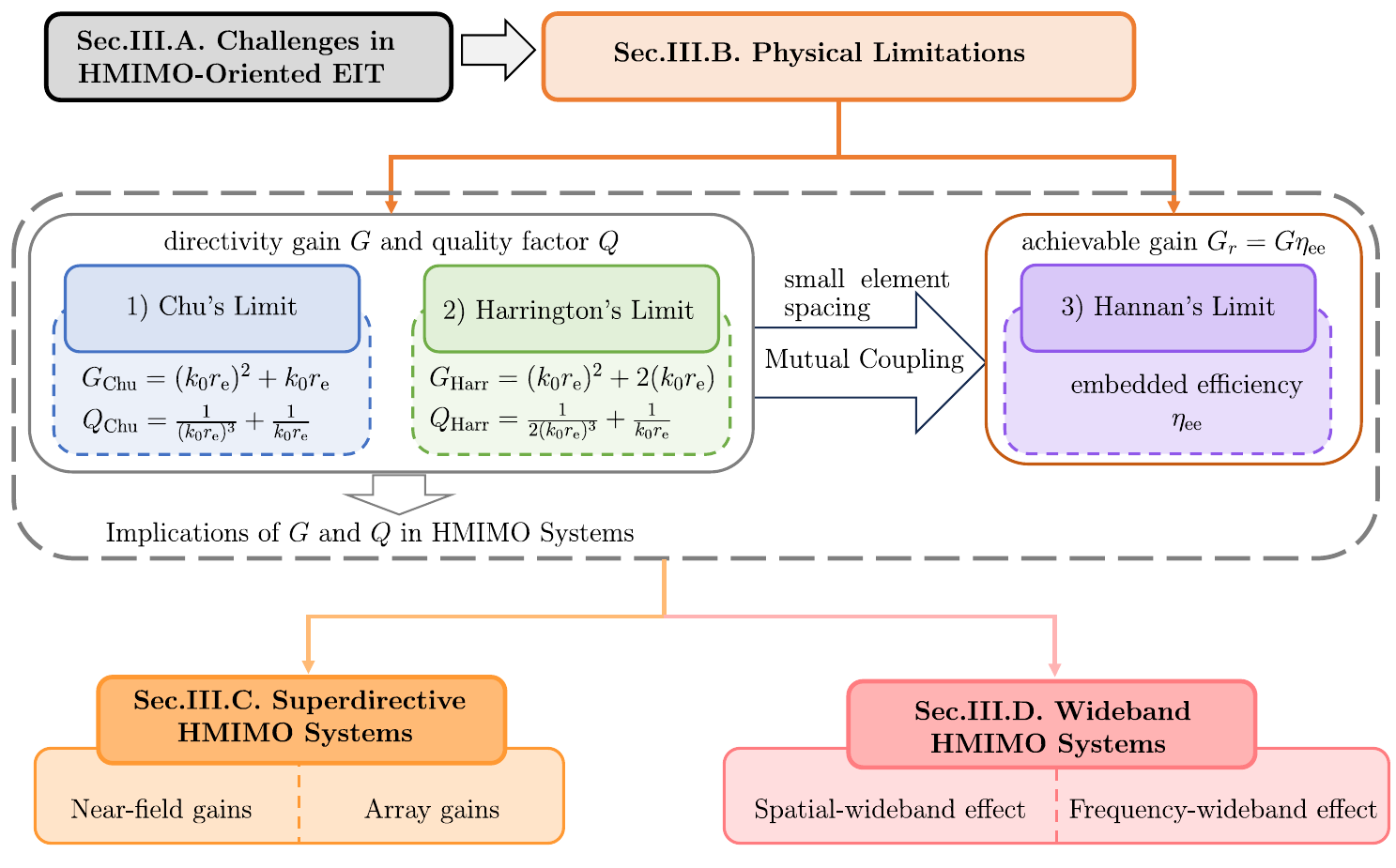}}  
		\caption{Framework of Sec.~\ref{sec:antennaDesign}.}  
		\label{fig:PhyLimi_graph} 
	\end{center}
\end{figure}

The antenna design of HMIMO systems imposes restrictions on intrinsic performance, such as the array gain and bandwidth. Therefore,  it becomes essential to explore the correlation between the physical antenna configuration and the underlying system's performance. For this purpose,  we present challenges and inherent physical limitations associated with the maximum directivity gain $G$,  the minimum $Q$ factor (approximately inversely proportional to bandwidth), and the achievable gain $G_r$ affected by energy efficiency $\eta_{\mathrm{ee}}$. Additionally, the superdirective HMIMO (mainly exploits the spatial DOF) and wideband HMIMO systems (primarily utilizes the spectral DOF) are discussed. The main structure of this section is depicted in Fig.~\ref{fig:PhyLimi_graph} for ease of reading. 

\subsection{Challenges in HMIMO-Oriented EIT} \label{subsec:challenges}
The following critical challenges arise in applying compact HMIMO systems:
\subsubsection{Frequency Sensitivity to High $Q$ Value}
In HMIMO systems, the excitation of higher-order radiation modes increases the quality factor $Q$ and results in higher frequency sensitivity \cite{GustafssonOptimalAntennaCurrents2013}. The quality factor $Q$ is commonly utilized to assess the inherent bandwidth and is mathematically approximated as inversely proportional to bandwidth, i.e.,  $B_n =\frac{f_h-h_l}{f_0} = \frac{1}{Q}$, where $f_h, f_l$ denote the upper and lower frequency, respectively, and center frequency $f_0=\frac{f_h+ f_l}{2}$. Therefore, a larger $Q$ corresponds to a narrower bandwidth or higher operating frequency, indicating a good fitting between the fixed resonant circuit to the antenna. Conversely, a lower $Q$ signifies a slow variation of input impedance with frequency, leading to a broader bandwidth for the antenna.  
 
\subsubsection{Reduced Energy Efficiency}
The energy efficiency $\eta_{\mathrm{ee}}$ quantifies the antenna's effectiveness in converting incoming electrical energy (denoted by directivity gain $G$) into radiated EM energy (denoted by achievable gain $G_r$). Mathematically, $G_r=\eta_{\mathrm{ee}} G$. Notably, there exists a trade-off between the element efficiency $\eta_{\mathrm{ee}}$ and the gain $G$. For instance, reducing the spacing between elements, typically to values less than $\lambda/2$, can lead to higher $G$. However, such a configuration also results in a compromise of the element efficiency (a lower $\eta_{\mathrm{ee}}$). Therefore, optimizing the array design requires striking a balance between these two factors to realize the desired performance characteristics for the given application.

\subsubsection{Deformed Antenna Pattern in Far Field}
Ideally, the radiation pattern of HMIMO systems should be stable over the whole antenna array at different frequencies in order to minimize the distortion of the transmitted signal \cite{4162465}. Despite being capable of large-size densely spaced arrays, HMIMO systems exhibit distinct antenna patterns across these elements. Specifically, the antennas in the central area display similar radiation patterns, while the antennas at the edge demonstrate severely distorted antenna patterns due to the mutual coupling effects \cite{Hannanelementgainparadoxphasedarray1964,YuanEffectsMutualCoupling2023}. This pattern deformation in the antenna array becomes more pronounced as the array size reduces and the element density increases, where mutual coupling introduces substantial deviations in the radiation patterns, leading to non-uniform and distorted array responses.

Additionally, achieving optimal excitation that leads to a desired radiation pattern with minimal power consumption is challenging. The authors in \cite{1139958} incorporated superdirectivity ratio constraint and applied spheroidal functions to derive an aperture distribution that yields maximum directivity for a continuous line source. Despite the promising results, the practical feasibility of achieving an optimal excitation setting is still in its early stages of exploration.

Clearly, two main performance factors are concerned in the above challenges: the directivity gain $G$ and the quality factor $Q$, as will be detailed in the following subsection.

\subsection{Physical Limitations in HMIMO Systems}
The maximum directivity gain $G$ and minimum quality factor $Q$ are given by Chu's limit and Harrington's limit,  providing insights into HMIMO system applications. Furthermore, the achievable array gain, impacted by mutual coupling effects, is described by Hannan's limit. 
\subsubsection{Chu's Limit and Harrington's Limit}
Both  Chu's limit and Harrington's limit derive the maximum directivity gain and minimum quality factor based on excited spherical modes, with minor differences in excited modes. 
 
\begin{itemize}
	\item \textbf{Chu's limit} \cite{ChuPhysicalLimitationsOmni1948}: For a large number of dominant spherical waves (denoted its number as $N_{\mathrm{e}}$) that contribute to radiation, the directivity gain is $G=2N_{\mathrm{e}}/\pi$. However, it is impractical to obtain an arbitrarily high gain (i.e., activating a large number of spherical modes) since the corresponding current distribution is hard to be achieved within an arbitrarily small antenna. Consequently, the maximum directivity of an antenna enclosed within the sphere with radius $r_{\mathrm{e}}$ is \cite{KildalFoundationsantennaengineering2015}
	\begin{equation}
		G_{\mathrm{Chu}}= (k_0 r_{\mathrm{e}})^2 + k_0 r_{\mathrm{e}},
	\end{equation}
    which is accurate for $k_0 r_{\mathrm{e}}>3$.
	
     The minimum $Q$ is obtained if there is only one spherical mode is activated, which is given by \cite{KildalFoundationsantennaengineering2015}
	\begin{equation}
		Q_{\mathrm{Chu}}=\frac{1}{(k_0 r_{\mathrm{e}})^3}+\frac{1}{k_0 r_{\mathrm{e}}}.
	\end{equation}
          
         \item  \textbf{Harrington's Limit} \cite{Harringtongainbeamwidthdirectional1958}: Taking both the transverse electric (TE) mode and transverse magnetic (TM) mode, the normal gain is given by 
           \begin{equation} \label{equ:HarringtonLimit}
	         G_{\mathrm{Harr}}=(k_0 r_{\mathrm{e}})^2+2 (k_0 r_{\mathrm{e}}).
            \end{equation} 
            The study in \cite{PigeonMiniaturedirectiveantennas2014} further indicated that compact antenna arrays can attain a higher directivity gain compared to Harrington's limit $G_{\mathrm{Harr}}$ in \cite{Harringtongainbeamwidthdirectional1958}, implying the potential superdirectivity achieved by HMIMO systems.

            The minimum $Q$ is  \cite{HarringtonEffectantennasize1960,KramerFundamentalLimitsDesign2009}
        \begin{equation}
	       Q_{\mathrm{Harr} }= \frac{1}{2(k_0 r_{\mathrm{e}})^3}+\frac{1}{k_0 r_{\mathrm{e}}}.
        \end{equation} 
\end{itemize}

\begin{figure}  [htp!]
	\begin{center}
		{\includegraphics[width=0.45 \textwidth]{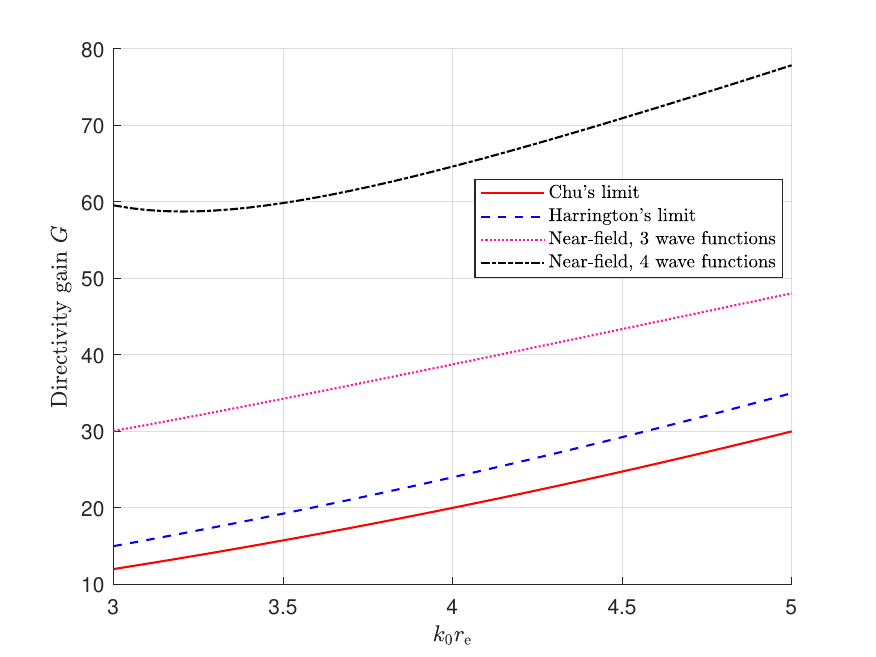}}  
		\caption{Maximum directivity gain $G$ for Chu's approach, Harrington's approach, and different number of spherical modes in near-field in \cite{HarringtonEffectantennasize1960}.}  
		\label{fig:maximumGain} 
	\end{center}
\end{figure}

The comparison of maximum directivity gain $G$ among Chu's limit, Harrington's limit, and different numbers of spherical modes in near-field \cite{HarringtonEffectantennasize1960} is given in Fig.~\ref{fig:maximumGain}.  It is noteworthy that the maximum directivity gains specified by Chu's limit and Harrington's limit are applicable in far-field regions. However, in the near-field region, the maximum directivity gain surpasses those obtained from Chu's and Harrington's limits, thus demonstrating the advantageous directivity gain in the near-field region compared to the far-field zone. 

\begin{figure}   [htp!]
	\begin{center}
		{\includegraphics[width=0.45 \textwidth]{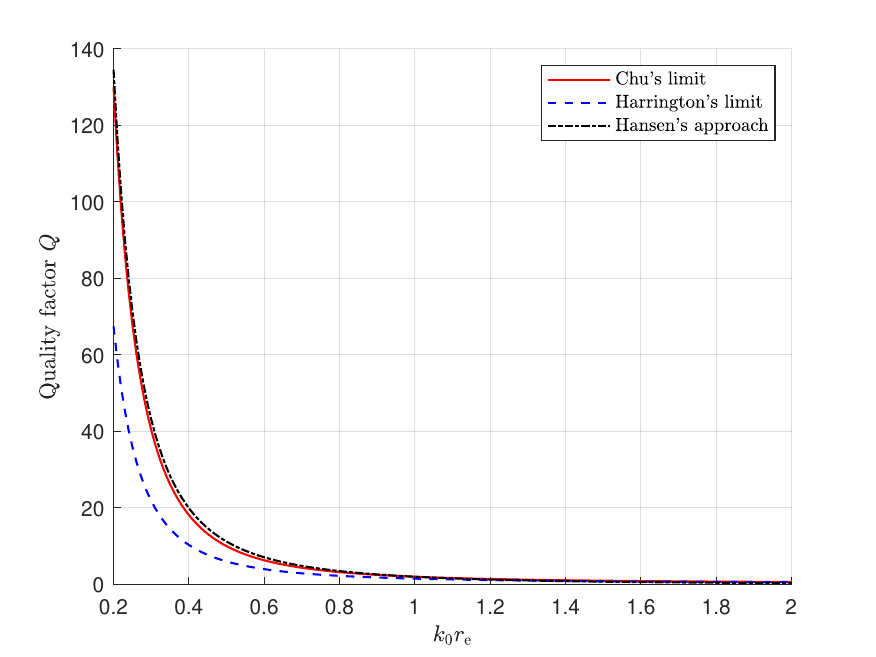}}  
		\caption{Minimum quality factor $Q$ for Chu's limit, Harrington's limit, and Hansen's approach.}  
		\label{fig:Qfactor} 
	\end{center}
\end{figure}

The comparison of quality factor $Q$ for Chu's limit, Harington's limit, and Hansen's approach \cite{HansenFundamentallimitationsantennas1981} (only the lowest TM mode with $k_0r_0<1$ is excited) are given in Fig.~\ref{fig:Qfactor}. Since either TE or TM mode is considered, $Q$ factor in Chu's limit is the highest, serving as a lower bound of practical $Q$ and corresponds to a narrower bandwidth $B_n$ \cite{McLeanreexaminationfundamentallimits1996}.

\subsubsection{Implications of $G$ and $Q$ in HMIMO Systems}
Clearly, the maximum directivity $G$ and minimum $Q$ cannot be achieved simultaneously. Therefore, their combined effects must be considered in HMIMO designs, particularly in practical scenarios where the antenna arrays exhibit directional, rather than ideal omnidirectional, beampatterns, necessitating the effective beam alignment scheme. For example, assuming a sectored gain pattern at both BS and user ends, the BS antenna gain in the foresight direction $\theta_b$ and user antenna gain in direction $\theta_u$ are given by \cite{8170329}
\begin{equation}\label{equ:TotalDirectivity}
    \begin{aligned}
        &G_b\left(\theta_b\right)= \begin{cases}G_b^{\max } & \text { if }\left|\theta_b\right| \leq B_b \\ G_b^{\min } & \text { otherwise }\end{cases}, \\
        & G_u\left(\theta_b\right)= \begin{cases}G_u^{\max } & \text { if }\left|\theta_b\right| \leq B_u \\ G_u^{\min } & \text { otherwise }\end{cases}, 
    \end{aligned}
\end{equation}
where $B_b$ and $B_u$ are the beamwidth of BS and user antenna array, respectively. 

Therefore, to achieve the maximal total directivity gain $G_b^{\max }G_u^{\max }$ in line-of-sight (LoS) scenarios,  the two angles, i.e., $\theta_b$ and $\theta_u$, must be well-aligned in the maximum gain direction. Similarly, in non-LoS (NLoS) scenarios, optimal beam alignment that maximizes directivity gain can be designed based on prior knowledge of the scattering environment, such as through the use of intelligent reflecting surfaces.   It should be noted that while the antenna gain described assumes a sectored pattern, practical beam patterns may exhibit multiple sidelobes and non-uniform gain distribution. As a result, interference from other users in multi-user scenarios is influenced by the directivity gain at undesired angles of arrival. This interference can be mitigated by minimizing sidelobes' power.
 
However, as shown in \eqref{equ:TotalDirectivity}, the maximum gain is only attainable within a specific beamwidth, and the beamwidth is approximately reversely to quality factor $Q$. Given the tradeoff between $G$ and $Q$, the system design for wireless communications should jointly optimize the directivity gain and $Q$, rather than focusing solely on optimizing $G$. Theoretically, there is a preferred $Q$ factor for optimal spectral efficiency, which varies with different antenna geometries and apertures \cite{EhrenborgCapacityBoundsDegrees2021}.  More importantly, while increasing the data rate for desired users can improve performance, it also exacerbates interference from other users, which must be carefully managed.

The good news is that implementing a specified directional antenna array can effectively mitigate interference leakage and significantly increase the signal-to-noise ratio (SNR) due to narrower main lobe beams. Such directional antenna arrays can be adopted in various scenarios, such as sub-6 GHz and sub-THz mobile access networks, to guarantee reasonable coverage at high frequencies and large bandwidths. The work \cite{10439223} demonstrated that little or no equalization is still attractive when adopting directive arrays in relieving the radio-frequency (RF) requirements, implying the hardware-efficient advantages of HMIMO systems over the traditional adaptive array processing that requires many RF chains.

The advantages of HMIMO systems that generate multiple narrow beams can be understood through spatial resolution and information transfer. Narrower beams offer a higher spatial resolution, determined by both the equivalent currents in the HMIMO system and the propagation environment. The maximum information transfer between the transmitter and receiver is limited by the number of distinguishable fields in the receiver space, which corresponds to the spatial DoF. In free-space, these distinguishable fields are linked to the beampatterns generated at the transmitter. If the HMIMO system can generate multiple non-overlapping beams, narrowing the beams increases the number of distinguishable field patterns, thereby enhancing information transfer, especially at the main lobe angles. Additionally, optimizing the current distribution or physical layout of the HMIMO system can boost spectral efficiency and reduce the need for extensive phase shifting in traditional beamforming. However, due to the finite size of the transmitter and the characteristics of the free-space channel, the spatial resolution of both the beampatterns and field patterns is inherently limited  (detailed in Sec. VI).

\subsubsection{Hannan's Limit}
Due to the unavoidable mutual coupling effects at small element spacings, the directivity gain $G$ is compromised and replaced by the achievable gain (also referred to as realized gain) in performance evaluation. The achievable gain $G_r$ is the combined impact of the directivity gain $G$ and the element efficiency $\eta_{\mathrm{ee}}$, as investigated in Hannan's limit \cite{Hannanelementgainparadoxphasedarray1964}. 

Specifically, the maximum directivity gain for an antenna is 
\begin{equation}
	\label{equ:element_gain_limit}
	G(\theta)=\frac{4\pi A}{\lambda^2}\cos \theta,
\end{equation}
where $A$ is the allotted area for each antenna element in the array, and $\theta$ is the scanning angle. The gain improvement brought by incremental antennas within fixed aperture is counteracted by closer spacing, resulting in stronger mutual coupling and a reduced allotted area $A$ for each antenna.  The achievable gain is then given by:
\begin{equation}\label{equ:realized_gain}
	G_{r}(\theta)=\frac{4 \pi A}{\lambda^2} \cos \theta \cdot \eta_{\mathrm{ee}},
\end{equation} 
where $\eta_{\mathrm{ee}}$ can be mathematically computed through two methods:  
\begin{itemize}
    \item Incorporating the reflection coefficient $R(\alpha,\beta)$ over the excitation phasings $\alpha$ and $\beta$, $\eta_{\mathrm{ee}}$ is given by \cite{Hannanelementgainparadoxphasedarray1964}:
\begin{equation}\label{equ:hannan_EmbededEfficiency}
	\eta_{\mathrm{ee}}=1-\frac{1}{\pi^2} \int_{\alpha=0}^\pi \int_{\beta=0}^\pi|R(\alpha, \beta)|^2 \mathrm{d} \alpha \mathrm{d} \beta,
\end{equation}
where $\eta_{\mathrm{ee}}=\pi/4$ for the ideal element in an infinite array with spacing $\Delta_s=\lambda/2$. Since the theoretical directivity gain $4$ reduces to the peak realized gain $\pi$ due to the reduced element efficiency and unavoidable mutual coupling \cite{Hannanelementgainparadoxphasedarray1964,kraus_antennas_1988}, the maximum element efficiency for a dense array (e.g., HMIMO systems) is given by \cite{KildalFundamentalDirectivityLimitations2016}
\begin{equation}
    [\eta_{\mathrm{ee}}]_{\max}=\frac{G_r(\theta)}{4}=\frac{\pi A}{\lambda^2} .
\end{equation}

\item Resorting to the scattering matrix $S$-parameter,  $\eta_{\mathrm{ee}}$ can be approximated by \cite{Pozarrelationactiveinput2003}  
\begin{equation}\label{equ:kildal_EmbededEfficiency}
	\eta_{\mathrm{ee} }=1-\sum |S_{ij}|^2,
\end{equation}
which is similar to \eqref{equ:hannan_EmbededEfficiency}.  
\end{itemize} 
Both $R(\alpha,\beta)$ in \eqref{equ:hannan_EmbededEfficiency}  and  $S_{ij}$ in \eqref{equ:kildal_EmbededEfficiency} reflect mutual coupling, $R(\alpha,\beta)$ is reflection coefficient that dependent on excitation phasings $(\alpha,\beta)$ while  $S_{ij}$ is the $S$-parameter between inactive $i$-th and active $j$-th ports.  Typically,  \eqref{equ:hannan_EmbededEfficiency} is an upper limit to \eqref{equ:kildal_EmbededEfficiency}, and \eqref{equ:kildal_EmbededEfficiency} is much more applicable than \eqref{equ:hannan_EmbededEfficiency} to the medium-size to small arrays \cite{KildalFundamentalDirectivityLimitations2016}.

So far, there are mainly two gains we should differentiate: one is maximum directivity gain $G(\theta)$ given by \eqref{equ:element_gain_limit}, it is the ideal case in an infinite antenna array; the other is the realized gain $G_r(\theta)$ given by \eqref{equ:realized_gain}, it takes losses (dissipation and impedance mismatch) into consideration.  In fact, the achievable gain in \eqref{equ:realized_gain} is more significant than the directivity gain given in \eqref{equ:element_gain_limit} since it considers the interaction between different antennas.

However, the loss in realized gain is undesirable, prompting researchers to seek feasible solutions to surpass Hannan’s limit. The work \cite{10542433} experimentally proved that the efficiencies in the 3D array are higher than that in 2D array due to the enlarged projection area, which breaks Hannan's efficiency limit by introducing the height difference.  Inspired by this finding, novel array structures are anticipated to enhance performance and potentially break the performance limit of traditional wireless communications. 

It is important to note that different array structures may respond distinctly to the spatial-wideband effects (detailed in the following sections), which should be properly addressed. For example, in uniform linear array (ULA) systems,  spatial-wideband effects lead to beam squint effects or even more significant split effects \cite{10399817}, while uniform circular array (UCA) systems experience beam defocus, where the beam pattern is severely distorted at non-central frequencies. With appropriate phase compensation techniques, UCA can provide a wider scan range and uniform beampattern along the entire azimuth plane compared to ULA, suggesting the potential of new array structures in improving performance.

\subsection{Superdirective HMIMO}
\begin{figure}  
	\begin{center}
		{\includegraphics[width=0.5 \textwidth]{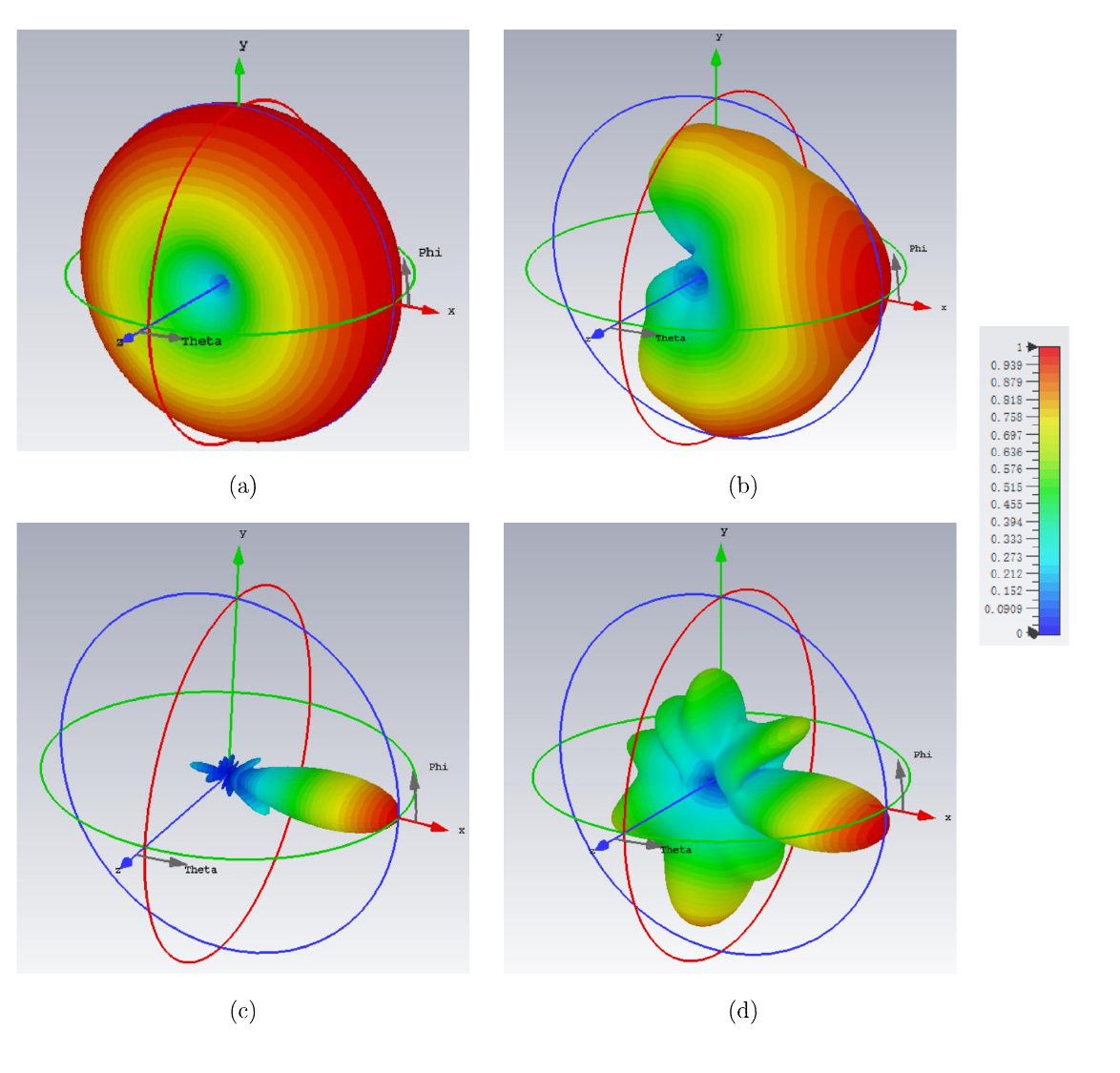}}  
		\caption{Radiation pattern for (a) an isolated antenna; (b) individual antenna in HMIMO  with spacing $0.3\lambda$; (c) HMIMO with spacing $0.3\lambda$;  (c) HMIMO with spacing $0.4\lambda$.}  
		\label{fig:SuperDirectBeam_graph} 
	\end{center}
\end{figure}

The unavoidable mutual coupling effects in HMIMO systems distort radiation patterns and generate directivity, as shown in Fig.~\ref{fig:SuperDirectBeam_graph}. Specifically, the radiation pattern for an isolated antenna (shown in Fig.~\ref{fig:SuperDirectBeam_graph} (a)) is distorted due to coupling effects induced by surrounding antennas, as shown in Fig.~\ref{fig:SuperDirectBeam_graph} (b). By adjusting the spacing between two adjacent antennas, HMIMO is expected to achieve superdirectivity, as shown in Fig.~\ref{fig:SuperDirectBeam_graph} (c-d).

As illustrated in Fig.~\ref{fig:SuperDirectBeam_graph}, superdirective HMIMO systems in the near-field zone exhibit significant potential in information transfer, which is attributed to the excited higher-order modes, enabling additional diversity gain and reactive energy that can carry more information. Information can be also encoded in both the electric and magnetic energy, however,  the conventional Rayleigh resolution limit (e.g., $\lambda/2$) may be surpassed, thus necessitating a detector designed to accommodate such deviations.

Additionally, superdirectivity is a promising approach for addressing distortion in edge radiation patterns, achieving minimal sidelobe energy leakage and enhancing array gain. Unlike traditional MIMO systems, where beamforming gain is proportional to $N$ (with half-wavelength spacing), superdirective antenna arrays can achieve a beamforming gain proportional to  $N^2$. The excitation coefficients (beamforming vector) can be designed based on the coupling matrix \cite{9838533}. Superdirective antenna arrays can also be applied to multi-user communications to improve spectral efficiency. The work \cite{han2023superdirectivity}  showcased that superdirective antenna arrays can achieve up to 9 times higher spectral efficiency compared to traditional MIMO systems, while reducing the array aperture by half.

Despite the superdirective HMIMO in the near-field zone admitting extremely high reactive energy and extra diversity gain,  it is accompanied by the potential drawback of low gain attributable to reduced energy efficiency.  Besides, achieving precise control over excitations to attain superdirectivity at the hardware level proves challenging due to the uneven distribution of current across patch antennas, as discussed in the text box ``Impacts of $G$ and $Q$ in Antenna Excitation Setting".

\begin{figure*}[!tbp]
    \centering
\begin{tcolorbox}[title = {Impacts of $G$ and $Q$ in Antenna Excitation Setting},
  fonttitle = \bfseries]  
La Paz and Miller \cite{1694726} put forward that there exists an optimum current distribution realizing the maximum directivity with some constraints, while Bouwkamp and De Bruijn \cite{Bouwkamp1945} mathematically proved that there is indeed no limit on theoretical directivity. Subsequently, the work in \cite{PritchardMaximumDirectivityIndex1954} presented the explicit expression of the directivity gain in terms of excitations (element coefficients), and derived the optimal excitations using the method of the Lagrangian multiplier. The numerical result revealed that for spacing $\Delta=\lambda/2$, the maximum directivity gain is achieved using uniform excitation.  It is found that the optimum excitation yields a directivity gain of $N^2$ at spacing $\Delta=\lambda/4$  while a directivity gain of $N$ at spacing $\Delta=\lambda/2$ \cite{KildalFundamentalDirectivityLimitations2016}. 
 
However, a much more complicated non-uniform excitation scheme is expected to achieve higher maximum directivity gain, especially for small element spacing (normally less than $\lambda/2$).  For example,  in a three-element array as the spacing approaches zero, one of the coefficients tends toward $-\infty$ while the other one approaches $+\infty$, resulting in fairly larger minor lobe and larger amplitude of nonuniform excitations \cite{PritchardMaximumDirectivityIndex1954}.  

The above work all designed the antenna excitation to achieve maximum directivity gain $G$. Nevertheless, the minimum quality factor $Q$ is also important in antenna design.  In this context, the study conducted by \cite{GustafssonOptimalAntennaCurrents2013} investigated the optimal current distribution for maximal directivity gain $G$ quotient and minimal quality factor $Q$ by solving the convex optimization problems. It is worth noting that the optimal current distribution is generally not unique since the field outside the field can be generated from multiple feasible excitations. 
\end{tcolorbox}
\end{figure*}

\subsection{Wideband HMIMO Systems} \label{subsec:WidebandHMIMO} 
Wideband systems offer substantial potential for improving wireless communication performance by enabling high data rates and strong robustness to multi-path environments. For example, wide bandwidth effectively enhances signal multiplexing and diversity by exploiting additional spectral DOFs in the frequency domain, i.e., multiple data streams are transmitted by independent subcarriers. As a result, wideband or ultra-wideband HMIMO systems become compelling. These systems undoubtedly provide significant benefits to wireless communications.  

Due to the wide bandwidth and large array aperture, the assumption in traditional narrowband communication systems{\textemdash}that the transmitted signal $s(t)$ with time delay $\tau_m$ for the $m$th antenna is approximately $s(t-\tau_m)\approx s(t)$ with symbol period $T_s\gg \tau_m${\textemdash}no longer holds. This is because the increased propagation differences and decreased $T_s$ lead to non-negligible time delay across the large aperture, resulting in intrinsic \textit{spatial-wideband effect} (or spatial selectivity), which differ from \textit{frequency-wideband effect} (or frequency selectivity) caused by multi-path delay spread.  It should be noted that both spatial- and frequency-wideband effects, i.e., \textit{dual-wideband effects}, impact the wideband HMIMO systems.  

However, as summarized in \cite{8443598}, the dual-wideband effects present new challenges in wireless communications. Specifically, the coherence bandwidth is reduced in frequency-wideband effect, causing inter-symbol interference. The extensive angular spreads in spatial-wideband effect generate beam squint/split or defocus \cite{10399817}, thus necessitates delay squint compensation and a complicated decoding process. To deal with squint issues in dual-wideband effects, there are mainly two types of mitigation methods, i.e., the algorithm-based method (e.g., precoding schemes or additional phase compensation operation) and the architecture-based method (e.g., true-time-delay device). These methods are incorporated in specific channel models, and current research on dual-wideband effects primarily adopts planar-wave and spherical-wave models.

Specifically, the work \cite{8354789} represented the dual-wideband channel using three coupled factors, i.e., spatial steering vector (the array response of each antenna towards the angle), frequency steering vector (the array response of different subcarrier frequencies to time delay), and the phase shift matrix (which describes diffusion effects of each path, disappearing or being all-ones matrix in narrowband systems). By taking the 2D inverse Fourier transformation of the spatial-frequency domain channel, one can obtain the angular-delay domain channel. Due to the difference between the planar-wave assumption and spherical-wave assumption, the angular-delay channel exhibits a rectangular shape in far-field regions \cite{8443598} and a diamond shape in near-field regions \cite{10663786}, indicating a stronger angular spread in angular-frequency domain. 

So far, dual-wideband effects have not been well-investigated in EM-compliant channel models (discussed in Sec.~\ref{sec:EMmodeling}). Current research remains within traditional channel assumptions. However, it is anticipated that these models will reveal similar, but more accurate, spatial-wideband effects, such as delay squint, beam squint, and diffusion effects. For instance, in Green's function-based channel models, propagation delay is incorporated into the exponential term. The notable angular spread at individual frequencies and its diminishing trend with increased distance is supposed to be observed through computations, i.e., wideband effects can be quantified and analyzed more accurately using EM-compatible channels.

In summary, leveraging the frequency and spatial dependency of HMIMO systems allows for the concentration of energy at desired locations, effectively suppressing interference and reducing signal processing complexity \cite{7087373}. For example, if each element in HMIMO systems is able to radiate a nonoverlapping segment of the entire frequency extent, the higher spatial resolution and multiplexing gain can be achieved without additional processors.  At the same time, through the combination of wide bandwidth and near-field gains, each subcarrier can accommodate more data streams through spatial-domain multiplexing enhanced by the near-field effect, further improving the performance of HMIMO systems.
 \color{black}

%%%%%%%%%%%%%%%%%%%%%%%%%%%%%%%%%%%%%%%%%%%%%%%%%%%%%%%%%%%%%%%%%%%%%%%%%%%%%%%%%%%%%%%%%%%%%%%%%%%%%%%%%%%%%%%%%%%%%%%%%%%%%%%%%%%%%%%%%%%%%%%%%%%%%%%%%%%%%%%%%%%%%%%%%%%%%%%%%%%%%%%%%%%%%%

\section{Field Sampling in HMIMO-Oriented EIT} \label{sec:fieldSampling}
The continuous field contains abundant information, thus, it presents challenges in efficiently processing the field with minimal overhead.
Resorting to EIT,  the field sampling theory is introduced. This not only offers insights into optimal antenna placement but also enables practical and quantitative evaluation of continuous fields.  Subsequently, the compact elements in HMIMO systems enable the oversampling technique and improve system performance, as detailed in this section.

\subsection{Field Sampling}
Field sampling theory is a computational method used to measure the information conveyed by continuous fields in both the temporal and spatial domains.  Theoretically, the optimal sampling schedule could retrieve information perfectly with the minimum sampling points. 

Classically, the Nyquist criterion is adopted in communications, facilitating interference-free symbol-by-symbol detection, where samples are nearly orthogonal with Nyquist sampling interval $\Delta_{\mathrm{Nyq}}=\lambda/2$ \cite{ShannonCommunicationPresenceNoise1949,UnserSampling50yearsShannon2000,AminehThreeDimensionalNearFieldMicrowave2011}.  The geometry of the antenna array exert impacts on the sampling. For example, linear arrays have a nonuniform sampling over the elevation direction; circular arrays have uniform sampling over the elevation directions and nonuniform sampling over the elevation direction; and spherical arrays have uniform sampling over both the elevation and the azimuth directions \cite{9737566}. The three-dimensional antenna aperture has a much larger sampling spacing compared with these one-dimensional and two-dimensional apertures \cite{loyka_information_2004}. Similarly, the sampling patterns also exert impacts on performance, for example, hexagonal sampling exhibits a higher efficiency compared with the rectangular sampling \cite{PizzoNyquistSamplingDegrees2022}.

The Nyquist sampling criterion indicates the minimum sampling points that characterize the information content of the signal, thus providing a good guideline in antenna design, especially the placement. For example, locating the transmitter (or receiver) antennas at these sampling points to recover all information conveyed in the field.  The sampling points carry the information content of the field, therefore, it is natural to connect information measurement (DOF) with the sampling process. Specifically, the Nyquist sampling density can be computed through DOF per temporal/spatial unit \cite{9798854}. The details of DOF are given in Sec. VI.  As suggested by the Nyquist sampling rate or DOF analysis, the far-field spatial sampling rate (resolution) is  $\lambda/2$ for infinite aperture. However, the number of sampling points in near-field scenarios is larger than that in far-field zones. As described in the previous section, the excited higher-order modes in the near-field region gradually vanish in the far field, thus more information content is embedded in the near field and requires more sampling points to retrieve information.    

\subsection{HMIMO and Oversampling}   
The number of antenna elements is extremely large in HMIMO systems, and this benefits both near-field communications and far-field communications. On the one hand,  the finer spatial sampling caused by closely spaced elements is capable of capturing abundant spatial information in near-field HMIMO communications. On the other hand, although the traditional Nyquist sampling is sufficient in far-field HMIMO communications, the compact antenna array in HMIMO systems still offers additional benefits due to its capability of faster-than-Nyquist (FTN) spatial signaling. 

    FTN technique, also termed oversampling, is first developed by Mazo \cite{6772210}, who proved that the minimum Euclidean distance remains unchanged for nonorthogonal FTN sampling pulses, thus providing extra bandwidth benefits. The maximum acceleration value is referred to as the Mazo limit, compared with the bandwidth in the Nyquist limit which orthogonality holds, the bandwidth in the Mazo limit is smaller, leading to substantial bandwidth savings \cite{6479673}.  It has proved that temporal/spatial FTN has a higher capacity and spectral efficiency compared with Nyquist schemes at the cost of complex detectors \cite{265506,5662127}. 

Therefore, the combination of HMIMO systems and FTN techniques is expected to exploit the spatial domain, which has merely been investigated before. Specifically, most initial work on FTN focuses on the time domain and is then extended to frequency FTN in 2005 \cite{1523482}, where the multi-carrier FTN signaling is expected to further enhance capacity \cite{5288497, 9174112}. Therefore, the spatial FTN in HMIMO systems is prospective to further improve available bandwidth, which is demanded to support explosive data growth in wireless communications. 

However, the unavoidable inter-symbol interference due to the non-orthogonal spatial pulses is one of the major issues in HMIMO-assisted oversampling,  increasing the detector complexity of signal reconstruction. To relieve the detection burden, specific pulse signals (e.g., non-sinc pulses \cite{7417358}) can be designed, or low-computational precoding schemes that eliminate interference are implemented. In addition, since the FTN signaling achieves a higher peak-to-average power ratio compared with Nyquist signaling with the same pulse shape, additional processing is required, e.g., clipping, constellation extension, selected mapping \cite{7043385}, and the errors in this processing cannot be ignored.

\color{black}
 
% \color{blue}
% The continuous eigenfunctions (bases) of space can be represented by a linear combination of finite sub-basis functions, and then the water-filling capacity solution can lead to bounded capacity \cite{4447351}. The increase of the sub-basis means a finer sampling, and the capacity would approach its upper bound. This number is exactly the NDF, which is strongly dependent of the aperture size, antenna element design, and surrounding environment. 
\color{black}

\section{Electromagnetic Channel Modeling}  \label{sec:EMmodeling}
The influence of the surrounding environment on communications is integrated into the channel model that describes the transmission between the transmitter and receiver. To demonstrate these impacts, this section focuses on exploring EM-compliant channel models in both far-field and near-field zones, within the EIT framework. Specifically, it introduces the Fourier plane wave expansion model in rich scattering environments, which is close to the Clarke model and inherently applicable to any planar scatterers. The dyadic Green's function model is then proposed to interpret LoS wireless communications, characterized by full polarizations. Subsequently,  stochastic Green's function model is presented to model general scattering environments based on the probabilistic model. Lastly, the continuous channel modeling method is also introduced.

\begin{figure*}[!tbp]
    \centering
\begin{tcolorbox}[ title = {Division of Green's Functions in Far-/Near-Fields},
  fonttitle = \bfseries]
  Through Weyl's decomposition, the scalar free-space Green's function in \eqref{equ:scalarGreen} which satisfies the inhomogeneous Helmholtz equation with a $\delta$-function source is expanded by a series of plane waves, i.e. \cite{PhysRevE.59.1200,PizzoSpatialCharacterizationElectromagnetic2022},  
\begin{equation}
	\frac{e^{-j k_0 r}}{r}\!\!=\!\!\frac{-j k_0}{2 \pi} \iint_{-\infty}^{\infty} \frac{1}{k_z} \exp [-j k_0(k_x x\!+\!k_y y\!+\!k_z|z|)] d k_x d k_y,
\end{equation}
where $k_z\!=\!\sqrt{1\!-\!k_x^2\!-\!k_y^2}$ for $\mathcal{D}(k)\!=\!\{(k_x,k_y)\mid k_x^2\!+\!k_y^2\! \leqslant \!1\}$, and $k_z\!=\! -j\sqrt{k_x^2\!+\!k_y^2\!-\!1}$ for $\bar{\mathcal{D}}(k)\!=\!\{(k_x,k_y)\mid k_x^2\!+\!k_y^2\! > \! 1\}$. 

Therefore, the scalar function $g(\mathbf{r})$ is the sum of integrals in two parts, i.e., $g(\mathbf{r})=g_H(\mathbf{r})+g_E(\mathbf{r})$ \cite{PhysRevE.59.1200}.  

The homogeneous part $G_H(\mathbf{r})$ contains all propagating plane waves in the far-field zone, which is given by
\begin{equation} \label{equ:GreenHomo}
	\begin{aligned}
		g_H(\mathbf{r})&= \frac{-j k_0}{2 \pi} \iint_{\mathcal{D}(k)} \frac{1}{k_z}  
	 \exp [-j k_0 (k_x x+k_y y+k_z|z|)] d k_x d k_y 
		\! \overset{(a)}{=}-j k_0 \int_0^1 \! \exp [-j \alpha(z) v] J_0\left[\beta(x, y) \sqrt{1-v^2}\right] d v,
	\end{aligned}
\end{equation}
where $(a)$ makes use of polar coordinates, $J_0$ is the Bessel function of the first kind and order zero, the coordinate-dependent parameters $\alpha(z)=k_0|z|$ and $\beta(x,y)=k_0\sqrt{x^2+y^2}$, and $v$ is radial integration variable \cite{PhysRevE.59.1200}. 

Similarly, the evanescent part $ G_E(\mathbf{r})$ contains all the exponentially decaying plane waves in the near-field zone, i.e., 
\begin{equation}
	\begin{aligned}
		g_E(\mathbf{r})&= \frac{-j k_0}{2 \pi} \iint_{\bar{\mathcal{D}}}  \frac{1}{k_z}  
		\times \exp [-j k_0(k_x x+k_y y+k_z|z|)] d k_x d k_y
		\!=\! k_0 \int_0^{\infty} \!\! \exp [-\alpha(z) v] J_0\left[\beta(x, y) \sqrt{v^2+1}\right] d v.
	\end{aligned} 
\end{equation}
\end{tcolorbox} 
\end{figure*}

\subsection{Physical Interpretation of EM Channel}

The current distribution $\mathbf{J}$ imposed by the transmit point source excites an EM wave $\mathbf{E}$ under the transfer function $\overline{{\mathbf{G}}}_0 $ (or $ \overline{{\mathbf{G}}}_{\text{S}}$). The concept of wireless channels, widely used in communications, has not been well established yet from the EM perspective. To gain a comprehensive understanding of EM channels, we interpret the EM channel as a continuous vector wave field excited by an impulse response. As a consequence, the (stochastic) dyadic Green's functions can be considered as the EM-domain wireless channel (denoted by $\mathbf{H}$) for systems with continuous apertures. In this context, the channel satisfies the vector wave equation, i.e., 
\begin{equation}
\nabla \times \nabla \times \mathbf{H} - k_0^2 \mathbf{H} =\mathbf{0}, 
\end{equation} 
where $\mathbf{H}$ can be alternatively chosen as one of its representations in different domains.
For example, the channel can be in a space-time representation, $\mathbf{H} = \mathbf{H} (\mathbf{r}, t)$ or in a wavenumber-time representation, $\mathbf{H} = \mathbf{H} (\mathbf{k}, t)$, with the following Fourier transformation relation
\begin{equation}
\mathbf{H} (\mathbf{r}, t)=\frac{1}{(2 \pi)^3} \int \mathbf{H} (\mathbf{k}, t) e^{j \mathbf{k} \cdot \mathbf{r}} d \mathbf{k},  
\end{equation}
where $\mathbf{k} = [k_x, k_y, k_z]^{T}$ is the wavevector, and the integration is performed over region $\mathcal{D}=\{k_x^2+k_y^2+k_z^2=k_0^2\}$. Furthermore, the integral region can be divided into region $k_x^2+k_y^2\leq k^2$ for supporting propagating waves existing in far-field, and region $k_x^2+k_y^2>k^2$ including evanescent waves in the near-field region. See the text box ``Division of Green's Functions in Far-/Near-Fields" for details of the scalar channel scenario. Such a division in the wavenumber domain is consistent with the communication region divided in the space domain.

\subsection{Fourier Plane Wave Expansion Model} \label{subsec:planeWave} 
The equivalent virtual channel, adopted to represent complex channel characteristics in beamspace or the wavenumber domains, has been widely investigated. In these models,  physical scattering is typically incorporated into an approximately uncorrelated virtual channel \cite{1033686}. Specifically, as the homogeneous part $G_H(\mathbf{r})$ of the far field is dominant and finite within the circular integration region $\mathcal{D}$, the finite sampling points are capable of containing the information inside this region \cite{PizzoNyquistSamplingDegrees2022}. Based on this observation, the plane wave expansion based model with finite wavenumber channel sampling points is proposed \cite{PizzoSpatialCharacterizationElectromagnetic2022}. Specifically, the channel response derived from \eqref{equ:GreenHomo} models a LoS propagation environment, where only upgoing plane waves are considered. Here are some details of the Fourier plane wave expansion model.   

\begin{figure}  
	\begin{center}
		{\includegraphics[width=0.45 \textwidth]{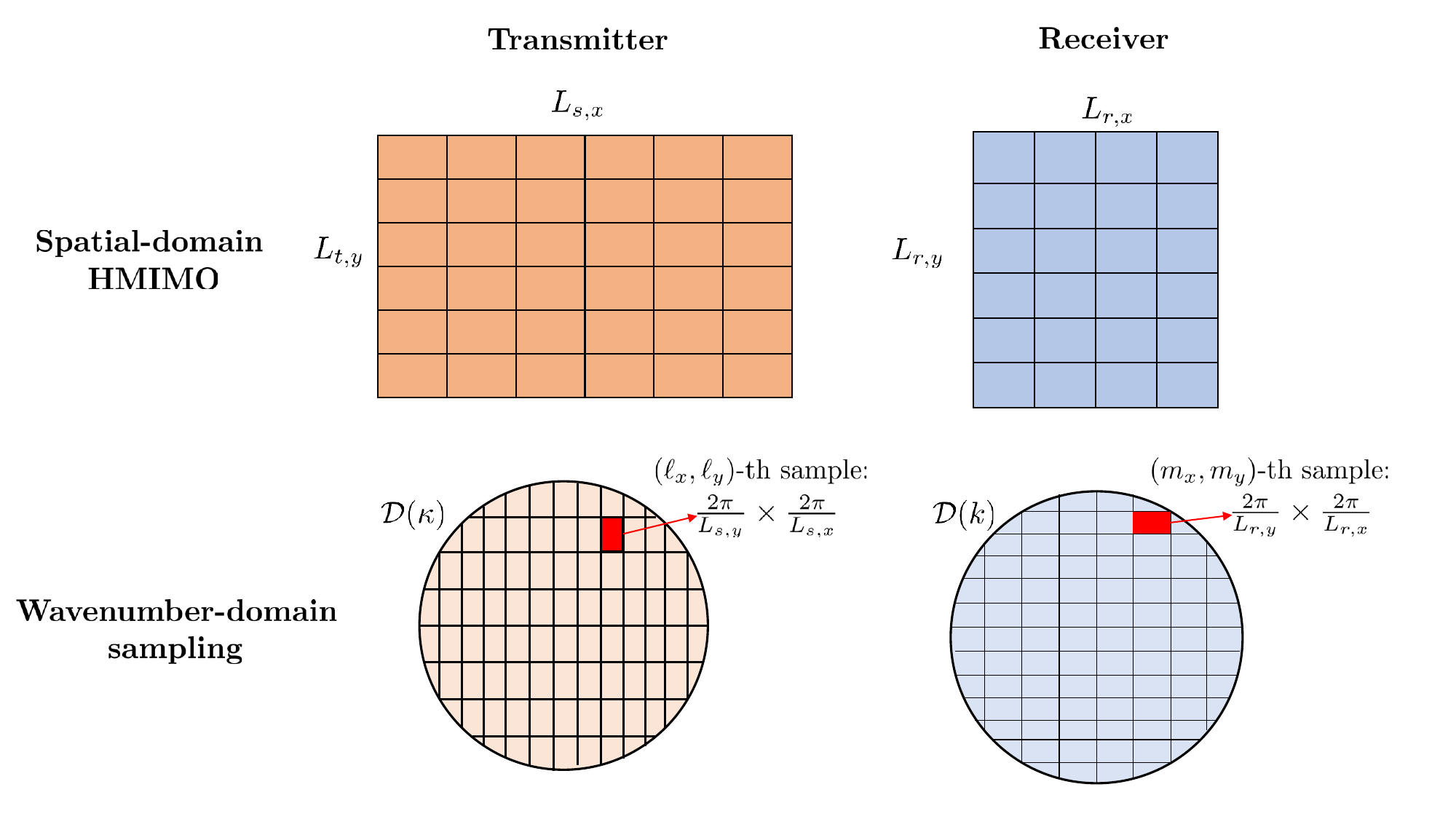}}  
		\caption{The wavenumber-domain sampling  at both transmitter and receiver HMIMO.}  
		\label{fig:WavenumberCh_graph} 
	\end{center}
\end{figure}

If the point source is placed towards $z$-axis, and the boundary condition is omitted, the whole system can be regarded as spatially stationary, i.e., a phase-shift of the two-dimensional wavenumber domain channel response $H_a\left(k_x, k_y, \kappa_x, \kappa_y\right) $ along the $z$-axis \cite{MarzettaSpatiallyStationaryPropagatingRandom2018}. Therefore, the spatial channel response and wavenumber channel are connected through the Fourier transform given by  \cite[Eq.(4)]{PizzoFourierPlaneWaveSeries2022}
\begin{equation}
	\begin{aligned}
		& h(\mathbf{r}, \mathbf{s})=\frac{1}{(2 \pi)^2} \iiiint_{\mathcal{D}(k) \times \mathcal{D}(\kappa)}   a_r\left(k_x, k_y, \mathbf{r}\right) \\
		& \quad \times H_a\left(k_x, k_y, \kappa_x, \kappa_y\right) a_s\left(\kappa_x, \kappa_y, \mathbf{s}\right) d k_x d k_y d \kappa_x d \kappa_y
	\end{aligned}
\end{equation}
where $a_r(k_x,k_y,\mathbf{r})$ and $a_s(\kappa_x,\kappa_y,\mathbf{s})$ are the receive response and source response, respectively,  which maps the receive propagation direction $ \boldsymbol{k}=(k_x\vec{x},k_y\vec{y})$ and transmit propagation direction $ \boldsymbol{\kappa}=(\kappa_x\vec{x},\kappa_y\vec{y})$ to the induced and excited current at receive point $\mathbf{r}$ and source point $\mathbf{s}$, respectively, acting as the basis functions in the transceiver end. The angular response $H_a\left(k_x, k_y,\kappa_x,\kappa_y\right)$ is computed from the bandlimited power spectral density, which is characterized by the variances of $N_{\mathrm{samp}}$ samples within the discretized region over the support $\mathcal{D}(k)\times\mathcal{D}(\kappa)$ \cite{PizzoFourierPlaneWaveSeries2022,PizzoSpatiallyStationaryModelHolographic2020}, as shown in Fig.~\ref{fig:WavenumberCh_graph}. 

\color{black}
 
Notably, the angular response  $H_a\left(k_x, k_y,\kappa_x,\kappa_y\right)\sim \mathcal{N}(0,\sigma_s^2)$ ($\sigma_s^2$ is the variance of sampling point) can be seen as the coupling coefficients between the receive response and transmit response. The isotropic propagation environment can be decomposed into two parts, one is dependent on the transmitter side and the other is dependent on the receiver side, leading to a separable computation, i.e., $H_a\left(k_x, k_y,\kappa_x,\kappa_y\right)=H_a\left(k_x, k_y \right)H_a\left( \kappa_x,\kappa_y\right)$.  However, in non-isotropic propagation environments, the angular response is dependent on both sides, which is much more complicated. Briefly speaking, the introduction of angular response that involves the stochastic characteristics of the surrounding environment facilitates complicated spatial channel modeling. What's more, in the far-field region, the low-computational algorithms (e.g., channel estimators and beamforming schemes) can be designed based on sparsity in the angular domain.

The angular response also accounts for the stochastic characteristics of the surrounding environment, which is implicitly embedded in the Gaussian distribution of variances for sampling points. Given the transmitter angle $\theta_{t}$ and receiver angtle $\theta_{r}$, the scattering contribution to the virtual angle pair $(\theta_{r},\theta_t)$ is proportional to the number of paths that lie in the rectangular virtual spatial bin of size $\frac{1}{N_r}\times \frac{1}{N_t}$ centered on $(\theta_{r},\theta_t)$ \cite{1033686}. Thus, the virtual channel representation (wavenumber domain channel) decomposes an arbitrarily clustered channel into independent subchannels whose structure is very similar to the i.i.d model, and the subchannels are represented by approximately uncorrelated entries derived from the distribution of sampling points.  

\begin{figure}   
	\begin{center}
		{\includegraphics[width=0.45 \textwidth]{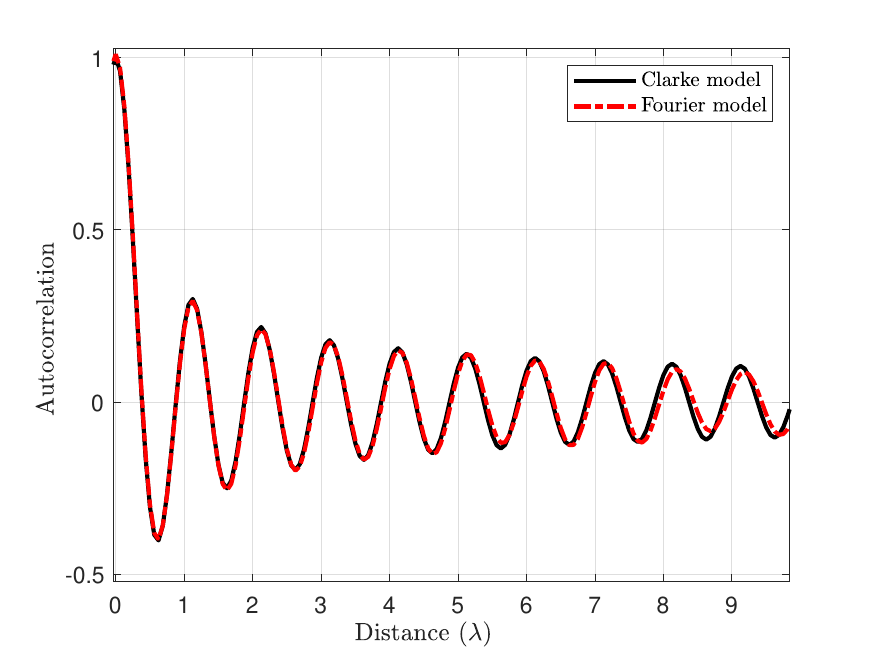}}  
		\caption{Comparison between Fourier plane wave expansion model and traditional Clarke model in terms of the autocorrelation function  vs. distance between two antennas \cite{PizzoSpatiallyStationaryModelHolographic2020} (source codes from \cite{Holographic-MIMO-Small-Scale-Fading}).}  
		\label{fig:fig_FourierCM} 
	\end{center}
\end{figure}

 The Fourier plane wave expansion model (refer to Fourier model) is compared with the traditional Clarke model \cite{ClarkeModel} in terms of the autocorrelation function, as illustrated in Fig.~\ref{fig:fig_FourierCM}. It is clearly shown that the two curves fit well in the small distance between two antennas, justifying the effectiveness of the Fourier plane wave expansion model.

\textit{\textbf{Limitations:}} Although the Fourier plane wave expansion model offers computational advantages by finite sampling points,  it is most applicable to regular scatterers, similar to the spherical or cylindrical harmonic wave models. Specifically, for regular scatterers (plane, cylinder, and sphere), we can expand the incident field into the corresponding harmonic functions, and then easily calculate the scattering field of each order of the harmonic function. However, for irregular scatterers, numerical methods like the Method of Moments (MoM) are the mainstream approach, where Green's function is typically used to construct the impedance matrix. In such cases,  the Fourier plane wave model may not be as convenient or straightforward, especially when irregular scatterers are present in the scattering environment.

\color{black}

\subsection{Dyadic Green's Function Model}
To delve into the more complicated EM phenomena in near-field communications, several approaches have been proposed \cite{DovelosIntelligentReflectingSurfaces2021,WeiChannelEstimationExtremely2022,YuanElectromagneticEffectiveDegree2022,WeiChannelModelingMultiUser2023,AlayonGlazunovSphericalVectorWave2009,glazunov2010physical}. One commonly employed method relies on spherical wave expansion \cite{DovelosIntelligentReflectingSurfaces2021}. Specifically, the channel is presented in the form of spherical waveforms, where the coefficients of spherical waves are modeled as Gaussian variables. Random scatters may be located far from or near the transceiver in practical scenarios, resulting in the coexistence of near-field and far-field communications. Thus,  a parameter is also introduced to control the portion of far-field and near-field path components to interpret this coexistence of near-field and far-field communications  \cite{WeiChannelEstimationExtremely2022}. However, such a method is still incapable of interpreting the polarization effects and acts as a mathematical tool to sample or interpolate the total radiation pattern without physical meaning, thus a new channel model is further proposed in \cite{AlayonGlazunovSphericalVectorWave2009,glazunov2010physical}. Unlike the spherical waves method, the study in \cite{AlayonGlazunovSphericalVectorWave2009,glazunov2010physical} introduced a physically accurate spherical vector wave expansion of the random electromagnetic field to characterize the propagation channel.

These channel models, based on spherical waves, are approximation methods. As such, they may not fully account for polarization effects in near-field communications and might underestimate the system capacity. To address this limitation, dyadic Green's function is suggested to be integrated into channel modeling.  The effectiveness of such an approach has been demonstrated in \cite{YuanElectromagneticEffectiveDegree2022}, where the DOF variation in near-field and far-field regions is captured. Subsequently, the channel model based on dyadic Green's function for HMIMO systems in LoS near-field communications has been presented in \cite{WeiChannelModelingMultiUser2023,10500751}. This integrated model offers a more inclusive representation of various EM characteristics in wireless communications, encompassing power gain, polarization diversity, and spatial diversity, leading to a more accurate and comprehensive understanding of the communication environment.
 
\begin{figure}  
	\begin{center}
		{\includegraphics[width=0.5 \textwidth]{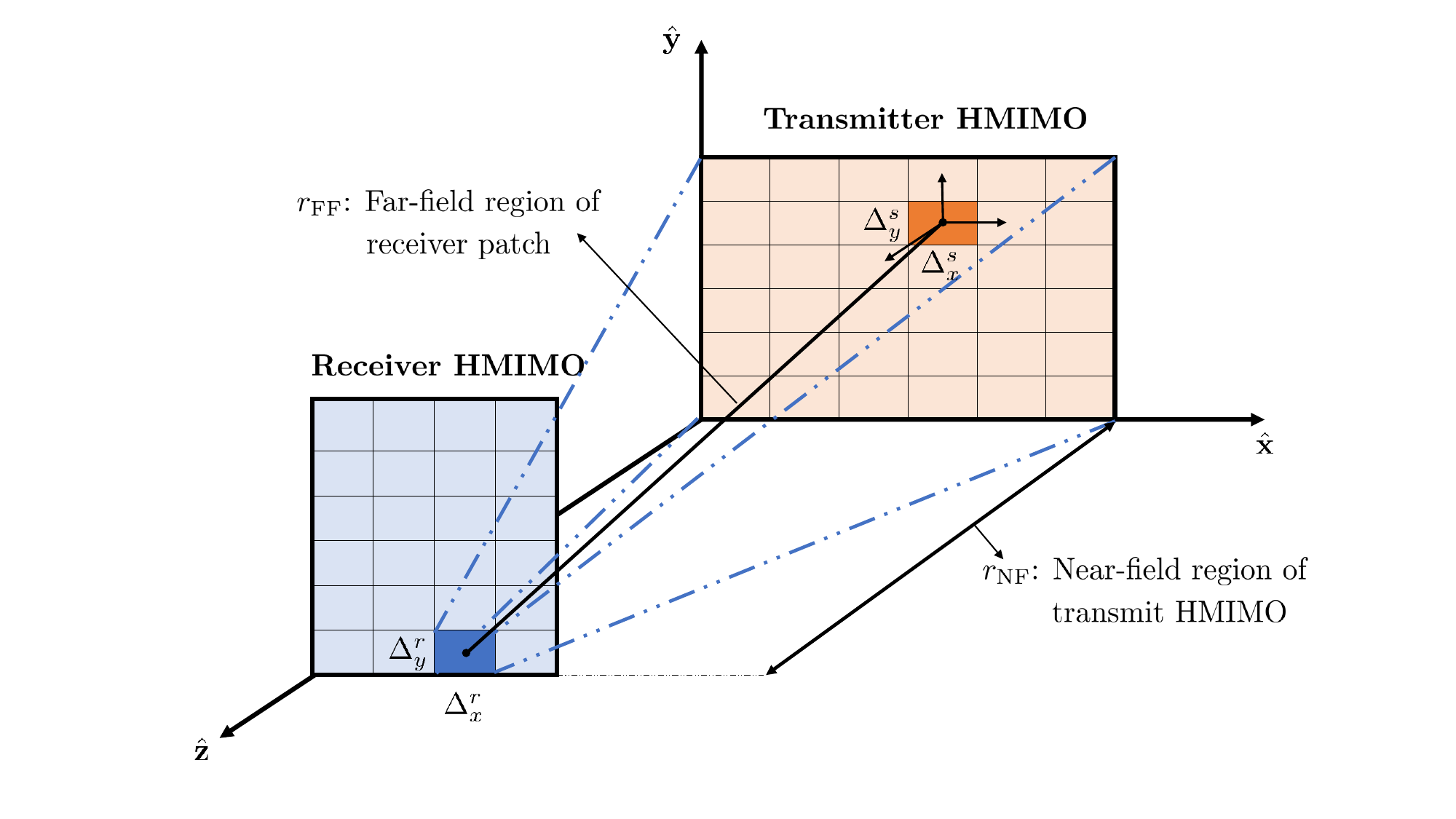}}  
		\caption{Field configuration of partitioned HMIMO at the transmitter and receiver.}  
		\label{fig:DivisionsGreenFunc_graph} 
	\end{center}
\end{figure}

Specifically, by partitioning HMIMO in many patches, the electric field in \eqref{equ:em_in_out} is discretized. Then, the whole electric field at the observation plane can be obtained through integration over these patches, as shown in Fig.~\ref{fig:DivisionsGreenFunc_graph}. Specifically, the HMIMO is divided into $N_s$ rectangles with size $ \Delta^s = \Delta_x^s \Delta_y^s$. Each patch is regarded as a point $(x'_n,y'_n)$ in a first coordinate system, and is further investigated in a second coordinate system defined within $(x'_0,y'_0)$. Under such consideration, the electric field  in \eqref{equ:em_in_out} can be rewritten in terms of integration over many patch antennas:
\begin{equation}
	\begin{aligned}
		&\mathbf{E}(\mathbf{r}) =\sum_{n=1}^{N_s}\! \int  \!\! \int d x'_0 d y'_0  \left[\overline{\mathbf{I}}\!\!+\!\!\frac{\nabla \nabla}{k_0^{2}}\right]\!\! \frac{e^{-j k_0 r_n }}{4 \pi r_n } \mathbf{J}\left(\mathbf{r}^{\prime}_n\right).
	\end{aligned}
\end{equation}
Since the antenna element size is comparably small to the transmitter-receiver distance, i.e., the distance to the observation point at the receiver is much greater than the size of the source. Hence, from the perspective of the whole transmitter, the receiver lies in the near-field zone of the transmitter, while for the small-size antenna element at the transmitter, the receiver is in the far-field region of the transmitter. Given such observation and integral operation, the discretized field is computed to obtain the summation over the whole transmitter and receiver region.  A more generalized EM channel model for arbitrarily placed surfaces was proposed in \cite{10500751, 10558821}.

\begin{figure}  
	\begin{center}
		{\includegraphics[width=0.45 \textwidth]{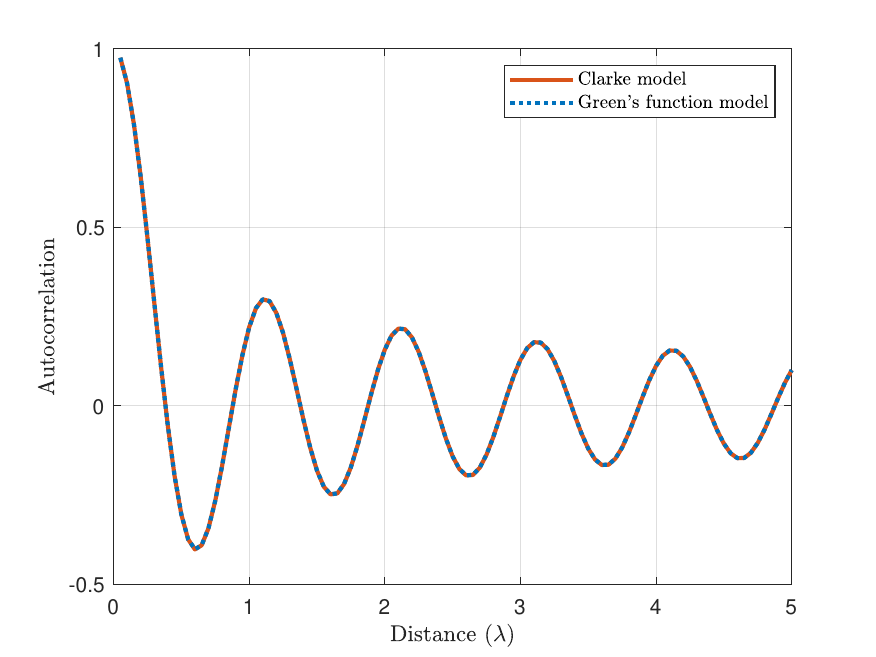}}  
		\caption{Comparison between Green's function model and traditional Clarke model in terms of the autocorrelation function vs. distance between two antennas \cite{YuanElectromagneticEffectiveDegree2022}. }  
		\label{fig:fig_GreenCM} 
	\end{center}
\end{figure}

The effectiveness of dyadic Green's function model is substantiated by the correlation performance comparison with the traditional Clarke model, as depicted in Fig.~\ref{fig:fig_GreenCM}. As observed from the figure, dyadic Green's function model is always coincident with the Clarke model for various distances between two antennas, thus supporting the validity of the proposed channel model.

\textit{\textbf{Limitations:}} While the dyadic Green's function model effectively captures polarization in wireless communications, it is primarily suited for LoS scenarios and incurs excessively high computational costs when accommodating random scatterers. In realistic settings, where scattering environments introduce multi-path effects, representing such a complex propagation environment using the dyadic Green's function model becomes prohibitively expensive due to its computational demands. Therefore, an efficient channel model is essential to characterize complex scattering scenarios.
\color{black}

\subsection{Stochastic Green's Function Model}
In order to effectively capture the randomness of scatterers in practical settings and provide a comprehensive description of multi-path wireless communication systems, it is anticipated that the stochastic Green's function introduced in Sec. \ref{subsubsec:StochasticGreen} will offer a potential solution. Different from the plane wave expansion method, this model is applicable to general scattering environment. Specifically, based on the random wave model and random matrix theory, this approach leverages probabilistic tools to model scattered waves in communications and derives the distribution of representative eigenvalues using the law of large numbers. The resulting channel, impacted by the presence of scatters, is then expressed in terms of eigenvalues and eigenvectors in the form of distribution. The stochastic Green's function in Sec. \ref{subsubsec:StochasticGreen} is well-suited for channel modeling in scenarios involving random scatters, allowing for characterization of the stochastic nature of wave propagation in wireless communications. 

\textit{\textbf{Limitations:}} Despite its potential advantages, the incorporation of stochastic Green's function in channel modeling poses some major issues. This is primarily because the stochastic Green's function was originally conceived within the context of a cavity model with well-defined parameters, such as volume and shape \cite{LinPredictingStatisticalWave2023}. Therefore, extending the stochastic Green's function to general wireless scenarios presents a formidable task. One critical difficulty arises from the lack of crucial information about model parameters in varying scenarios within broader wireless contexts. This presents a valuable opportunity for further research and exploration.  

The comparison between different channel modeling methods is provided in Table~\ref{table:com} with respect to applicable scenarios and limitations.

\begin{table*}  
	\caption{Channel Modeling Methods}
	\vspace{-0.5cm}
	\begin{center}
		\renewcommand{\arraystretch}{1.5}
		\small  
         \begin{tabular} {|p{4cm}|p{2.3 cm}|p{5.5cm}|p{4cm}|} 
			\hline 			
			{\textbf{Channel Modeling Methods}}  & {\textbf{Related   Works}}    &{\textbf{Applicable Scenarios}} &{\textbf{Limitations}}  \\    \hline

              Fourier plane wave expansion model&  \cite{PizzoFourierPlaneWaveSeries2022,PizzoSpatiallyStationaryModelHolographic2020}  & Approximately represent the spatial NLoS path with less wavenumber-domain samples at low cost.  & Difficult to handle irregular scatterers.     \\     \hline

              Dyadic Green’s function model &  \cite{YuanElectromagneticEffectiveDegree2022,WeiChannelModelingMultiUser2023,10500751}&  Exactly describe both near-field and far-field LoS scenarios using dyadic Green's functions.  & Computationally prohibitive in NLoS modeling.  \\    \hline

              Stochastic Green’s function model &  \cite{LinVectorialPropertyStochastic2022,LinPredictingStatisticalWave2023,LinStochasticGreenFunction2018, LinStochasticGreenFunction2020}&  Statistically depict the general scattering environment using random wave model and random matrix theory.  & High complexity; difficult to obtain statistics of model parameters.  \\    \hline
			
		\end{tabular} \label{table:com} 
	\end{center}
\end{table*}

\subsection{Continuous Channel Modeling}
To unveil the achievable performance limit, analyzing the continuous aperture is important because it is independent of the number of antenna elements and the spacing between adjacent antennas. Unlike discrete cases, where matrix operations are used (as in previous channel models), the analysis in the continuous realm employs square-integrable functions and Hilbert-Schmidt operators for mathematical computation. Specifically, both the channel operator $k(\mathbf{r},\mathbf{s})$ and input signal $f(\mathbf{s})$ are square-integrable functions: $f: \mathbb{R}^{d}\rightarrow\mathbb{C}$ satisfying $\int |k(\mathbf{r},\mathbf{s})|^2<\infty$ and $\int |f(\mathbf{s})|^2<\infty$. Define the Hibert-Schmidt operator $\mathcal{H}: L^2(\mathbb{R}^d) \rightarrow L^2(\mathbb{R}^d)$, the input-output mapping is given by $\mathcal{Y}=(\mathcal{H}f)(\mathbf{s})=\int k(\mathbf{r},\mathbf{s}) f(\mathbf{s}) d\mathbf{s}$. The details of integral operators in continuous cases and their matrix counterparts in discrete cases can be found in \cite[Table I]{9973178}.

The orthonormal eigenfunctions of the transmitter space $\mathcal
{S}_T$ and receiver space $\mathcal{S}_{R}$ can effectively characterize the continuous channel. Specifically, the input signal $f(\mathbf{s})$ is represented by a set of orthonormal basis functions $\{\phi_n(\mathbf{s})\}_{n\in\mathbb{N}}$ and the radiated field is expressed using eigenfunction set $\{\psi_n(\mathbf{r})\}_{n\in\mathbb{N}}$. A real sequence $\{\xi_n\}_{n\in\mathbb{N}}$ (i.e., the singular values of the channel operator), converging to zero, satisfies the following relation: 
\begin{equation}
   \mathcal{H} f=\sum_{n=1}^{\infty} \xi_n\left\langle f, \phi_n (\mathbf{s})\right\rangle \psi_n (\mathbf{r}),
\end{equation}
where the Hilbert-Schmidt kernel $k(\mathbf{r}, \mathbf{s})=\sum_{n=1}^{\infty} \xi_n \phi_n \psi_n^*$ is connected to Green's function $\overline{ {\mathbf{G}}}_0(\mathbf{r},\mathbf{s})$ \cite{9475156}. 

Consequently, the eigenfunctions of the transmitter and receiver surfaces form the basis for continuous channel modeling. The derivation of these orthonormal eigenfunctions (or basis functions) is summarized as follows.
\subsubsection{Optimal Eigenfunctions} 
In the point-to-point transmission, the receiver basis function $\psi(\mathbf{r})$ and transmitter basis function $\phi(\mathbf{s})$ are connected via the integral of the Green's function $\overline{ {\mathbf{G}}}_0(\mathbf{r},\mathbf{s})$ over the transmit space $\mathcal{S}_T$, i.e., $\psi(\mathbf{r})=\int_{\mathcal{S}_T} \overline{ {\mathbf{G}}}_0(\mathbf{r}, \mathbf{s}) \phi(\mathbf{s}) d \mathbf{s}$. Therefore, the optimal eigenfunctions can be derived by solving a coupled eigenfunction problem given by \cite[Eq.(12)-(15)]{9139337,9641865}
\begin{equation}
    \begin{aligned}
\xi_n^2 \phi_n(\mathbf{s}) & =\int_{\mathcal{S}_T} K_T\left(\mathbf{s}, \mathbf{s}^{\prime}\right) \phi_n\left(\mathbf{s}^{\prime}\right) d \mathbf{s}^{\prime}, \\
\xi_n^2 \psi_n(\mathbf{r}) & =\int_{\mathcal{S}_R} K_R\left(\mathbf{r}, \mathbf{r}^{\prime}\right) \psi_n\left(\mathbf{r}^{\prime}\right) d \mathbf{r}^{\prime},
\end{aligned}
\end{equation}
where
\begin{equation}
    \begin{aligned}
K_T\left(\mathbf{s}^{\prime}, \mathbf{s}\right) & =\int_{\mathcal{S}_R} \overline{ {\mathbf{G}}}_0^*(\mathbf{r}, \mathbf{s}) \overline{ {\mathbf{G}}}_0\left(\mathbf{r}, \mathbf{s}^{\prime}\right) d \mathbf{r} \\
K_R\left(\mathbf{r}, \mathbf{r}^{\prime}\right) & =\int_{\mathcal{S}_T} \overline{ {\mathbf{G}}}_0(\mathbf{r}, \mathbf{s}) \overline{ {\mathbf{G}}}_0^*\left(\mathbf{r}^{\prime}, \mathbf{s}\right) d \mathbf{s} .
\end{aligned}
\end{equation}

However, finding these optimal eigenfunctions analytically is generally intractable. Typically, one has to resort to EM simulations, e.g., Galerkin's method. Yet, these numerical methods become impractical for large antenna arrays due to unacceptable computational complexity. Consequently,  it is recommended to approximate the continuous channel using discretization methods or alternative approximate eigenfunctions.

\subsubsection{Integral Over Discretized Spaces} 
Discretizing the transmitter and receiver spaces is an effective solution to address eigenfunction problems. This transforms the continuous eigenfunction problem into an SVD problem with a finite and discretized dimension  \cite{9139337}.  
 
Specifically, for each discretized patch, the integral counterpart of channel matrix $\mathbf{H}$ is $\mathcal{H}: L^2(\mathbb{R}^2)\rightarrow L^2(\mathbb{R}^2)$, is expressed by 
\begin{equation}
    \begin{aligned}
      (\mathcal{H}x)(\mathbf{r})=\int h(\mathbf{r},\mathbf{t}) x(\mathbf{s}) \mathrm{d}  \mathbf{s},
    \end{aligned}
\end{equation}
where $h(\mathbf{r},\mathbf{t})$ is the coupling coefficient between the transmitter and receiver surfaces, and $h(\mathbf{r}_m,\mathbf{t}_n)=\mathbf{H}_{mn}$ in the discrete case. 

With the aid of Riemann sum approximation, the singular value relation between the continuous and discretized representations is given by \cite{9973178}
\begin{equation}
    \sigma_k(\mathbf{H})\approx \sqrt{\frac{N_t N_r}{\mathcal{A}_t \mathcal{A}_r}}  \sigma_k  (\mathcal{H}),
\end{equation}
 where $\mathcal{A}_t$ and $\mathcal{A}_r$ are the Lebesgue measure of transmitter and receiver on $\mathbb{R}^2$. 
 
 However, this method relies on the small-aperture approximation and assumes an unblocked direct path between the transmitter and receiver. Therefore, extending this approach to large apertures imposes significant challenges on the computational costs. 

 \subsubsection{Other Choices of Approximate Eigenfunctions}
There are many choices for approximate eigenfunctions, including prolate spheroidal wave functions, Fourier modes \cite{10753352}, and focusing functions that enable transmitter antennas to focus energy efficiently \cite{9641865}. The selection of the appropriate approximate eigenfunctions depends on the specific communication scenario. 

By leveraging the characteristics of compact sets and integrable functions, the asymptotic behavior of singular values can be derived, along with the DOF analysis (detailed in Sec. VI).  

\textit{\textbf{Limitations:}} The continuous channel modeling relies heavily on the selection of eigenfunctions. However, solving for the optimal eigenfunctions by addressing the coupled eigenfunction problems is often intractable. Even though discretizing the transmitter and receiver into small segments and treating them as piecewise constant basis functions can reduce the computational burden to some extent, this approach becomes impractical for large apertures. Therefore, it is crucial to design feasible and near-optimal approximate eigenfunctions that balance computational efficiency with communication performance requirements.
\color{black}
%%%%%%%%%%%%%%%%%%%%%%%%%%%%%%%%%%%%%%%%%%%%%%%%%%%%%%%%%%%%%%%%%%%%%%%%%%%%%%%%%%%%%%%%%%%%%%%%%%%%%%%%%%%%%%%%%%%%%%%%%%%%%%%%%%%%%%%%%%%%%%%%%%%%%%%%%%%%%%%%%%%%%%%%%%%%%%%%%%%%%%%%%%%%%%
 
\section{Theoretical Study}
In this section, we will delve into SIT and KIT for HMIMO-oriented EIT, where both are powerful analysis tools for information measurement. As a traditional information theory, SIT adopts a probabilistic model to explore the interplay between input symbols and output \cite{Shannonmathematicaltheorycommunication1948}, while KIT employs a deterministic model to measure the information carried by continuous fields \cite{MiglioreRoleNumberDegrees2006}. These two theories possess distinct methodologies but share some common traits in performance evaluation, both contributing to the development of EIT. In this section, we will explore the fundamentals and the underlying relationship between Shannon's and Kolmogorov's information theory. By examining these two fundamental theories, we will provide the evaluation and improvement methods of DOF along with its applications in HMIMO systems.

\subsection{Shannon's Information Theory}
\begin{figure*}  
	\begin{center}
		{\includegraphics[width=0.8 \textwidth]{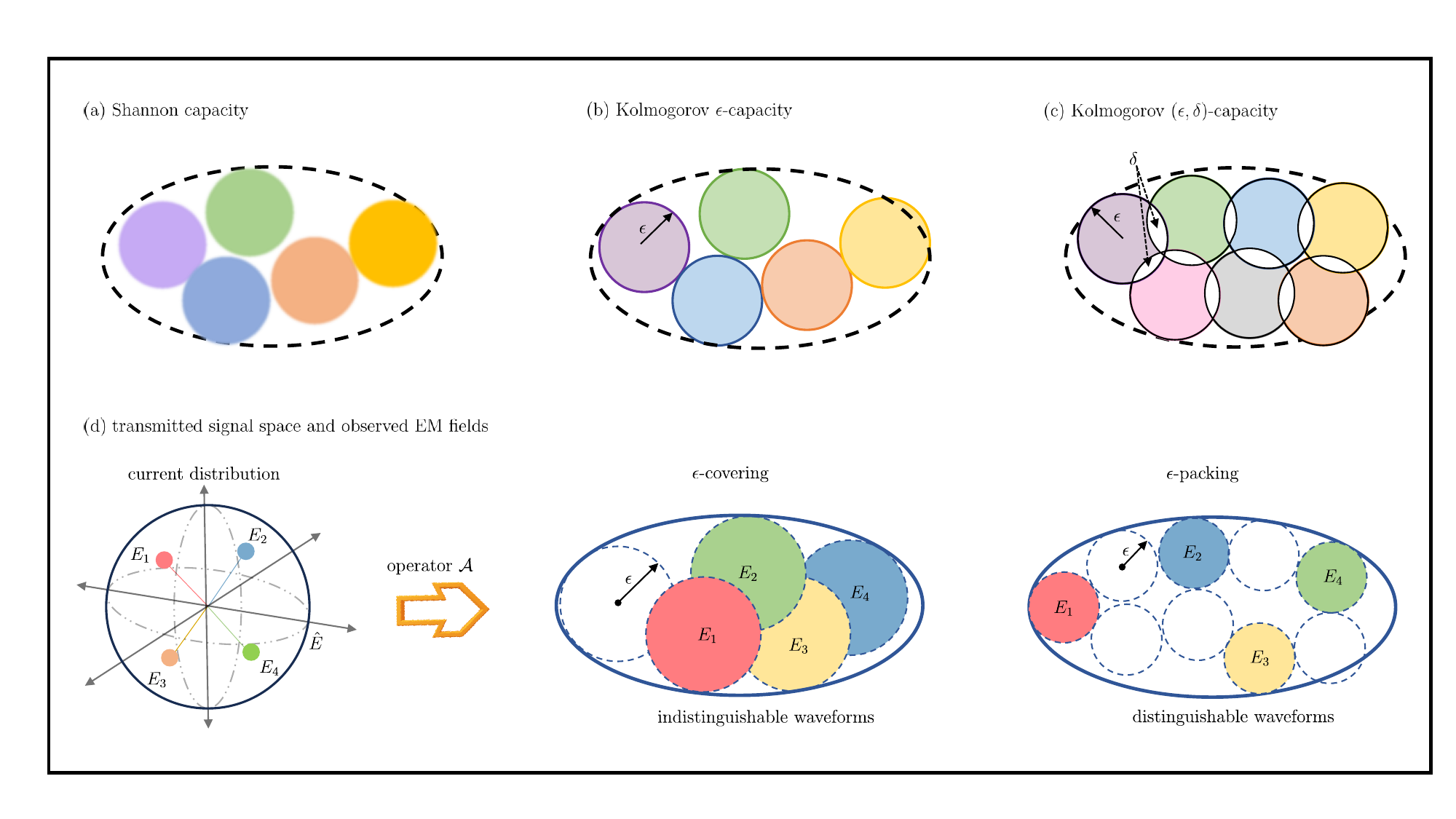}}  
		\caption{Geometric illustration of an abstract concept: (a) Shannon capacity; (b) Kolmogorov $\epsilon$-capacity; (c) Kolmogorov $(\epsilon,\delta)$-capacity; and (d) transmitted signal space and observed EM field.}  
		\label{fig:SIT_KIT_graph} 
	\end{center}
\end{figure*}

In SIT, the information measure, \textit{Shannon capacity}, for wireless communication systems is performed within a probabilistic framework \cite{Shannonmathematicaltheorycommunication1948}.   Shannon capacity can be reformulated as a packing problem, as shown in Fig.~\ref{fig:SIT_KIT_graph} (a). Specifically, the ${N}_{\text{SIT}}$-dimensional space has ${N}_{\text{SIT}}$ orthonormal basis functions, then the embedded signal represented by ${N}_{\text{SIT}}$ coefficients corresponds to a point in this functional space. Due to the probabilistic property, each point is surrounded by a soft ball, and the soft ball transforms into a hard ball for certain conditions (e.g., observation time $T\rightarrow \infty$ or bandwidth $B\rightarrow \infty$). The Shannon capacity is then given by $\log_2 {N}_{\text{SIT}}$, with $N_{\text{SIT}}$ being the number of balls.  The SNR is given by \cite{7805306}
\begin{equation}
\mathrm{SNR}_{\text{SIT}}=( P {N}_{\text{SIT}} )/{(\sigma \sqrt{{N}_{\text{SIT}}})^2},
\end{equation}
where the power of time-limited signal is $P{N}_{\text{SIT}}$ ($P$ is the transmitted power of each symbol) and $\sigma \sqrt{{N}_{\text{SIT}}}$ is the radius of uncertainty ball.  

SIT necessitates knowledge of the probabilities of all variables within the set in measuring information, therefore,  it becomes difficult to individually investigate the contribution of each variable within the context of probabilistic SIT, which encourages the utilization of deterministic KIT, a functional analysis-based theory. 

% \subsection{ Kolmogorov's Information Theory} \label{subsec:KIT} 
% In KIT, Kolmogorov entropy/capacity aligns with Shannon's entropy/capacity, as both are quantitative approaches to information analysis \cite{MiglioreElectromagneticsInformationTheory2008,GrunwaldShannonInformationKolmogorov2004}. Clearly, each equation within the realm of Kolmogorov complexity finds its counterpart in Shannon entropy.
 
\subsection{Kolmogorov's Information Theory} 
In KIT, the square-integrable and band-limited signals are analyzed in deterministic models \cite{MiglioreRoleNumberDegrees2006,7805306}.  Specifically, the infinite-dimensional functional space is approximated by a finite-dimensional space with uncertainty $\epsilon$, involving a negligible information leak. The resulting information measurement is denoted by Kolmogorov $\epsilon$-capacity. 

\subsubsection{Kolmogorov $\epsilon$-capacity}
Similar to Shannon capacity, Kolmogorov $\epsilon$-capacity can also be recast as a sphere packing problem, as shown in Fig.~\ref{fig:SIT_KIT_graph} (b).  Different from the probabilistic model in Shannon’s theory, the deterministic setting is adopted. Therefore, the signal point in $N_{\text{KIT}}$-dimensional space is surrounded by a hard ball with radius $\epsilon$, where the uncertainty $\epsilon$ mainly comes from imperfect antennas and imprecise measurements. Similarly, the Kolmogorov $\epsilon$-capacity is $\log_2 N_{\text{KIT}}$, with $N_{\text{KIT}}$ being the number of balls. If the transmitted signals are square-integrable with maximal energy $\hat{E}$, any transmitted signal can be seen as a point in the hypersphere with radius $\sqrt{\hat{E}}$. The SNR ratio of the band-limited signal is then given by 
\begin{equation}
\mathrm{SNR}_{\text{KIT}}=\hat{E}/{\epsilon^2},
\end{equation}
with uncertainty sphere of radius $\epsilon$ enclosing each signal point.  

 \subsubsection{Kolmogorov $(\epsilon,\delta)$-capacity} 
The geometric representation of Shannon capacity and Kolmogorov is the same but the error probability in SIT is not described in deterministic KIT. Consequently, to connect SIT and KIT, the normalized error region of size $\delta$ is introduced to emulate the vanishing error in the deterministic model, i.e., there exists overlap between $\epsilon$-balls, resulting in $(\epsilon,\delta)$-capacity \cite{7805306}, as shown in Fig.~\ref{fig:SIT_KIT_graph} (c). In this setting, the ratio between the overlap region to the $\epsilon$-ball is $\delta$, representing the decoding/detection error.

 In multi-user cases, the inter-user interference is introduced, thus the noise $\epsilon$ plus interference term $\sqrt{\mathcal{I}}$ has a larger uncertainty radius $\epsilon_{\text{MU}}=\epsilon +\sqrt{\mathcal{I}} $, and the signal-to-interference-noise ratio (SINR) becomes 
\begin{equation}
\mathrm{SINR}_{\text{KIT}}=\hat{E}/({\epsilon^2+\mathcal{I}}),
\end{equation}
where the interference term $\mathcal{I}$ is independent of the noise $\epsilon$. Equivalently, it requires partial DOF in Kolmogorov $\epsilon$-capacity to eliminate the additional unwanted interference $\mathcal{I}$. 

\textit{Remarks:} Both the probabilistic  SIT and deterministic KIT  are applicable to the temporal domain and spatial domain, attributing to the symmetry between space and time. Graphically speaking, they can assess the information of EM systems through the minimal number of unit balls that cover the ellipsoid, i.e., the solution of the $\epsilon$-packing problem. There exists little distinction between SIT and KIT in the formulated problem. Specifically, SIT is derived based on the probability model, resulting in soft balls in the $\epsilon$-packing without fixed radius $\epsilon$. As the observation measurement (e.g., time and bandwidth) approaches infinity, each ball is asymptotically fixed to a radius $\epsilon$  due to the law of large numbers. In contrast, KIT relies on functional analysis, and each ball within the  $\epsilon$-packing maintains a fixed radius of  $\epsilon$ without necessitating a large number of variables. This inherent difference necessitates both SIT and KIT for information measurement in various scenarios of wireless communications.   

% Both the probabilistic SIT and deterministic KIT are applicable to the temporal domain and spatial domain, attributing to the symmetry between space and time. However, since there is a slight asymmetry between the temporal domain and spatial domain, the KIT is still necessary as a complementary tool to SIT. For example, the DOF is limited by the effective rank of the time-frequency limiting operator resulting in a limited-length spatial block, thus, the optimal rate with the fixed error probability is difficult to obtain using SIT from a probabilistic perspective. Under such a circumstance, the deterministic KIT could provide a rigorous treatment of information.   

\subsection{Visualization of SIT and KIT in EM Systems} \label{subsec:Visual_EM}
 
The input-output relationship of the current distribution and electric field in \eqref{equ:em_in_out} is visualized in Fig.~\ref{fig:SIT_KIT_graph} (d). The relationship is characterized by the integral operator $\mathcal{A}: X \rightarrow Y$, which is an analytic and compact operator. The compact channel operator $\mathcal{A}$ is dependent on the environment, and it can be expanded using the Hibert-Schmidt decomposition, i.e. \cite{MiglioreElectromagneticsInformationTheory2008},
\begin{equation}
	A x=\sum_{k=1}^{\infty} \sigma_k\left\langle x, v_k\right\rangle u_k, 
\end{equation}
where  $u_k,v_k,\sigma_k$ are the left singular functions, the right singular functions, and the singular values of the operator respectively. $\langle\cdot, \cdot\rangle$ denotes the inner product in $L^2(X)$. The singular values $\sigma_1\geq \sigma_2 \geq \ldots$, and $\lim _{k\rightarrow\infty} \sigma_k \rightarrow 0$ due to the compactness of the operator. The eigenvalues of the operator exhibit step-like behavior and the transition region between the dominant eigenvalues and almost zero eigenvalues. Therefore, through the truncation method, the infinite eigenvalues can be approximated by the finite dominant eigenvalues, and this is equivalent to the number of packed balls ($\mathcal{N}_{\text{SIT}}$ or $\mathcal{N}_{\text{KIT}}$ in the former subsections).   

Graphically speaking, if the energy of the source is bounded, then the set of source currents is contained in a hypersphere with finite radius $\hat{E}$.  As shown in Fig.~\ref{fig:SIT_KIT_graph} (d),  the compact operator $\mathcal{A}$ transforms a sphere with radius $\hat{E}$ having infinite dimensions into a finite-dimensional ellipsoid with the $k$th semi-axis $\sigma_k\hat{E}$. Therefore, each point in the hypersphere that represents one current distribution generates a specific waveform in the field through the channel operator $\mathcal{A}$.  
 
Each ball in the ellipsoid represents one radiation pattern, if the two balls do not intersect, then these two patterns are distinguishable, otherwise, yield indistinguishable patterns or error-tolerant resolution. If all balls within the ellipsoid are non-overlapping, the $\epsilon$-covering case becomes the $\epsilon$-packing case, indicating error-free transmission systems. In a physical interpretation, the number of distinguishable balls in $\epsilon$-packing of $Y$ is exactly the maximum number of distinct waveforms,  where $\epsilon$ also presents the minimal distance between two different waveforms (disjoint balls).

Obviously, the information measurement in SIT (Shannon capacity) and KIT (Kolmogorov $\epsilon$-capacity) depends on the number of packed balls in the signal space, and this is exactly equivalent to DOF. Therefore, DOF provides insight into available resources in wireless communications, and thus becomes one of the important measures in performance evaluation, so we will delve into DOF in the following part. 

\subsection{Evaluation of DOF}
There are mainly two methods in deriving DOF: one is artificially truncating the dominant singular values of the operator \cite{MiglioreElectromagneticsInformationTheory2008, 7805306, 4447351} (as shown in Sec.~\ref{subsec:Visual_EM}), and the other is related to the number of nonzero optimal source $L^2$ norms \cite{4685903}. The former DOF is adopted in the majority of works and easily visualized as a sphere-packing problem, while the latter DOF appears as a byproduct of maximizing capacity. Consequently, if the singular values decay smoothly without a clear gap (e.g., near-field communications), truncating DOF is challenging and the latter method may be more appropriate. To ease the understanding of DOF in a more general case, we mainly discuss the former truncation-based evaluation of DOF. 

\begin{figure*}[!tbp]
    \centering
\begin{tcolorbox}[title = {Scatterers and Spatial DOF Gain},
  fonttitle = \bfseries]
 It is a well-accepted fact that additional spatial gain is achieved in the scattering environment.  However, the justification for such a spatial gain is controversial. 
 
 Specifically, R. Janaswamy \cite{5976389} proved that the presence of scatterers will slow the decaying rate of higher mode powers in the near field (e.g, decays algebraically), and the slowing is much more dramatic if the scatterers are closer to the observation circle, which further increases the DOF. Therefore, R. Janaswamy argues that much of the benefit of increased NDOF comes from scatters. However, Franceschetti et al. present a contradictory conclusion in \cite{7101845} that the information gain is purely a near-field effect that is independent of scatterers, especially for large-sized communication systems. 

 In fact, the conclusions drawn from \cite{5976389} and \cite{7101845} are valid in their respective regimes, as investigated in \cite{9737566} from the following aspects: 
\begin{enumerate}
    \item If a finite aperture is considered (as assumed in \cite{5976389}) in a rich scattering environment, the scatterers would subtend all possible angles on the observation plane. Therefore, the additional DOF is induced and attributed to backscattering effects, which supports the conclusion in \cite{5976389}. 
    \item On the other hand, if an infinite aperture (as assumed in \cite{7101845}) is considered in LOS-only propagation, the aperture also subtends all possible angles on the observation plane. Therefore, the scatterers cannot increase angular extent beyond LOS-only propagation, and the additional DOF gain is a purely near-field gain, which assures the conclusion in \cite{7101845}. 
    \item However, if the aperture is finite in the LOS-only case, the angles subtended by the aperture are much less than that in an infinite aperture, and the additional DOF is also decreased \cite{9737566}. 
\end{enumerate}
 Consequently, based on the relationship between the subtended angles and DOF, the contradictory conclusions drawn from \cite{5976389} and \cite{7101845} are explainable.  
\end{tcolorbox}
\end{figure*}

\subsubsection{Eigenvalues and DOF}
In wireless communications, the DOF is associated with singular values of the operators, and operators in different domains are connected with the Fourier transformation. Landau \cite{landau1975szego} and Splepian \cite{slepian1983some} investigate the asymptotic behavior of eigenvalues of the functions on different sets, providing novel insights into DOF computation of the signal space. Based on the variations of Landau's eigenvalue theorem in \cite{7159085}, we will present the DOF evaluation of systems in this part. 

Considering a circular source with radius $\tilde{r}$ (is normalized by the speed of light $c$, i.e., $\tilde{r}=r/c$, with $r$ being the physical radius that encompasses both source and scatterers) and there exists a cut-set boundary between the source and receiver, the radiated field is dependent on the measured field on the cut-set boundary \cite{7159085}. The duration period $t\in [-T/2,T/2]$ and angular frequency bandwidth is $f \in [-B, B]$. The radiated field is observed over the spatial interval $\phi\in [-\pi, \pi]$ and occupies a wavenumber bandwidth $w\in [-W,W]$, where the spatial bandwidth is $w=f\tilde{r} + o(f\tilde{r})$ as $f\tilde{r}\rightarrow \infty$. With the scaling parameters $\tau \rightarrow \infty$ and $\rho \rightarrow \infty$, we could obtain the band-limited signals as $T'=\tau T\rightarrow \infty$ and time-limited signals as $B'=\rho B \rightarrow \infty$. 

Following the variation of Landau's eigenvalue theorem in \cite{7159085}, the DOF of the observed field radiated by the band-limited source and time-limited source are \cite{7159085}
\begin{equation} \label{equ:band-limited}
\mathcal{N}_{\epsilon}^{\text{BL}}= \frac{BT}{\pi} \cdot \frac{2\pi \tilde{r} B}{2\pi}+ o(\tilde{r}T), \quad \tilde{r}, T\rightarrow \infty,
\end{equation}
and \cite{7159085}
\begin{equation} \label{equ:time-limited}
    \mathcal{N}_{\epsilon}^{\text{TL}}=\frac{BT}{\pi} \cdot \frac{2\pi \tilde{r} B}{2\pi}+ o(B^2), \quad B \rightarrow \infty,
\end{equation}
respectively. The uncertainty $\epsilon$ is embedded in the phase transition of the eigenvalues. This transition region is centered at $N_{\epsilon}$ with a width that becomes negligible as $\tilde{r}, T\rightarrow \infty$ (for band-limited signal) or $B\rightarrow \infty$ (for time-limited signal), regardless of the value of $\epsilon$.   The above two DOFs are the same but the negligible terms. 

The first term in product is the temporal DOF of a one-dimensional signal, i.e., time-limited signal $s(t)$ with period $[-T/2, T/2]$ or band-limited signal $S(f)$ with bandwidth $[-B, B]$, given by \cite{7159085}
\begin{equation} \label{equ:temporalDOF}
    \mathcal{N}_{\epsilon}^{\text{time}}=\frac{BT}{\pi}+o(B T), \quad B T \rightarrow \infty.
\end{equation}
It is evident that there are primarily two approaches to enhancing temporal DOF: one is to improve bandwidth (i.e., constrained by the minimum quality factor $Q$) and the other is to increase the observation time. Theoretically, temporal DOF is unbounded due to infinite observation time. 

The second term in product is the spatial DOF of space-wavenumber signal $s(\phi)$ radiated with frequency $f$ and observed on the circular perimeter boundary of angel $\phi \in [-\pi, \pi]$ given by \cite{7159085}
\begin{equation} \label{equ:spatialDOF}
    \mathcal{N}_{\epsilon}^{\text{space}}=\frac{f 2\pi \tilde{r}}{\pi}+o(f \tilde{r}), \quad f \tilde{r} \rightarrow \infty. 
\end{equation}
There are mainly two methods to increase the spatial DOF of HMIMO systems: one is to increase the physical size of HMIMO, and the other is to exploit the multi-path effects in the scattering environment. The former method is costly and unachievable in various scenarios while the latter method is much more practical. It should be noted that the spatial gain due to scatterers deserves to be discussed in-depth, as discussed in the text box ``Scatterers and Spatial DOF Gain".

The product of the temporal DOF in \eqref{equ:temporalDOF} and spatial DOF in \eqref{equ:spatialDOF} is the total DOF of the space-time field given in \eqref{equ:band-limited} (or \eqref{equ:time-limited}), which incorporates the total number of spatial DOF that all frequency components carry. However, the accumulation of the error $\epsilon$ is not evident in the product of temporal-spatial signals, which necessitates the joint description. More importantly, decreasing the minimum quality factor results in reduced directivity gain $G$ (detailed in Sec. III), leading to capacity decrease despite an improved temporal DOF. This capacity loss can be explained by the joint contributions of temporal and spatial DOF. Specifically, although the temporal DOF improves with a lower $Q$, the spatial DOF is decreased greatly due to lower directivity gain. Therefore, increasing the bandwidth may not always improve the total DOF.  

\textit{Remarks:} Although the temporal DOF and spatial DOF share similarities, the practical enhancement of these DOFs exhibits distinct challenges. As previously mentioned,  improving the temporal DOF through bandwidth expansion is costly, while the spatial DOF is limited by physical antenna constraints. Therefore, the spatial DOF is imposed to a deterministic limit and it is upper-bounded, whereas the temporal number of DOF remains unbounded for an arbitrarily long temporal interval. Even if the observation domain spans the whole space, the number of DOF is still upper-bounded due to the knee behavior exhibited by the compact operator $\mathcal{A}$,  where the singular values of $\mathcal{A}$  decrease rapidly after
the knee value \cite{MiglioreElectromagneticsInformationTheory2008}.  Specifically, for singular values, the operator demonstrates smooth variation (slow decreasing rate) in the near-field region and steep variation (fast decreasing rate) in the far-field region.

\subsubsection{Spatial DOF of HMIMO systems}
The HMIMO systems are characterized by continuous apertures in the space domain, so we will discuss spatial DOF derivation that applies to HMIMO systems with arbitrary shapes. 

As proposed in \cite{1386525}, the spatial DOF is dependent on the array size $ \mathcal{A} $ (i.e., the array domain of the excitation distributions) and solid angles $|\Omega|$ (i.e., the angular domain of the radiated field patterns).  Specifically, the spatial DOF of HMIMO systems is $ \mathcal{A} |\Omega|$ for $ \mathcal{A}  |\Omega|\rightarrow \infty$, with $\mathcal{A}$ being the area of the HMIMO systems. $ \mathcal{A} =2L$ for a $2L$-length linear HMIMO, $ \mathcal{A} =2R$ for a circular HMIMO with radius $R$, and $\mathcal{A}=\pi R^2$ for a spherical HMIMO with radius $R$.  The aperture of the transmitter and receiver are $\mathcal{A}_t$ and $\mathcal{A}_r$, respectively. The solid angles observed from the transmitter and receiver are $ {\Omega}_t$ and $ {\Omega}_r$, respectively. Then, the spatial DOF of the communication systems is dependent on the minimum spatial DOF at transmitter and receiver \cite{1386525}, i.e., 
\begin{equation}
    \mathcal{N}^{\text{space}}=\min \left\{ \mathcal{A}_t |\Omega_t|,  \mathcal{A}_r |\Omega_r| \right\},
\end{equation}
for $\mathcal{A}_t |\Omega_t|\rightarrow \infty, \mathcal{A}_r |\Omega_r|\rightarrow \infty$. It should be noted that the spatial DOF could be further increased for dual-polarization or tri-polarization antenna arrays.

\begin{figure}  
	\begin{center}
		{\includegraphics[width=0.5 \textwidth]{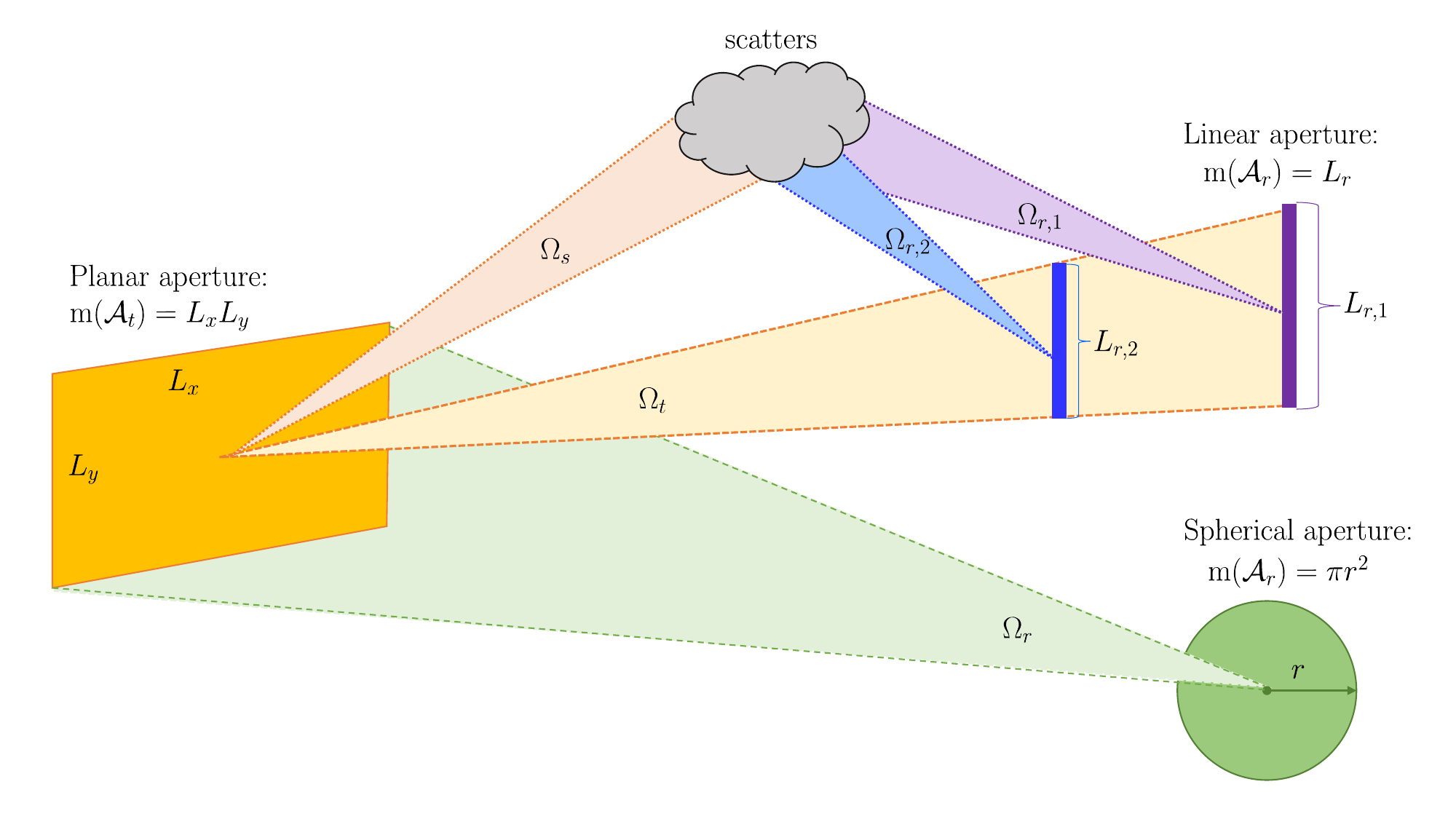}}  
		\caption{Illustration of the angular domain and array domain for transmitter, receiver, and scatters.}  
		\label{fig:DOF_aperture} 
	\end{center}
\end{figure}

This provides insights in analyzing DOFs of HMIMO systems with different settings. For example, as shown in Fig.~\ref{fig:DOF_aperture}, we consider two linear receivers with apertures $L_{r,1}>L_{r,2}$ and solid angles $|\Omega_1| = |\Omega_2|$ communicating with the same planar HMIMO systems with aperture $\mathcal{A}_t$ and solid angle $|\Omega|_t$. Then, the spatial DOFs of the two cases are 
\begin{equation}
    \begin{aligned}
        & \mathcal{N}^{\text{space}}_{(1)}=\min \left\{ \mathcal{A}_t |\Omega_1|,  L_{r,1} |\Omega_t| \right\},\\
&\mathcal{N}^{\text{space}}_{(2)}=\min \left\{ \mathcal{A}_t |\Omega_2|,  L_{r,2} |\Omega_t| \right\},
    \end{aligned}
\end{equation}
where $ \mathcal{A}_t  |\Omega_1| =  \mathcal{A}_t  |\Omega_2|  $ and  $L_{r,1} |\Omega_t| > L_{r,2} |\Omega_t|$. Therefore, if the DOF at the transmitter is smaller than that at the receiver, the larger aperture $L_{r,1}$ would not bring additional spatial DOF compared with aperture $L_{r,2}$. Otherwise,  the larger observation lengths or larger subtended angles could increase the amount of independent information that can be retrieved with minimal error \cite{pierri1998information}.  

As illustrated in the former subsections, the spatial DOF can be transformed into sphere-packing problems in the signal space.  Given that the volume of the $d$-dimensional unit ball is \cite{9737566}
\begin{equation}
    \beta_d=\frac{\pi^{\frac{d}{2}}}{\Gamma\left(\frac{d}{2}+1\right)},
\end{equation}
where $\Gamma(\cdot)$ is the gamma function, so the volume of 1D ball, 2D ball, and 3D ball are $2, \pi$, and $\frac{4}{3}\pi$, respectively. Then, the spatial DOF is the number of the packed unit balls in the hyper-ellipsoids of radii $(\delta_x,\delta_y,\delta_z)$  along the $x-,y-$ and $z-$ directions, which is dependent on the physical size of antennas and frequency. According to Definition 5 in \cite{9737566}, the measure of the space $\mathcal{A}$ and spectra $\Omega$ is denoted by $\mathrm{m} (\mathcal{A} \times {\Omega})$, then the spatial DOF is mathematically computed as  
\begin{equation}
   \mathcal{N}^{\text{space}} =\frac{ \mathrm{m}(\mathcal{A} \times \Omega)}{\beta_d \cdot\left(\frac{\delta_x}{2}\right) \cdot\left(\frac{\delta_y}{2}\right) \cdot\left(\frac{\delta_z}{2}\right)}.
\end{equation}

Therefore, increasing the array aperture $\mathcal{A}$ and solid angle $\Omega$ are two approaches to improve the spatial DOF of HMIMO systems. However, the former is usually infeasible due to the confined application scenarios, thus increasing solid angles is much more favorable. For example, the introduction of relay systems or cooperative techniques could effectively increase solid angles and DOF. Additionally, the multi-path effects can also be deployed for diversity and multiplexing, attributing to the larger solid angles, as discussed in the text box ``Scatterers and Spatial DOF Gain". Notably, such an increase in solid angels is up to $2\pi$ for 2D scenarios and $4\pi$ for 3D scenarios.  

The spatial DOF in the scattering environment can also be interpreted in terms of distinct paths, considering antenna array configurations and clusters. Paths are considered distinct if they can be differentiated in either transmit or receive virtual angles, and these distinct paths contribute to both capacity and diversity \cite{1033686}. Specifically, the number of transmit and receive virtual angles is equal to the rank of the channel matrix, and the level of diversity associated with each parallel channel is determined by the number of virtual receive angles that couple with each virtual transmit angle. Theoretically, increasing antenna spacing results in more virtual angles interacting with the scatters, which enhances the level of diversity. However, this improvement also introduces interference due to overlapping clusters, providing a physical interpretation of spatial DOF in HMIMO systems.

\subsection{Applications of DOF} 
The number of spatial DOF provides guidelines for antenna design from the perspective of spatial sampling.  For example, in the far-field LOS communications, the lower bound on spatial sampling is $\lambda/2$ for infinite aperture and the optimal aperture sampling is larger than the $\lambda/2$ for finite size \cite{9737566}. Therefore, if we adopt HMIMO in the far-field sampling case, it is always over-sampling, enabling higher data transmission. If HMIMO systems are located in sufficiently rich scattering and the NLOS-only environment,  the available spatial DOF further increases as the solid angles increase and the sampling spacing improves as well, thus more information can be carried in HMIMO systems.

 In addition to the guidelines in the efficient sampling schemes or antenna configurations, we can also allocate DOF for different purposes. For example, in the joint imaging and communications systems, assuming the DOF is $N_0=N_0^{ \text{comm} }+N_0^{\text{imag}}$, we can adopt $N_0^{ \text{comm} }$ for transmitting data, and the rest $N_0^{\text{imag}}$ DOFs in imaging resolution, as proposed in \cite{9724249}. In addition, if the spatial overlap exists between the imaging plane and the channel clusters, the imaging performance can be improved at the cost of a smaller decoding rate since the partial DOF in communications is allocated to image reconstruction.

Similarly, the DOF can also be exploited in multi-user communications, some DOFs are used for data transmission, and the rest are for interference-cancellation. Moreover, the active illumination on receivers and intelligently controlled scatterers can be adopted to induce additional DOF in both near-field and far-field zones. Therefore, the multiple HMIMOs can be configured in wireless communications to artificially increase DOF and achieve higher data transmission.

\color{blue}

% We can also interpret DOF in another way. The sequential filter consisting of temporal filter $Q(t)$ and spectral filter $R(\omega)$ is adopted, i.e.,
% \begin{equation}
%     \int \mathrm{d} \omega \frac{|R(\omega)|^2}{2\pi}=B, \qquad \int \mathrm{d} t |Q(t)|^2=T,
% \end{equation}
% resulting in the time-bandwidth product \cite{raymer2020time}
% \begin{equation}
%     \sum_{n=0}^{\infty} \lambda_n^2 = \int \mathrm{d} t |R(t)|^2 \int \mathrm{d} t |Q(t)|^2 =BT,
% \end{equation}
% characterizing the bandwidth and duration of the filters. 

% There are mainly two types of constraints to be considered in information capacity derivation: the source energy constraint (source $L^2$ bound) and the radiated power constraint. The bounded source energy result in bounded radiated power, but the opposite does not hold. The unconstrained source $L^2$ norm may result in non-convergent capacity (due to the unconstrained superdirectivity) for most configurations of antennas. The bounded radiation power alone may also result in divergent capacity.   

% Artificially limiting the NDF is a way to constrain the source $L^2$ norm. 

\color{black}

% Consequently, for a temporal signal modulated by different functional points in signal space (current distribution), both temporal DoF and spatial DoF contribute to the overall DoF, i.e., improving the observation time and radiation pattern modes can enhance the DoF. The radiation pattern mode is primarily controlled by the EM characteristics of transceivers, such as optimal current density design and field generation. Moreover, the environment significantly impacts performance improvement. Since scatters can reflect EM waves and act as secondary sources, thus, complex environments with abundant scatters tend to exhibit higher DoF compared to environments lacking such secondary sources.  Consequently, controllable scattering objects are also expected to induce higher DoF. For example, adopting HMIMO to transmit EM waves to the desired direction with less energy consumption while increasing DoF simultaneously.

% \color{red}
% In information theory, it is typically believed that the power gain (or channel gain) is proportional to the increase in the number of antennas, this is one of the motivations in HMIMO. However, 

\color{black}
%%%%%%%%%%%%%%%%%%%%%%%%%%%%%%%%%%%%%%%%%%%%%%%%%%%%%%%%%%%%%%%%%%%%%%%%%%%%%%%%%%%%%%%%%%%%%%%%%%%%%%%%%%%%%%%%%%%%%%%%%%%%%%%%%%%%%%%%%%%%%%%%%%%%%%%%%%%%%%%%%%%%%%%%%%%%%%%%%%%%%%%%%%%%%%
\section{Numerical Evaluation} 
In this section, the performance analysis of the EM-compliant HMIMO models is conducted. Specifically, two performance matrices, namely DOF and capacity, are employed to assess HMIMO systems across different scenarios.  In addition, the sphere-packing solution is also simulated.  
\subsection{DOF and Eigenvalues Evaluation}
Considering a fixed surface area, constantly increasing antennas would not bring the expected DOF improvement, and the DOF would finally reach the limit along with the deformed antenna pattern due to the strong coupling effect.  As shown in Fig.~\ref{fig:DoF_FixedSurface}, which depicts the DOF of the generated channel with varying numbers of transmit antennas in LoS scenarios. The HMIMO  surface area is $ 64\lambda^2$ (square shape), and we consider transmitter-receiver distance at $r=5\lambda,7\lambda$ and $9\lambda$, respectively.  From the figure, it is evident that HMIMO in the near field exhibits the highest DOF, primarily because the third polarization component possesses the strongest power. However, as the distance increases, the near-field region moves to the far-field region, resulting in a considerable decrease in DOF. Additionally, the DOF does not continuously improve with an increasing number of transmit antennas. This behavior can be attributed to the distorted antenna pattern and the strong coupling effects. 

An intriguing finding from \cite{YuanEffectsMutualCoupling2023} suggests that the strong mutual coupling can also slightly enhance DOF, likely due to efficiency reduction and pattern deformation. Furthermore, the spatial DOF of aperture-constrained HMIMO is proportional to the surface area for both far-field and near-field HMIMO systems \cite{PizzoDegreesFreedomHolographic2020,WeiMultiUserHolographicMIMO2022}. This relationship can be explained by the larger transmitter/receiver area's ability to transmit/collect more power, leading to a larger DOF and more dominant eigenvalues of channels.

%\iffalse
\begin{figure}  
	\begin{center}
		{\includegraphics[width=0.45 \textwidth]{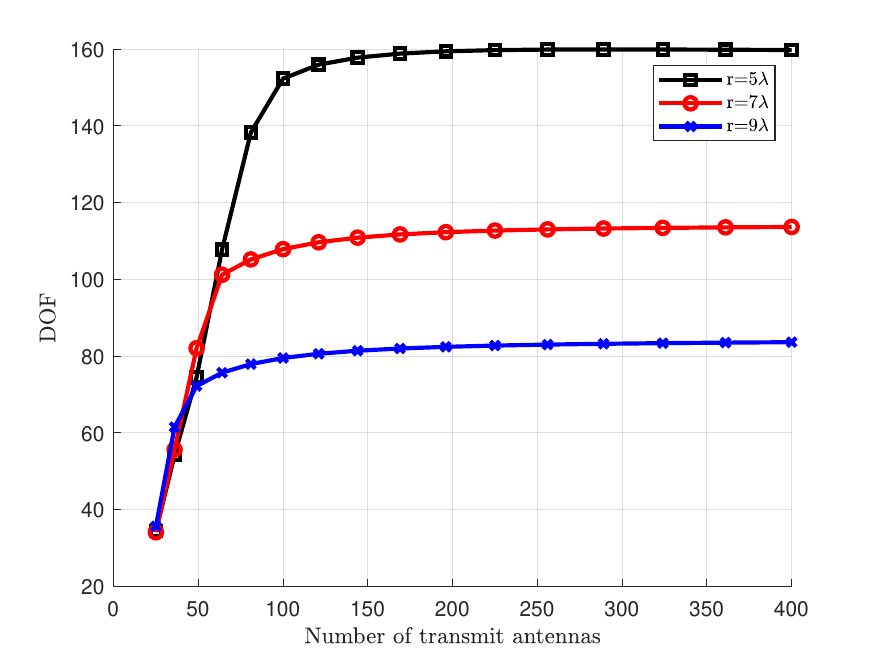}}  
		\caption{DOF v.s. number of antennas within the fixed transmit surface at different distances $r$.}
		\label{fig:DoF_FixedSurface} 
	\end{center}
	\vspace{-0.5cm}
\end{figure}  
%\fi

For a more comprehensive analysis of DOF evaluation in both far-field and near-field communications, we investigate channel eigenvalues under these two conditions, since eigenvalues clearly reflect the number of independent channels (i.e., DOF) and the gain of each independent channel.

\iffalse
\begin{figure*}[!tbp]
	\centering
	\subfloat[]{\label{fig:DoF_FixedSurface}\includegraphics[width=0.92\columnwidth]{Figures//DoF_FixedSurface.pdf}} \quad
	\subfloat[]{\label{fig:FarField_Eigenvalue}\includegraphics[width=0.92\columnwidth]{Figures//FarField_Eigenvalue.pdf}} \quad
	\subfloat[]{\label{fig:NearField_Eigenvalue}\includegraphics[width=0.92\columnwidth]{Figures//eigenvalue1lambda.pdf}} \\
	%\subfloat[Arabic numerals]{\label{fig1:c}\includegraphics[width=0.6\columnwidth]{Figures//nMSE_ElementNum.pdf}}
	%\vspace{-0.3em}
	\caption{Eigenvalue demonstration of channel matrices with respect to different transmit and receive elements at the TX-RX distance of (a) $2.4427 \lambda$, (b) $10.1803 \lambda$, and (c) $39.2 \lambda$, respectively.}
	\label{fig:Main_EigenValue_Demon}
	%\vspace{-0.4em}
\end{figure*}
\fi

The eigenvalue of far-field NLoS HMIMO channels for different spacing at the receiver is depicted in Fig.~\ref{fig:FarField_Eigenvalue}. The figure shows an uneven distribution of coupling coefficient intensities. The initial eigenvalues are substantial but rapidly converge towards zero, indicating the emergence of spatial correlation within the far-field HMIMO channel. The more uneven the coupling coefficients and the steeper the eigenvalues decay, which implies a stronger correlation and less DOF.  In addition, smaller spacing between antennas results in a stronger coupling effect. Consequently, fewer dominant eigenvalues are present, and their corresponding strengths are diminished. To provide an example, the HMIMO configuration with a spacing of  $\lambda/6$ exhibits a smaller number of nonzero eigenvalues and weaker strengths when compared to the  $\lambda/4$ and  $\lambda/3$ cases.  
%\iffalse
\begin{figure}  
	\begin{center}
		\includegraphics[width=0.45 \textwidth]{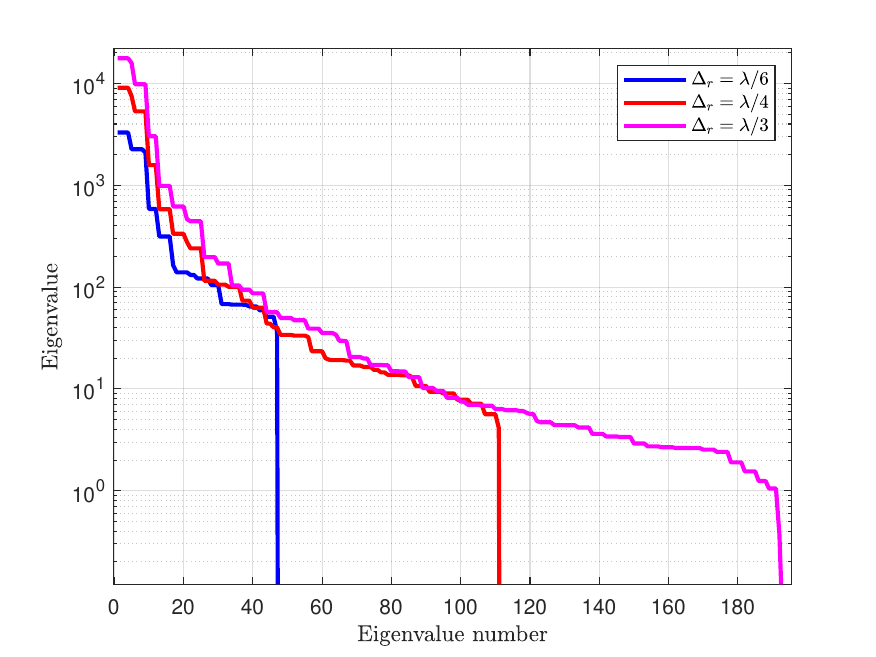} 
		\caption{Eigenvalues of far-field NLoS HMIMO systems for different spacing at the receiver.}
		\label{fig:FarField_Eigenvalue} 
	\end{center}
 \vspace{-0.5cm}
\end{figure}
%\fi

The eigenvalue characteristics of near-field LoS HMIMO channels, as depicted in Fig.~\ref{fig:NearField_Eigenvalue}, are notably more intricate compared to those of far-field HMIMO channels illustrated in Fig.~\ref{fig:FarField_Eigenvalue} due to the involvement of polarized components, which involves co-polarized channels (XX, YY, and ZZ polarized components) and cross-polarized channels (XY, YZ, and XZ polarized components). It can be observed from the figure that the third co-polarized component (ZZ) in the near field displays dominant values similar to the other two co-polarized components (XX and YY). In addition, the eigenvalues of cross-polarized components retain non-zero values, which degrades the performance (e.g., DOF and capacity).   
%\iffalse
\begin{figure}  
	\begin{center}
		{\includegraphics[width=0.45 \textwidth]{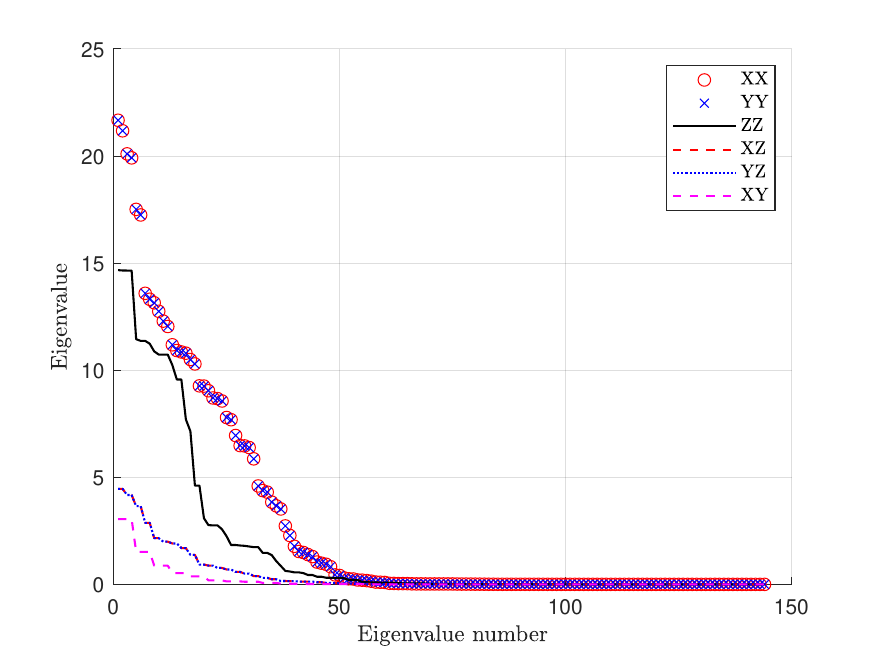}}  
		\caption{Eigenvalues of near-field LoS HMIMO systems for co-polarized and cross-polarized channel components at distance $r=\lambda$.}
		\label{fig:NearField_Eigenvalue} 
	\end{center}
	\vspace{-0.5cm}
\end{figure}  
%\fi

%\subsection{Capacity and Spectral Efficiency Analysis}  
 
\subsection{Capacity Evaluation} 
The capacity is defined as the maximum mutual information, and the mutual information is the uncertainty reduced by observations \cite{Shannonmathematicaltheorycommunication1948}. In order to measure the capacity of continuous-space communication channels, the basis functions can be employed to approximate continuous space in transmitter and receiver \cite{4447351}, then the input-output relationship is represented in discrete forms with mappings between transmit and receive eigenfunctions, thus facilitating the capacity with finite eigenfunctions.  A similar method is adopted in \cite{WanMutualInformationElectromagnetic2023}, where the current density and generated field are decomposed in orthogonal basis functions. It analyzed the mutual information and the capacity through spatial spectral density for random EM fields between two continuous regions, and simulation results proved the suboptimality of the half-wavelength sampling.

To improve the feasibility of capacity analysis, multiple physical constraints are imperative. For example,  the radiated power constraint alone is insufficient, and the source current constraint is also imperative in deriving practical capacity \cite{4685903}. This can be attributed to the non-uniqueness of the mapping between the radiated power and source excitation, i.e., a constrained radiated power is not necessarily generated from a constrained source $L^2$ norm. By incorporating the both source $L^2$ norm constraint and power constraint, the work in \cite{4685903} maximizes the mutual information between the source distributions and the observed field for practical estimates of capacity. Recently, the physical bounds of capacity computation for several shapes  (i.e., plate, disk, cylinder, and sphere) in wireless communications are further investigated in \cite{8998551}. Subsequently, \cite{EhrenborgCapacityBoundsDegrees2021} takes the $Q$-factor constraint into capacity analysis to impose bandwidth requirements of the antenna array. However, the current capacity analysis in wireless communications is still environment-central and ignores the impacts of antenna design, thus the integration of these physical constraints in wireless communications is still in exploration.  

In the following, we will provide numerical capacity analysis for far-field and near-field HMIMO systems. 

\color{black}

\subsubsection{Capacity of Far-Field NLoS HMIMO} 
\begin{figure}  
	\begin{center}
		\includegraphics[width=0.45 \textwidth]{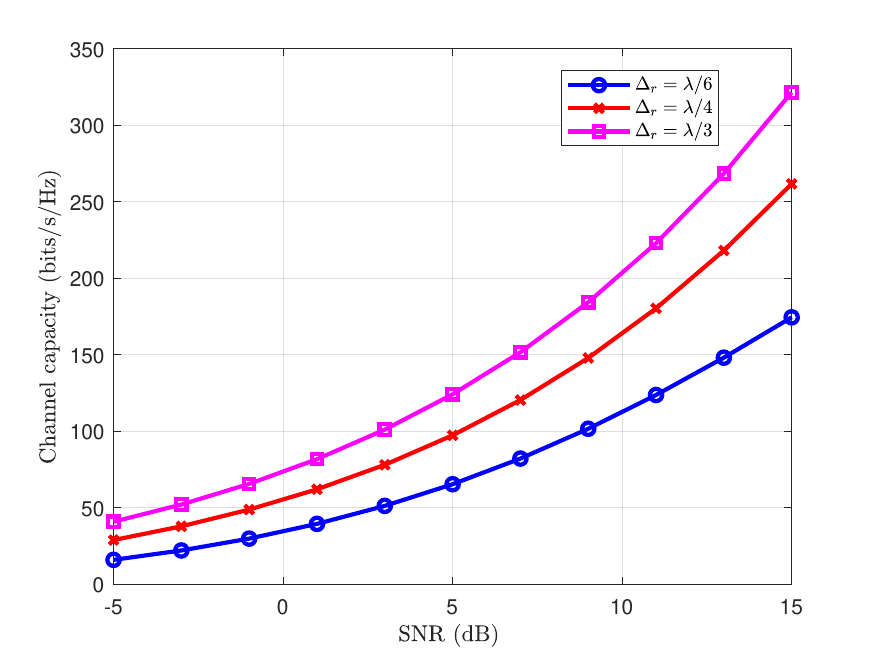} 
		\caption{Capacity of far-field NLoS HMIMO scheme for different spacing at the receiver.}
		\label{fig:FarField_Capacity} 
	\end{center}
 \vspace{-0.5cm}
\end{figure}

\begin{figure}  
	\begin{center}
		\includegraphics[width=0.45 \textwidth]{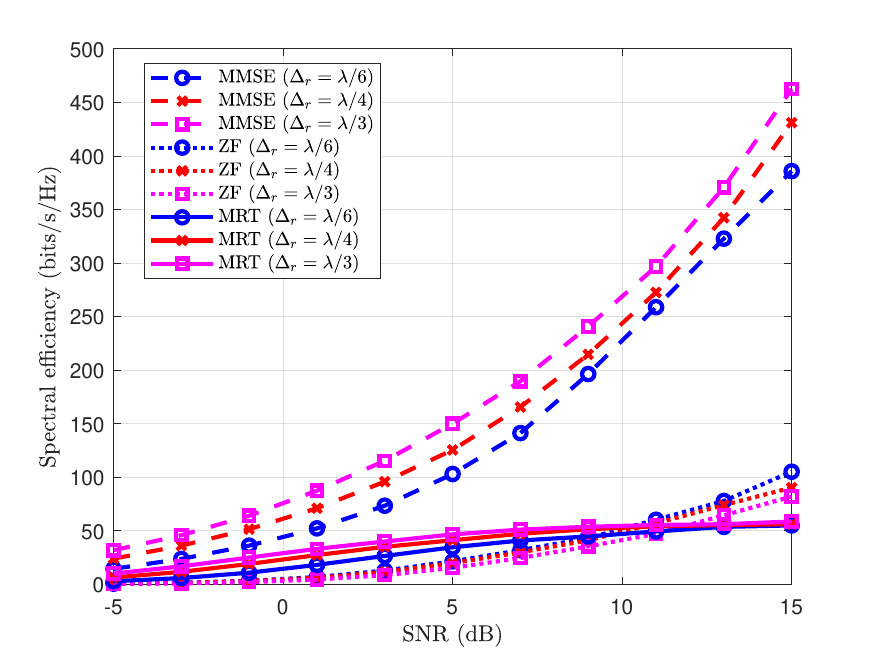} 
		\caption{Spectral efficiency of far-field NLoS HMIMO scheme with MMSE, ZF, and MRT precoding schemes for different spacing at the receiver.}
		\label{fig:Ns3600Nr144r3_MMSE_ZF_MRT} 
	\end{center}
 \vspace{-0.5cm}
\end{figure}

The channel capacity for far-field HMIMO systems with varying spacing at the receiver is depicted in Fig.~\ref{fig:FarField_Capacity}. The less spacing has a lower capacity due to the stronger coupling effects. For example, the curve with spacing $\lambda/6$ achieves $180$ bits/s/Hz at a SNR $15$dB. To mitigate the impact of spatial correlation, effective precoding techniques become imperative. Three predominant conventional methods are commonly employed: minimum mean square error (MMSE), zero-forcing (ZF), and maximum ratio transmission (MRT) schemes.  The spectral efficiency of HMIMO systems equipped with these three linear techniques are shown in Fig.~\ref{fig:Ns3600Nr144r3_MMSE_ZF_MRT}. Clearly, incorporating such schemes could greatly enhance spectral efficiency. However,  it's important to note that due to the compact antennas utilized in HMIMO systems, the implementation of precoding schemes entails resource-intensive operations, particularly in cases involving matrix inversions, such as MMSE and ZF precoding methods, and the exploration of efficient precoding designs remains an ongoing area of investigation.

\subsubsection{Capacity of Near-Field LoS HMIMO} Based upon the relation between the electric field and current distribution in \eqref{equ:em_in_out}, the communication model can be reformulated as \cite{10500751}
\begin{align}
	\mathbf{E} = \mathbf{H} \mathbf{J} + a_{R} \mathbf{W},
\end{align}
where $\mathbf{E}$, $\mathbf{J}$ and $\mathbf{W}$ are vectors consisting of the electric field, current distribution, and noise from all antenna elements, respectively; $a_{R}$ denotes the receive element area. Specially, the $m$-th noise element of $\mathbf{W}$ follows from the complex Gaussian distribution, i.e., $\mathbf{W}_{m} \sim \mathcal{CN}(0, \sigma_{w}^{2} {{\mathbf{I}}_3})$, where $\sigma_{w}^{2}$ is the noise variance and ${{\mathbf{I}}_3}$ is a $3 \times 3$ identity matrix. The $(m,n)$-th element of channel matrix $\mathbf{H}$ is given by $\mathbf{H}_{mn} \triangleq -j \omega \mu \int\nolimits_{a_R} \int\nolimits_{a_T} {\bar{\mathbf{G}}\left( {{{\mathbf{r}}_m},{{\mathbf{s}}_n}} \right)} {\rm{d}}{{\mathbf{s}}_n} {\rm{d}}{{\mathbf{r}}_m} \approx -j \omega \mu \cdot a_R a_T \bar{\mathbf{G}} \left( \mathbf{d}_{mn} \right)$ \cite{10500751}, where $a_T$ denotes the transmit element area. 
Assume that the current distribution $\mathbf{j}$ is expressed as a weighted sum of transmit patterns $\mathbf{u}_{p} \in \mathbb{C}^{N \times 1}$, and that the electric field is expressed as a weighted sum of receive patterns $\mathbf{v}_{p} \in \mathbb{C}^{M \times 1}$, where the weights correspond to the transmit symbols $c_{p}$ and receive symbols $\hat{c}_{p}$, respectively. Denoting $\mathbf{T} = [\mathbf{u}_{1}, \mathbf{u}_{2}, \cdots, \mathbf{u}_{P}]$, $\mathbf{R} = [\mathbf{v}_{1}, \mathbf{v}_{2}, \cdots, \mathbf{v}_{P}]$, $\mathbf{c} = [c_{1}, c_{2}, \cdots, c_{P}]^{T}$ and $\hat{\mathbf{c}} = [\hat{c}_{1}, \hat{c}_{2}, \cdots, \hat{c}_{P}]^{T}$, we get the communication model between the transmit and receive symbols as follows \cite{10500751}
\begin{align}
	\hat{\mathbf{c}} = -j \omega \mu \cdot a_R a_T \mathbf{R}^{H} \bar{\mathbf{G}} \mathbf{T} \mathbf{c} + a_R \mathbf{R}^{H} \mathbf{W}.
\end{align}
Given the bilinear decomposition of $\bar{\mathbf{G}} = \mathbf{R} \mathbf{D} \mathbf{T}^{H}$, where $\mathbf{D}$ is a diagonal matrix with its $p$-th element given by $\gamma_{p} \approx a_{R} a_{T} \mathbf{v}_{p}^{H} {\bar{\mathbf{G}}} \mathbf{u}_{p}$, the channel capacity can be written as and upper bounded by \cite{10500751}
\begin{align}
	\nonumber
	C &= \sum\limits_{p = 1}^P {\log_{2} \left( {1 + \rho  \mathrm{SNR}  \gamma_{p}^2} \right)} \\
    \nonumber
    &\le P \cdot \log_{2} \Bigg( 1 + \frac{\rho \cdot \mathrm{SNR}}{P} \cdot a_{R} a_{T} \cdot \sum_{m=1}^{M} \sum_{n=1}^{N}  \Bigg. \\
		\label{eq:Cupperbound1}
		&\qquad \qquad \qquad \qquad \qquad \Bigg. \left( \frac{\varepsilon_1}{{\bar{d}_{mn}^2}} + \frac{\varepsilon_2}{{\bar{d}_{mn}^4}} + \frac{\varepsilon_3}{{\bar{d}_{mn}^6}} \right) \Bigg),
\end{align}
by considering a uniform power allocation scheme for all independently parallel transmission channels, where $\rho = \omega^2 \mu^2$ and $\mathrm{SNR} = \frac{\mathcal{P}_{t}}{P a_{R} \sigma_w^2}$ denotes the average transmit SNR with $\mathcal{P}_{t}$ being the total transmit power; the coefficients $\varepsilon_1$, $\varepsilon_2$ and $\varepsilon_3$ are given by
	\begin{align}
		\varepsilon_1 &= \frac{1}{{16 {\pi ^2} }} \left[ 3 - {\rm{trace}} \left( {\frac{{{\bar{\mathbf{d}}_{mn}}{\bar{\mathbf{d}}}_{mn}^T}}{{\bar{d}_{mn}^2}}} \right) \right], \tag{\ref{eq:Cupperbound1}{a}} \label{eq:coefficients(a)} \\
		\nonumber
		\varepsilon_2 &= \frac{1}{{16 {\pi ^2} k_0^2}} \left[ {5 \cdot {\rm{trace}} \left( {\frac{{{\bar{\mathbf{d}}_{mn}}{\bar{\mathbf{d}}}_{mn}^T}}{{\bar{d}_{mn}^2}}} \right) - 3} \right], \tag{\ref{eq:Cupperbound1}{b}} \label{eq:coefficients(b)} \\
		\nonumber
		\varepsilon_3 &= \frac{1}{{16 {\pi ^2} k_0^4}} \left[ {3 \cdot {\rm{trace}} \left( {\frac{{{\bar{\mathbf{d}}_{mn}}{\bar{\mathbf{d}}}_{mn}^T}}{{\bar{d}_{mn}^2}}} \right) + 3} \right]. \tag{\ref{eq:Cupperbound1}{c}} \label{eq:coefficients(c)}
	\end{align}
This upper bound can be generalized to the far-field regions as a special case. where the communication distance exceeds the Rayleigh distance. As such, the high order distance terms ($d_{mn}^{4}$ and $d_{mn}^{6}$) tend to reduce to zeros. More details can be found in \cite{10500751}. 

\begin{figure}[t!]
	\centering
	\includegraphics[width=0.45 \textwidth]{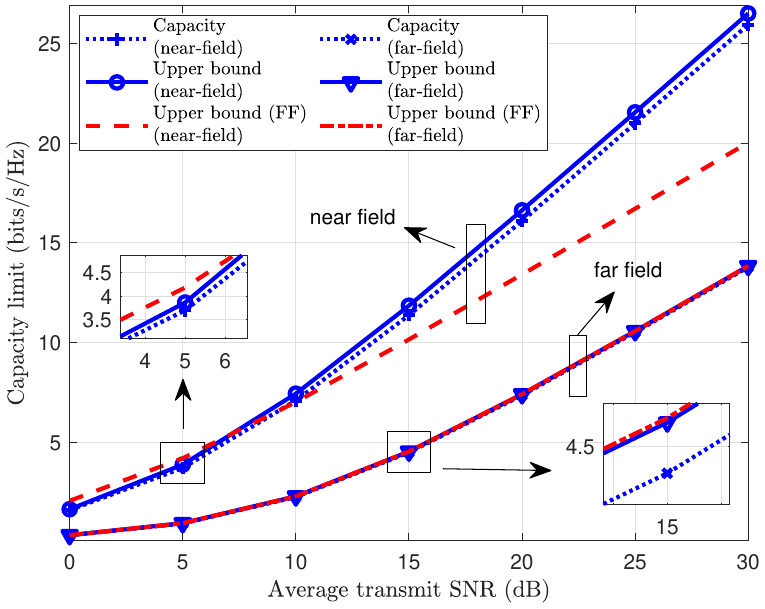}
	%\vspace{-0.5em}
	\caption{Capacity limit versus the transmit average SNR in LoS scenarios.}
	\label{fig:AvePowerAlloc_Main_Capacity_SNR_3DistancesComp}
\end{figure}

The demonstration of channel capacity and its upper bound in both near field and far field cases are given in Fig. \ref{fig:AvePowerAlloc_Main_Capacity_SNR_3DistancesComp}, where ``Capacity" denotes the real value, ``Upper bound" indicates the upper bound obtained by \eqref{eq:Cupperbound1}, and ``Upper bound (FF)" represents the far-field upper bound reduced from \eqref{eq:Cupperbound1}. As seen from the figure the capacity in the near-field region is higher than that in the far-field region, and it is proportional to the SNR. We then observe that the ``Upper bound" provides an effective and tightness bound for the exact capacity in both near-field and far-field scenarios. Besides, the ``Upper bound (FF)" is incapable of providing an effective bound on the exact capacity, especially for the near-field region.

In addition, the complex spatial and polarization correlation in near-field communications requires effective beamforming design \cite{DeSenaIRSAssistedMassiveMIMONOMA2021}. It is proved that the power gain using conventional beamforming is suboptimal to that in near-field beamfocusing using the spherical wave channel model  \cite{DovelosIntelligentReflectingSurfaces2021}. When considering to exploit the polarization diversity in near field, the work \cite{WeiTriPolarizedHolographicMIMO2022} proposed a two-layer precoding scheme. Specifically, a Gaussian elimination method was adopted to eliminate polarization interference and a block diagonalization method was employed to remove inter-user interference. However, such a method requires higher computational complexity, and an unequal efficient power allocation scheme is necessary due to the power imbalance in the three polarizations. 
\color{black}

As demonstrated above, the DOF and capacity of both far-field and near-field HMIMO can be substantially enhanced through various approaches. These include expanding the surface area of transmitters/receivers, integrating near-field communication strategies, and leveraging advanced precoding/beamforming techniques. However, it is important to acknowledge certain limitations. For example,  the compact antennas installed in HMIMO systems require sophisticated and costly precoding/beamforming strategies, which necessitate in-depth exploration.

\subsection{Sphere-Packing Solution} 
As presented in the former section, both DOF and capacity measurement in SIT/KIT framework can be transformed into sphere-packing problems. Therefore, we will evaluate the performance with power constraints through the sphere-packing solution in this part. 

Specifically, the emitted power within a specified area is constrained to remain below a certain threshold.  Consequently, parts of DOF are allocated to limit the energy radiation range, i.e., pattern constraint, while the remaining DOF is utilized for information transmission. Notably, such power constraint decreases the overall capacity of the system. Thus, the balance between energy transmission and information transmission becomes critical in optimizing the system's performance with the power constraint.

\color{black}

\begin{figure}  
	\begin{center}
		{\includegraphics[width=0.5 \textwidth]{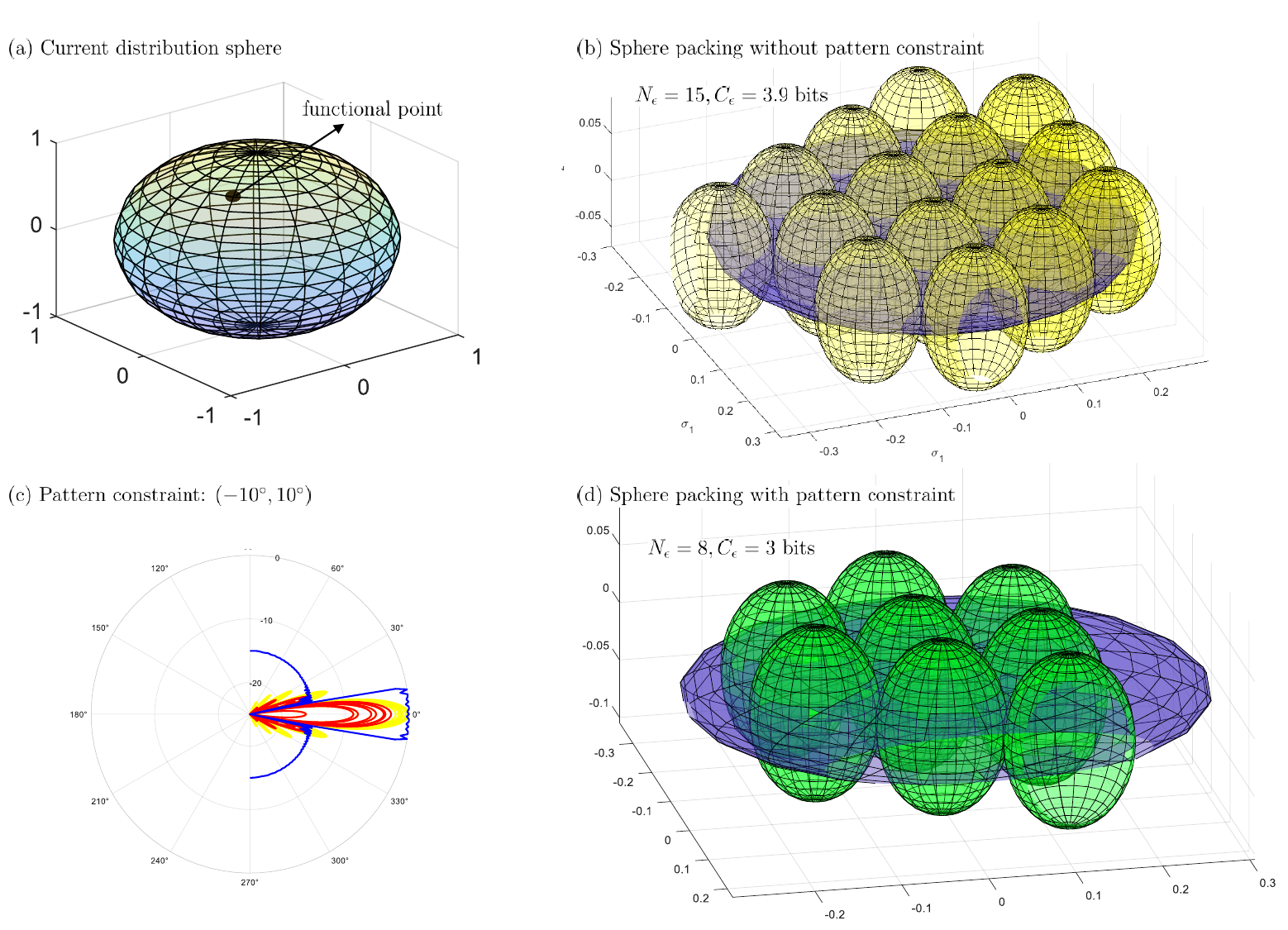}}  
		\caption{The electric line with length $10\lambda$ and the linear observation line with length $6\lambda$ for distance  $10\lambda$. (a) current distribution sphere; (b) Sphere packing without pattern constraint; (c) Pattern constraint: $(-10^\circ, 10^\circ)$; and (d) Sphere packing with pattern constraint  (source codes from \cite{ElectromagneticInformationKolmogorov})}  
		\label{fig:Kolmogorov_graph} 
	\end{center}
\end{figure} 

As shown in Fig.~\ref{fig:Kolmogorov_graph} (source codes from \cite{ElectromagneticInformationKolmogorov}), considering the transmission between the linear electric line with length $10\lambda$ and the observation line with length $6\lambda$ for distance $10\lambda$. In Fig.~\ref{fig:Kolmogorov_graph} (a), each functional point in the sphere represents the specific current distribution. Through the compact operator $\mathcal{A}$, each functional is transformed into a sphere in the field ellipsoid, and there are $15$ non-overlapping balls, thus the capacity is $3.9$ bits, as shown in Fig.~\ref{fig:Kolmogorov_graph} (b). However, if imposing pattern constraints, e.g., the radiation pattern outside $(-10^\circ,10^\circ)$ is set to below a certain value, as shown in Fig.~\ref{fig:Kolmogorov_graph} (c), then the capacity would be reduced to $3$ bits since there are $7$ distinguishable waveforms do not meed pattern constraint, as shown in Fig.~\ref{fig:Kolmogorov_graph} (d).

It is imperative to acknowledge that superdirective HMIMO systems are exceptions.  Specifically, a superdirective source permits the presence of arbitrarily high reactive energy. Consequently, the resulting radius of the source hyper-ball $\hat{E}$ could become substantially large. In such a configuration, even the small singular values of the integral operator $\mathcal{A}$ may make significant contributions to communications, especially for large semi-axis $\sigma_k \hat{E}$ within the ellipsoid $Y$ \cite{MiglioreElectromagneticsInformationTheory2008}. As a result, the superdirective source has the potential to achieve theoretically high $\epsilon$-entropy.

%%%%%%%%%%%%%%%%%%%%%%%%%%%%%%%%%%%%%%%%%%%%%%%%%%%%%%%%%%%%%%%%%%%%%%%%%%%%%%%%%%%%%%%%%%%%%%%%%%%%%%%%%%%%%%%%%%%%%%%%%%%%%%%%%%%%%%%%%%%%%%%%%%%%%%%%%%%%%%%%%%%%%%%%%%%%%%%%%%%%%%%%%%%%%%

%%%%%%%%%%%%%%%%%%%%%%%%%%%%%%%%%%%%%%%%%%%%%%%%%%%%%%%%%%%%%%%%%%%%%%%%%%%%%%%%%%%%%%%%%%%%%%%%%%%%%%%%%%%%%%%%%%%%%%%%%%%%%%%%%%%%%%%%%%%%%%%%%%%%%%%%%%%%%%%%%%%%%%%%%%%%%%%%%%%%%%%%%%%%%%

\section{Future Research Directions}
The HMIMO-oriented EIT introduces novel opportunities for exploring performance within the framework of practical physical constraints. In this context, we offer insights for prospective research directions in EIT for HMIMO systems. 

\subsection{Mechanism of EM Environment Control and Interactions}
Utilizing HMIMO systems to offer significant flexibility and adaptability for achieving multiple communication goals across various scenarios is a complex task, necessitating a thorough examination of the mechanisms that govern favorable environment regulation and interactions. 

To begin, it is important to note that HMIMO systems are expected to improve performance (e.g., DOF, capacity, and channel rank) substantially while bringing difficulty in environment control. However, such improvement introduces complexities in environmental control.  For example, rotating the HMIMO plane may bring more benefits in terms of capacity compared to spreading it over a wide area, as shown in \cite{ChengRISAidedWirelessCommunications2022}. In addition, harnessing the unevenly spaced patch antennas over HMIMO to control complicated EM environments is also anticipated while remaining untapped. 
 
Moreover, investigating the optimal number of patch antennas for environment regulation over HMIMO systems appeals to research interests, since constructing a continuous structure is impractical.   Specifically, the performance gain (e.g., DOF, capacity) ceases to increase constantly with the increase of the number of patch antennas \cite{WeiChannelModelingMultiUser2023,YuanEffectsMutualCoupling2023,10083296}.  However, theoretically analyzing the optimal number of antennas given a fixed area is not fully investigated yet.    

In essence, the intricacies of managing the EM environment, along with its associated wave interactions, remain unresolved, not to mention the related real-time control and security issues. 

\subsection{Design and Implementation of 3D HMIMO-based Superdirective Antennas}
Newly envisioned architectures, i.e., 3D HMIMO-based superdirective antennas, are expected to break EM limits such as the Hannan limit, and demonstrate remarkable potential in capacity enhancement. Nevertheless, the pursuit of such designs encounters several formidable challenges.

Primarily, combating the serious energy efficiency loss inherent in such a structure is imperative. Different from traditional 1D (linear) and 2D (planar) antenna arrays, the individual patch antenna in 3D HMIMO systems suffers from much more severe coupling effects stemming from surrounding patch antennas in 3D space, resulting in excessively low energy efficiency and array gain. Consequently, devising 3D HMIMO-based superdirective antennas remains an unresolved quandary. 

Moreover, while the promising superdirectivity is anticipated to cater to practical scenarios in the context of 3D HMIMO-based antennas, a suitable matching network is difficult to implement. More precisely, within such a 3D framework, perfectly matching the input impedance and output impedance is problematic in transferring almost all power to the load.  Therefore, from the perspective of hardware realization, designing an effective feed circuit and matching network are challenging in achieving 3D HMIMO-based superdirective antennas.

\subsection{Efficient Field Sampling Schemes}
Despite the existence of various field sampling methods, such as uniform and nonuniform, planar and spherical sampling, an optimal sampling method that can accurately approximate the field with the minimum number of sampling points has not been determined yet. 

First, it should be noted that the optimal sampling method in the far-field zone may differ from that in the near-field region. While the conventional half-wavelength rectangular sampling method is less effective than (elongated) hexagonal sampling methods in the far-field zone due to the inclusion of extra information (reactive energy) \cite{PizzoNyquistSamplingDegrees2022}, this does not necessarily imply that the conventional rectangular sampling method is also ineffective in the near-field zone. Furthermore, it is important to explore other sampling values in the rectangular sampling method. For instance, considering a sampling interval less than $\lambda/2$ outside a specified region may prove to be more effective in capturing critical information in certain scenarios. 

Next,  the utilization of a nonuniform sampling method holds promise for reducing the number of samples required. The author in \cite{MiglioreSamplingElectromagneticField2015} proposed a nonuniform random sampling method for sparse EM sources to reduce the number of measure data with the aid of a priori information. The space dimension is initially reduced, and the amount of information is then evaluated using Kolmogorov $\epsilon$-entropy. The findings substantiate that nonuniform sampling proves more effective than the conventional $\lambda/2$ equispaced sampling strategy when employing geometrical data.  Although nonuniform sampling has demonstrated its effectiveness, the problem of designing sampling patterns in the near-field and far-field zones is still in the early stages of research. It is suggested that incorporating geometrical information of the source and surrounding environment may prove beneficial in designing a cost-efficient yet effective nonuniform sampling strategy. 

What's more, the transformation between the far field and near field is also fascinating,  especially reconstructing the near field with the less far-field measurements. From a technical standpoint, the near-field information contains more abundant data compared to the far-field context. However, the sampling density in such a case is high. Even if the bandlimited field is adopted to approximate the near field with lower sampling points, it still induces substantial approximation error due to truncation error.   Therefore, sampling the near field with the minimum number of measurements to reconstruct the far field becomes a meaningful pursuit.

\subsection{Impacts of Scatters and EM Noise} 
The current performance analysis for HMIMO systems is mainly categorized into two types: conventional MIMO analysis and antenna analysis. In traditional MIMO analysis, the mutual coupling and polarization effects are normally modeled as a random process, simplifying the analysis using probabilistic assumption while neglecting the physical influences, such as multi-path and scattering environments, resulting in the underestimation of HMIMO performance, such as capacity and power gain.   On the other hand, antenna analysis takes radiation patterns and efficiencies into consideration. However, this approach falls short of offering a complete interpretation of communications in a more explicable manner. For example, although the $Q$ factor and directivity gain describe the performance of antenna systems in practical deployment, the information content from the perspective of communications in such a setting is still unclear. Therefore, integrating the above concerns into EIT is still challenging, and the complex environment further increases the difficulty of related analysis.  

Primarily, multiple scatterers can be present, resulting in the propagation of EM waves from the transmitter to the receiver through various paths. Each scatterer also acts as a secondary source, further contributing to the radiation of EM waves. However, many existing works in antenna analysis tend to focus solely on the direct link between the transmitter and receiver, disregarding the presence and impact of scatterers. In these analyses, dyadic or scalar Green's functions are often employed to model the transmission process. The involvement of secondary sources or higher-order sources is only discussed in very few works \cite{electronics11193232,10016303,10012628}, despite their common occurrence in wireless communications.

Furthermore, noise modeling in traditional MIMO systems typically assumes Gaussian variables. However, in the context of EM noise, the scenario becomes more intricate, encompassing radiation interference from undesired sources and measurement noise stemming from physical antenna configurations. Given the complexity of EM noise, it is essential to properly address noise modeling in performance analysis and applications. Accurate noise modeling allows for a more realistic evaluation of system performance, leading to the development of effective encoding and modulation schemes that can mitigate the impact of noise in HMIMO systems. 

\subsection{Accurate Capacity Region Evaluation}
The precise representation of the capacity region is significant in quantifying HMIMO systems, however,  several unsolved problems persist in the evaluation process, especially for near-field HMIMO communications.

First of all,  the EM waves interaction in the near-field region is complicated to be numerically described. As previously discussed, unlike the planar waves in the far-field zone, there are numerous spherical waves in the near-field region, incorporating rich information on phase and distance. Therefore, measuring such information accurately is laborious. Typically, the truncation methods are employed to represent the near-field region with a finite number of dominant waves, aiming at reducing the computational complexity. However, determining the number of dominant modes in near-field communications is normally infeasible, and improper truncation would introduce notable approximation errors. Under such a circumstance, the capacity region depiction of near-field HMIMO systems becomes a formidable task.

In addition, near-field HMIMO is anticipated to exhibit excellent superdirectivity, yielding excessively high radiation power, and enabling abundant encoding/decoding schemes. Nevertheless, the traditional capacity region technique is not applicable due to the consideration of inexhaustible scheduling policies in the derivation of optimal capacity region.  

Lastly, the channel in traditional communications is modeled as a probabilistic process characterized by means and variances, resulting in direct computation of the channel's eigenvalues and facilitating capacity region evaluation. Nonetheless, this modeling approach may overlook the antenna configuration and fail to fully interpret the physical environment, leading to an erroneous estimation of capacity.

\subsection{Excitation/Field Encoding and Modulation}
To minimize temporal costs, it is essential to fully exploit the available spatial resources. The information content carried by the field can be measured by DOF, representing the number of distinguishable patterns. Each pattern corresponds to an independent information component that is generated through the manipulation of current density. Therefore, exploring the linear or nonlinear relationship between the excitation current and the resulting radiated field is of particular interest. However, there are three main issues that need to be addressed in this research.

First, it is essential to note that the optimal current distribution is generally not unique, i.e.,  each pattern can be excited by multiple possible excitations. As a result, the mapping from the current distribution to the specified antenna pattern is not definitively defined. To address this issue, the problem can be divided into two steps. In the first step, the goal is to find the available excitations based on the kernel functions (e.g., $\mathcal{A}$) and the desired pattern. Subsequently, in the second step, the objective current distribution is derived by employing mathematical methods and considering constraints related to directivity or the $Q$ value.  

Then, the distinguishable field also depends on the surrounding environment and physical configuration. The radiation pattern, for instance, can exhibit variations in the near-field zone compared to the far-field region, especially for evanescent waves. Additionally, closely spaced antennas may lead to radiation pattern deformations, causing certain patterns to become indistinguishable from others. As a result, it is crucial to take these physical constraints into account during antenna design. By incorporating these constraints, the effectiveness of the encoding and modulation process can be enhanced.   Once we obtain the current distribution that produces the distinguishable field pattern, the information bits can be encoded with excitations. Through such an encoding scheme, the received signal can be retrieved from the observed field pattern.

Furthermore, distinguishable patterns can be leveraged in spatial modulation techniques. For example, similar to index modulation, the index of excited field patterns can carry additional information beyond what is embedded in the individual excited patterns. Additionally, the polarization domain can be exploited in spatial modulation by combining different polarization states to convey more information. The spatial information itself can also contribute to the modulation process. For example, users at different locations may observe distinct waveforms, allowing for location-based information transmission. By incorporating these spatial modulation strategies, HMIMO systems can achieve more efficient and versatile data transmission, making use of the diverse characteristics of the spatial domain to enhance communication performance. 

It should be noted that the superdirectivity source with arbitrarily high reactive energy is a unique case. Geometrically, consider an antenna enclosed within a sphere with an arbitrarily large energy radius $\hat{E}$. In this context, the ellipsoid generated from the operator $\mathcal{A}$ would have arbitrary large semi-axes, admitting high $\epsilon$-entropy, i.e., $C_{\epsilon}$. Theoretically, such a source could achieve an arbitrarily long length $2^{C_{\epsilon}}$-codebook without delay \cite{MiglioreRoleNumberDegrees2006}.

In addition to the aforementioned directions, there are still several other unmentioned research avenues to be explored. For example, there is a need for low-cost beamforming designs, which can lead to cost-effective and efficient HMIMO systems. Image reconstruction techniques are also vital to enable accurate representation and visualization of EM fields, facilitating better decision-making processes. Additionally, activity detection methods can be developed to detect and monitor signal transmissions in the HMIMO system, enhancing system performance and security.

These unexplored areas present exciting opportunities for further research and development in HMIMO systems, promising advancements in the design, performance, and practical applications of these advanced wireless communication systems.

\section{Conclusions} \label{sec:conclusions}
We have conducted a comprehensive review of the HMIMO-oriented EIT framework. This framework seamlessly integrates the information-theoretical constraints and the EM constraints into HMIMO systems. Regarding the physical design,  the presence of inevitable mutual coupling in HMIMO systems results in reduced directivity gain and an incremental quality factor. To address this oversight and account for these physical constraints, the mathematical assessment of coupling effects is undertaken through reflection coefficients or the scattering matrix. 

To further illustrate the input-output relationship in HMIMO scenarios, we presented three EM-compliant channel models: the Fourier plane wave expansion model, the dyadic Green's function model, and the stochastic Green's function model. The implementation and subsequent performance evaluation based on these models are presented. Notably, all models consistently demonstrated that a larger transmit/receive surface, greater spacing, and shorter distance contribute to the DOF and capacity gains. 

{ Both the probabilistic SIT and deterministic KIT are applicable to the temporal domain and spatial domain, attributing to the symmetry between space and time. However, since there is a slight asymmetry between the temporal domain and spatial domain, the KIT is still necessary as a complementary tool to SIT. Based on these two methods, the temporal and spatial DOF can be derived, and it is observed that increasing the aperture and solid angles are beneficial to spatial DOF improvement.  }

In addition to reviewing and evaluating HMIMO systems with practical constraints, here we are going to share some crucial insights and lessons that we have learned in the research and development:
\begin{itemize} 
    \item \textbf{Optimizing superdirectivity with strong coupling:} Although small element spacing generates serious coupling, such strong coupling effects prove advantageous in achieving superdirectivity. Consequently, adjusting the spacing and distributions of excitation facilitates the realization of superdirective HMIMO systems. However, this design may lead to lower element efficiency, necessitating careful consideration in practical design. 
    \item \textbf{Applicable scenarios for channel models:} The three presented channel models accommodate distinct scenarios, each demanding varying computational costs. The Fourier plane wave expansion model efficiently models rich scattering scenarios at low computational expense. The dyadic Green's function model simulates LoS communications at high computational cost. The stochastic Green's function model effectively models arbitrary scattering propagation environments but incurs a higher computation cost.   
    \item \textbf{Information measurements in near-field communications:} The most of research delves into the DOF of far-field communications. Considering that the near field has a larger spatial DOF and eigenvalue distribution is different from the far-field case, the information measurement in near-field communications is necessary.  
    \color{black}
    % \item \textbf{Information measurements in spatial and temporal domain:} KIT is capable of measuring spatial domain information, whereas SIT mainly focuses on time domain information measurement. Technically speaking, the number of distinguishable patterns in the radiated field is the spatial DoF, which is closely related to spatial bandwidth and observation length in KIT. Given constrained antenna size, spatial DoF is also limited, elucidating that it cannot constantly increase.  At the same time, since the temporal bandwidth in SIT is embedded in physical constraints, how to jointly improve spatial and temporal DoF is still challenging in HMIMO systems design.  
\end{itemize}
To conclude, the exploration of HMIMO within the EIT structure is expected to establish a consistent connection between the practical applications and theoretical analysis, integrating physical constraints. The probabilistic KIT and functional SIT methodologies are involved in laying the theoretical foundations of HMIMO systems.  Such an incorporation opens up new avenues for investigating HMIMO performance. Despite there being substantial challenges in making the HMIMO-oriented EIT fully operational, our survey stands as a robust presentation, affirming feasibility and providing potential directions for further exploration.

% \newpage


%\bibliographystyle{Exported Items}
\bibliographystyle{IEEEtran.bst}
\bibliography{EIT_PIEEE}
%\bibliography{Exported Items}

\end{document}